\documentclass[aps,twocolumn,showpacs,superscriptaddress]{revtex4}

\usepackage{graphicx}
\usepackage{amsfonts}
\usepackage{amssymb}
\usepackage{amsbsy}
\usepackage{amsmath}
\usepackage{mathrsfs}
\usepackage{latexsym}
\usepackage{natbib}
\usepackage{bm}
\usepackage{subfigure} 
\usepackage{color}
\usepackage{wasysym}
\usepackage{mathbbol}
\usepackage{bigints}

\usepackage{hyperref}
\hypersetup{pdftex,colorlinks=true,allcolors=blue}
\usepackage{hypcap}

\allowdisplaybreaks

\def\rh{r_H}
\def\ts{t_s}
\def\rstar{r_{\star}}

\def\tsj{t_{s_{j}}}
\def\tsl{t_{s_{l}}}

\def\rstarj{r_{\star_{j}}}
\def\rstarl{r_{\star_{l}}}

\def\ca{c_{_{A}}}
\def\cb{c_{_{B}}}

\begin{document}

\title{Entanglement harvesting from conformal vacuums between two Unruh-DeWitt detectors moving 
along null paths}

\author{Subhajit Barman}
\email{subhajit.b@iitg.ac.in}

\author{Dipankar Barman}
\email{dipankar1998@iitg.ac.in}

\author{Bibhas Ranjan Majhi}
\email{bibhas.majhi@iitg.ac.in}

\affiliation{Department of Physics, Indian Institute of Technology Guwahati, 
Guwahati 781039, Assam, India}


\date{\today}

\begin{abstract}
It is well-known that the $(1+1)$ dimensional Schwarzschild and spatially flat FLRW spacetimes are conformally flat. This work examines entanglement harvesting from the conformal field vacuums in these spacetimes between two Unruh-DeWitt detectors, moving along outgoing null trajectories. In $(1+1)$ dimensional Schwarzschild spacetime, we considered the Boulware and Unruh vacuums for our investigations. 
%
%
In this analysis, one observes that while entanglement harvesting is possible in $(1+1)$ dimensional Schwarzschild and $(1+3)$ dimensional de Sitter spacetimes, it is not possible in the $(1+1)$ dimensional de Sitter background for the same set of parameters when the detectors move along the same outgoing null trajectory. The qualitative results from the Boulware and the Unruh vacuums are alike. 
%
%
Furthermore, we observed that the concurrence depends on the distance $d$ between the two null paths of the detectors periodically, and depending on the parameter values, there could be entanglement harvesting shadow points or regions.
We also observe that the mutual information does not depend on $d$ in $(1+1)$ dimensional Schwarzschild and de Sitter spacetimes but periodically depends on it in $(1+3)$ dimensional de Sitter background.
We also provide elucidation on the origin of the harvested entanglement.

\end{abstract}

\maketitle

\section{Introduction}\label{Introduction}

The fascinating phenomenon of quantum entanglement has garnered significant interest in the 
scenarios of relativistic particles in flat and curved spacetimes \cite{Reznik:2002fz, Lin:2010zzb, 
Ball:2005xa, Cliche:2009fma, MartinMartinez:2012sg, Salton:2014jaa, Martin-Martinez:2015qwa, 
Cai:2018xuo, Menezes:2017oeb, Menezes:2017rby, Zhou:2017axh, Floreanini:2004, Pan:2020tzf, 
Menezes:2015veo, Cong:2020nec, Chowdhury:2021ieg, Kane:2021rhg, Barman:2021igh}. Studying the 
dynamics of entangled particles in flat and curved spacetimes \cite{FuentesSchuller:2004xp, 
Menezes:2017oeb, Hu:2015lda, Barman:2021oum} has presented many enthralling perspectives. Another 
interesting facet is the possibility of entanglement extraction \cite{SUMMERS1985257, Summers-jmp, 
VALENTINI1991321, Reznik:2002fz, Reznik:2003mnx, Salton:2014jaa, Henderson:2017yuv, 
Henderson:2020ucx, Stritzelberger:2020hde} from the quantum field into atoms or other suitable 
systems interacting with the field, which is known as entanglement harvesting. The prospect of 
utilizing this harvested entanglement in quantum information-related purposes \cite{Hotta:2008uk, 
Hotta:2009, Frey:2014} to solidify experimental verification of many theoretical predictions has 
made the harvesting a desirable arena to venture further.

Reznik solidified the possibility of harvesting entanglement from the vacuum of the background 
quantum field in \cite{Reznik:2002fz, Reznik:2003mnx} where he recognized entanglement extraction in 
a system of two anti-parallelly accelerated two-level atomic detectors. Reznik's work signifies the 
role of the quantum field vacuum in entanglement extraction as one observes entanglement harvesting 
between two causally disconnected accelerated detectors with no possibility of classical 
correlation. His works and the subsequent works \cite{VerSteeg:2007xs, Pozas-Kerstjens:2015gta, 
Kukita:2017etu, Martin-Martinez:2015qwa, Pozas-Kerstjens:2016rsh, Martin-Martinez:2013eda, 
Lorek:2014dwa, Brown:2014pda, Sachs:2017exo, Trevison:2018ear, Li:2018xil} usually deal with a 
system composed of two detectors that interact with the background field and are in an initial 
uncorrelated state. One can perceive any entanglement harvested between the two detectors by 
checking whether the partial transposition of the final system reduced density matrix has negative 
eigenvalues \cite{Peres:1996dw, Horodecki:1996nc}. We mention that one realizes this reduced density 
matrix as the system's final density matrix with the field degrees of freedom traced out. This 
formulation was further improved in \cite{Koga:2018the, Ng:2018ilp, Koga:2019fqh, Tjoa:2020eqh, 
Foo:2021gkl, Gallock-Yoshimura:2021yok}, where the authors considered proper time ordering into the 
picture. It resulted in the introduction of the Feynman propagator rather than the Wightman function 
in some places of the estimated eigenvalues.
We mention that the entanglement harvesting condition and the measure of the harvested entanglement 
depends on the background geometry \cite{VerSteeg:2007xs, Pozas-Kerstjens:2015gta, Kukita:2017etu, 
Henderson:2018lcy, Ng:2018drz, Gallock-Yoshimura:2021yok}, boundary conditions 
\cite{Henderson:2018lcy, Cong:2018vqx, Cong:2020nec}, the two detectors' trajectories 
\cite{Salton:2014jaa, Gallock-Yoshimura:2021yok, Barman:2021bbw}, etc.

Recently, entanglement harvesting in black hole spacetimes \cite{Henderson:2017yuv, Tjoa:2020eqh, 
Robbins:2020jca, Gallock-Yoshimura:2021yok} has gained much interest. In this context, which vacuum 
to choose in these spacetimes to formulate quantum field theory and obtain the Green's functions 
corresponding to observers in different trajectories has a general notion 
\cite{Gallock-Yoshimura:2021yok}. Namely, in an $(1+1)$ dimensional Schwarzschild background, these 
are the \emph{Boulware}, \emph{Unruh}, and \emph{Hartle-Hawking vacuums} 
\cite{Gallock-Yoshimura:2021yok}. They denote conformal vacuums corresponding to different 
coordinate choices in the background spacetime, as all $(1+1)$ dimensional spacetimes are 
conformally flat \cite{book:Birrell, Das:2019aii}. We mention that the spatially flat 
\emph{Friedman-Lema\^itre-Robertson-Walker} (FLRW) metric is also conformally flat. One should also 
note that nontrivial findings for entanglement harvesting is closely related to semi-classical 
particle creation. However, in a $(1+1)$ dimensional Schwarzschild background, a static observer 
does not see the Boulware vacuum to be particle creating.  Likewise, with static detectors, one 
obtains entanglement harvesting related observations from the Boulware vacuum that are similar to 
the flat spacetime \cite{Gallock-Yoshimura:2021yok}.  Furthermore, in \cite{Scully:2017utk}, it was 
shown that a radially in-falling detector in a time-like trajectory observes the Boulware vacuum to 
be thermal. These freely-falling detectors also keep nontrivial entanglement harvesting profiles 
from the Boulware vacuum, as was pointed out in \cite{Gallock-Yoshimura:2021yok}. Subsequently, in 
\cite{Chakraborty:2019ltu, Dalui:2020qpt}, one also observes the thermal nature of the Boulware 
vacuum and that of the conformal vacuum in the FLRW spacetime, with detectors infalling in null-like 
trajectories. These facts motivated us to study entanglement-related phenomena with detectors in 
null paths in the $(1 + 1)$ dimensional Schwarzschild and FLRW spacetimes. In particular, we shall 
investigate the entanglement harvesting from the above-mentioned conformal vacuums in these 
backgrounds. This consideration of null trajectories for detectors in black hole spacetimes is 
interesting from the point of view that it may shed light on the entanglement harvested between 
light-like particles emitted from astrophysical bodies along null paths. In this regard, we mention 
that these are situations related to the Hawking effect, as originally, Hawking, in his pioneering 
work \cite{hawking1975} elucidated the black hole evaporation in terms of field modes in ingoing and 
outgoing null trajectories. Thus our consideration of entanglement harvesting along null 
trajectories may open up directions to provide new insights towards the understanding of the black 
hole information loss paradox and also remains relevant from the cosmological point of view.

As we have already stated, in this work, we study the entanglement harvesting conditions with 
Unruh-DeWitt detectors in null trajectories in the background of $(1+1)-$dimensional Schwarzschild 
black holes and the FLRW spacetime. The Unruh-DeWitt detectors conceptualized to understand the 
Unruh effect \cite{Unruh:1976db, Unruh:1983ms, 2010grae.book}, are point-like two-level hypothetical 
detectors. In particular, we have considered estimating the entanglement harvesting condition in the 
de Sitter era of the FLRW universe. We also mention that we obtain the entanglement harvesting 
condition for each specific field mode frequency.
%
%
We observe that entanglement harvesting is indeed possible from the conformal vacuums in $(1+1)$ 
dimensional Schwarzschild, $(1+1)$ and $(1+3)$ dimensional FLRW spacetimes with two detectors in 
outgoing null paths. In all these cases except the $(1+1)$ dimensional de Sitter spacetime, the concurrence, which measures the harvested entanglement, 
peaks at a particular field mode frequency when the detectors move along the same path. Furthermore, the interesting phenomenon is that this 
concurrence is, in all these cases, periodically dependent on the distance $d~(\neq0)$ between different 
outgoing null paths of the two detectors. 
%
%
In particular, we observe that in $(1+1)$ dimensional Schwarzschild and $(1+3)$ dimensional de Sitter backgrounds, there are periodic zero entanglement harvesting regions and points in the distance $d$ corresponding to low and high values of the field frequency respectively. In contrast, for $(1+1)$ dimensional de Sitter spacetime, one only perceives zero entanglement harvesting regions. Thus providing the notion of the so-called entanglement shadow points and shadow regions \cite{Henderson:2017yuv, Robbins:2020jca}, from where one cannot harvest entanglement. However, unlike \cite{Henderson:2017yuv, Robbins:2020jca} where the entanglement shadow region appears near the black hole event horizon, in our case, these regions are periodic in the distance $d$.
On the other hand, we observe that the mutual information in $(1+1)$ dimensional 
Schwarzschild and de Sitter spacetimes are independent of this distance $d$ between the null paths 
of the two detectors. However, in $(1+3)$ dimensional de Sitter spacetime, the mutual information is 
dependent on $d$. We also observe that it is periodic with the distance $d$ between different 
outgoing null paths of the two detectors.

\color{blue}
\color{black}

We organize this paper in the following way. In Sec. \ref{Sec:model-set-up} we start with a brief overview of the model set-up for entanglement harvesting with two two-level point-like atomic detectors interacting with the background massless real scalar field through monopole couplings. This section elucidates the entanglement harvesting condition, the measure of the harvested entanglement (the concurrence), and the total correlation (the mutual information). In Sec. \ref{Sec:conformal-vacuum} we illuminate the importance of conformal vacuum in curved spacetimes to formulate quantum field theory and briefly discuss the construction of these vacuums in  FLRW and $(1+1)$ dimensional Schwarzschild spacetimes. Subsequently, in Sec. \ref{Sec:Greens-fn-null-paths} we consider a massless conformally invariant scalar field in these spacetimes and construct the necessary Green's functions for observers in null trajectories. In Sec. \ref{Sec:entanglement-harvesting} we study the entanglement harvesting condition from the conformal vacuum in the spacetimes as mentioned earlier and also study the entanglement measure concurrence.
Furthermore, in Sec. \ref{Sec:Entanglement-estimator} we provide a discussion on the origin of the harvested entanglement. In particular, we try to understand how much of the harvested entanglement is from the background field state, considered truly harvested, and how much is due to communication between the detectors.
In the following section \ref{Sec:Mutual-information} we investigate the mutual information in the considered spacetimes. We conclude this work in Sec. \ref{Sec:discussion} with a discussion of our findings.\vspace{0.5cm}

\section{Model set-up}\label{Sec:model-set-up}

In this section, we will briefly discuss the model set up for entanglement harvesting, emphasizing 
the necessary notations and symbols of the different system parameters. This model was introduced 
initially in \cite{Koga:2018the, Ng:2018ilp, Koga:2019fqh}, which keeps into consideration the 
necessary time ordering in the construction of the Green's functions.

In this model set-up, one considers two point-like two-level Unruh-DeWitt detectors, each carried by 
a distinct observer. One of these observers is Alice denoted by $A$, and another one is Bob denoted 
by $B$. We denote the detector states as $|E_{n}^{j}\rangle$, with the symbols delineating the 
$n^{th}$ state of $j^{th}$ detector, i.e., we have $j=A,B$ and $n=0,1$. We also consider these 
states to be non degenerate so that $E_{1}^{j}\neq E_{0}^{j}$, and the difference $\Delta E^{j} = 
E_{1}^{j}- E_{0}^{j}>0$ signifies the transition frequency. Furthermore, we consider a massless, 
minimally coupled real scalar field $\Phi(X)$ interacting with these detectors through monopole 
couplings $m^{j}(\tau_{j})$. One can express the corresponding interaction action as
\begin{eqnarray}\label{eq:TimeEvolution-int}
 S_{int} &=& \int_{-\infty}^{\infty} \bigg[ 
\ca\kappa_{A}(\tau_{A})m^{A}(\tau_{A}) \Phi\left(X_{ A}(\tau_{A})\right) 
d\tau_{A} \nonumber\\ 
&& +~ \cb \kappa_{B}(\tau_{B})m^{B}(\tau_{B}) \Phi\left(X_{ 
B}(\tau_{B})\right) d\tau_{B}\bigg]~,
\end{eqnarray}
where, $c_{j}$, $\kappa_{j}(\tau_{j})$, and $\tau_{j}$ respectively denote the couplings between the 
individual detectors and the scalar field, the switching functions, and the individual detector 
proper times. We consider the initial detector field state in the asymptotic past to be $|in\rangle 
= |0\rangle |E_{0}^{A}\rangle |E_{0}^{B}\rangle$, where $|0\rangle$ denotes the field's ground 
state. Then the final detector field state in asymptotic future will be $|out\rangle = T\left\{e^{i 
S_{int}}|in\rangle\right\}$, where $T$ signifies time ordering. One can get the explicit expression 
of this final state by treating the coupling constants $c_{_{j}}$ perturbatively. In this way and by 
tracing out the final field degrees of freedoms one obtains the final detector density matrix in the 
basis of $\big\{|E_{1}^{A}\rangle |E_{1}^{B}\rangle, |E_{1}^{A}\rangle |E_{0}^{B}\rangle, 
|E_{0}^{A}\rangle |E_{1}^{B}\rangle, |E_{0}^{A}\rangle |E_{0}^{B}\rangle\big\}$ as
\begin{widetext}
\begin{equation}\label{eq:detector-density-matrix}
 \rho_{AB} = 
 {\left[\begin{matrix}
 0 & 0 & 0 & \ca \cb\varepsilon\\~\\
0 & \ca^2P_{A} & \ca\cb P_{AB} & \ca^2 W_{A}^{(N)}+\ca\cb W_{A}^{(S)}\\~\\
0 & \ca\cb P_{AB}^{*} & \cb^2P_{B} & \cb^2 W_{B}^{(N)}+\ca\cb W_{B}^{(S)}\\~\\
\ca\cb\varepsilon^{*} & \ca^2 W_{A}^{(N){*}}+\ca\cb W_{A}^{(S){*}} & \cb^2 
W_{B}^{(N){*}}+\ca\cb W_{B}^{(S){*}} & 
1-(\ca^2 P_{A}+\cb^2 P_{B})
 \end{matrix}\right]}
 +\mathcal{O}(c^4)~,
\end{equation}
\end{widetext}
where, $P_{j}$, $\varepsilon$, $W_{j}^{(N)}$, and $W_{j}^{(S)}$ 
are explicitly expressed as 
\begin{eqnarray}
P_{j} &=& |\langle E_{1}^{j}|m_{j}(0)|E_{0}^{j}\rangle|^2~
\mathcal{I}_{j}\nonumber
\end{eqnarray}
\begin{eqnarray}
\varepsilon &=& \langle E_{1}^{B}|m_{B}(0)|E_{0}^{B}\rangle\langle 
E_{1}^{A}|m_{A}(0)|E_{0}^{A}\rangle 
\mathcal{I}_{\varepsilon}\nonumber
\end{eqnarray}
\begin{eqnarray}
P_{AB} &=& \langle E_{1}^{A}|m_{A}(0)|E_{0}^{A}\rangle \langle 
E_{1}^{B}|m_{B}(0)|E_{0}^{B}\rangle^{\dagger} 
\mathcal{I}_{AB}\nonumber
\end{eqnarray}
\begin{eqnarray}
W_{j}^{(N)} &=& \langle E_{1}^{j}|m_{j}(0)|E_{0}^{j}\rangle\Big[\left(\langle 
E_{1}^{j}|m_{j}(0)|E_{1}^{j}\rangle -\right. \nonumber\\
~&& \left.\langle 
E_{0}^{j}|m_{j}(0)|E_{0}^{j}\rangle\right)
\mathcal{I}_{j,1}^{(N)}-i \langle 
E_{0}^{j}|m_{j}(0)|E_{0}^{j}\rangle
\mathcal{I}_{j,2}^{(N)}\Big]\nonumber
\end{eqnarray}
\begin{eqnarray}\label{eq:all-PJs}
W_{j}^{(S)} &=& -i\langle E_{1}^{j}|m_{j}(0)|E_{0}^{j}\rangle\langle 
E_{0}^{j'}|m_{j'}(0)|E_{0}^{j'}\rangle
\mathcal{I}_{j}^{(S)}~,
\end{eqnarray}
where $j'\neq j$ and the quantities $\mathcal{I}$'s are given by 
\begin{eqnarray}
 \mathcal{I}_{j} &=& \int_{-\infty}^{\infty}d\tau'_{j} 
\int_{-\infty}^{\infty}d\tau_{j}~e^{-i\Delta E^{j}(\tau'_{j}-\tau_{j})} 
G_{W}(X'_{j},X_{j}),\nonumber
\end{eqnarray}
\begin{eqnarray}
\mathcal{I}_{\varepsilon} &=& -i\int_{-\infty}^{\infty}d\tau'_{B} 
\int_{-\infty}^{\infty}d\tau_{A}~\scalebox{0.91}{$e^{i(\Delta 
E^{B}\tau'_{B}+\Delta E^{A}\tau_{A})} G_{F}(X'_{B},X_{A}),$}\nonumber
\end{eqnarray}
\begin{eqnarray}
\mathcal{I}_{AB} &=& \int_{-\infty}^{\infty}d\tau'_{B} 
\int_{-\infty}^{\infty}d\tau_{A}~\scalebox{0.91}{$e^{i(\Delta 
E^{A}\tau_{A}-\Delta E^{B}\tau'_{B})} G_{W}(X'_{B},X_{A}),$}\nonumber
\end{eqnarray}
\begin{eqnarray}
\mathcal{I}_{j,1}^{(N)} &=& \int_{-\infty}^{\infty}d\tau'_{j} 
\int_{-\infty}^{\infty}d\tau_{j}~\scalebox{0.91}{$e^{i\Delta 
E^{j}\tau_{j}}~ \theta(\tau'_{j}-\tau_{j})G_{W}(X'_{j},X_{j}),$}\nonumber
\end{eqnarray}
\begin{eqnarray}
\mathcal{I}_{j,2}^{(N)} &=& \int_{-\infty}^{\infty}d\tau'_{j} 
\int_{-\infty}^{\infty}d\tau_{j}~\scalebox{0.91}{$e^{i\Delta 
E^{j}\tau_{j}}~G_{R}(X_{j},X'_{j}),$}\nonumber
\end{eqnarray}
\begin{eqnarray}\label{eq:all-integrals}
\mathcal{I}_{j}^{(S)} &=& \int_{-\infty}^{\infty}d\tau'_{j'} 
\int_{-\infty}^{\infty}d\tau_{j}~\scalebox{0.91}{$e^{i\Delta 
E^{j}\tau_{j}}~G_{R}(X_{j},X'_{j'})~.$}
\end{eqnarray}
Here we have considered $\kappa_{j}(\tau_{j})=1$, i.e., the detectors are 
eternally interacting with the field. The expressions of the quantities 
$G_{W}(X_{j},X_{j'})$, $G_{F}(X_{j},X_{j'})$, and $G_{R}(X_{j},X_{j'})$; which  
respectively denote the positive frequency Wightman function with 
$X_{j}>X_{j'}$, the Feynman propagator, and the retarded Green's function; are 
\cite{Koga:2018the}
\begin{eqnarray}\label{eq:Greens-fn-gen}
 G_{W}\left(X_{j},X_{j'}\right) &\equiv& \langle 
0_{M}|\Phi\left(X_{j}\right)\Phi\left(X_{j'}\right)|0_{M}
\rangle~,\nonumber\\
 G_{F}\left(X_{j},X_{j'}\right) &\equiv& -i\langle 
0_{M}|T\left\{\Phi\left(X_{j}\right)\Phi\left(X_{j'}\right)\right\}|0_{M}
\rangle~,\nonumber\\
 G_{R}\left(X_{j},X_{j'}\right) &\equiv& i\theta(t-t')\langle 
0_{M}|\left[\Phi\left(X_{j'}\right),\Phi\left(X_{j}\right)\right]|0_{M}
\rangle.\nonumber\\
\end{eqnarray}
We refer the readers to  \cite{Koga:2018the} for a complete analysis. 

From the general study \cite{Peres:1996dw, Horodecki:1996nc} of bipartite systems, it is observed 
that one must have a negative eigenvalue of the partial transposition of the reduced detector 
density matrix for entanglement harvesting. Here, with the reduced density matrix 
(\ref{eq:detector-density-matrix}) this condition results in
\begin{equation}\label{eq:EH-cond1}
 P_{A}P_{B}<|\varepsilon|^2~. 
\end{equation}
Moreover, in terms of the integrals (\ref{eq:all-integrals}) this condition 
(\ref{eq:EH-cond1}) takes the form \cite{Koga:2018the, Koga:2019fqh}
\begin{equation}\label{eq:cond-entanglement}
 \mathcal{I}_{A}\mathcal{I}_{B}<|\mathcal{I}_{\varepsilon}|^2~.
\end{equation}
We mention that the Feynman propagator and the Wightman functions are related 
among themselves \cite{Koga:2018the} as $iG_{F}\left(X_{j},X_{j'}\right) = 
G_{W}\left(X_{j},X_{j'}\right) + i G_{R}\left(X_{j'},X_{j}\right) = 
G_{W}\left(X_{j},X_{j'}\right) + \theta(T'-T) \left\{G_{W} 
\left(X_{j'},X_{j}\right)-G_{W}\left(X_{j},X_{j'}\right)\right\}$, which can be 
used to further simplify the expression of the integral 
$\mathcal{I}_{\varepsilon}$ of (\ref{eq:all-integrals}) as
\begin{eqnarray}
  \mathcal{I}_{\varepsilon} &=& -\int_{-\infty}^{\infty}d\tau_{B} 
\int_{-\infty}^{\infty}d\tau_{A}~e^{i(\Delta 
E^{B}\tau_{B}+\Delta E^{A}\tau_{A})} \nonumber\\
~&&\times \Big[G_{W}(X_{B},X_{A})+ \theta(T_{A}-T_{B})\nonumber\\
~&& \times 
\left\{G_{W}\left(X_{A},X_{B}\right)-G_{W}\left(X_{B},X_{A}\right)\right\}
\Big].\label{eq:Ie-integral}
\end{eqnarray}
This particular expression is advantageous because all of the integrals $\mathcal{I}_{A}$, 
$\mathcal{I}_{B}$  and $\mathcal{I}_{\varepsilon}$, imperative for the verification of the 
entanglement harvesting condition (\ref{eq:cond-entanglement}), are now expressed in terms of the 
Wightman functions. Furthermore writing the $\mathcal{I}_{\varepsilon}$ in this form also enables 
one to identify the separate contributions with or without the considered time ordering.
Note that condition (\ref{eq:cond-entanglement}) is constructed at the order $c^2$ in perturbation 
series. Later on our whole analysis will be done at this order.

After the condition for entanglement harvesting (\ref{eq:cond-entanglement}) is met, one is prompted 
to quantify its measures. The common entanglement measures are negativity and concurrence 
\cite{Zyczkowski:1998yd, Vidal:2002zz, Eisert:1998pz, Devetak_2005}. Negativity signifies the upper 
bound of the distillable entanglement, and one obtains it from the sum of all negative eigenvalues 
of the partial transpose of $\rho_{AB}$. Concurrence $\mathcal{C}(\rho_{AB})$ is another convenient 
entanglement measure \cite{Koga:2018the, Koga:2019fqh, Hu:2015lda}, which enables one to find the 
entanglement of formation \cite{Bennett:1996gf, Hill:1997pfa, Wootters:1997id, Koga:2018the, 
Koga:2019fqh}.  In the two qubits system, the concurrence is given by \cite{Koga:2018the}
\begin{eqnarray}\label{eq:concurrence-gen-exp}
 \mathcal{C}(\rho_{AB}) &=& 
max\bigg[0,~ 2c^2 
\left(|\varepsilon|-\sqrt{P_{A}P_{B}}\right)+\mathcal{O}(c^4)\bigg]\nonumber\\
~&\approx& max\bigg[0,~ 2c^2|\langle E_{1}^{B}|m_{B}(0)| E_{0}^{B}\rangle| |\langle 
E_{1}^{A}|m_{A}(0)| E_{0}^{A}\rangle|\nonumber\\
~&& ~~~~~~\times 
\left(|\mathcal{I}_{\varepsilon}|-\sqrt{\mathcal{I}_{A}\mathcal{I}_{B}}
\right)\bigg]~,
\end{eqnarray}
where, we have assumed $c_{A}=c_{B}=c$, i.e., both detectors have equal couplings with the scalar 
field. Now the quantities $|\langle E_{1}^{j}|m_{j}(0)| E_{0}^{j}\rangle|$ are obtained from the 
detectors' internal structure; the spacetime and background scalar fields do not contribute in them. 
Then to understand the effects of the trajectories and the spacetime in the harvested entanglement, 
we shall only study the relevant quantity
\begin{equation}\label{eq:concurrence-I}
\mathcal{C}_{\mathcal{I}} = \left(|\mathcal{I} 
_{\varepsilon}| -\sqrt{\mathcal{I}_{A}\mathcal{I}_{B}} \right)
\end{equation} 
with the intention of studying the concurrence. Note in the symmetric case 
$\mathcal{I}_{A}=\mathcal{I}_{B}$, this quantity becomes $\mathcal{C}_{\mathcal{I}} = 
\left(|\mathcal{I} _{\varepsilon}| - \mathcal{I}_{j} \right)$, see 
\cite{Koga:2018the, Koga:2019fqh}.

Another measure of correlation is mutual information $\mathcal{M}$, which signifies the total of 
classical and quantum correlations, defined as
\begin{equation}\label{eq:MI-general}
 \mathcal{M}(\rho_{AB}) \equiv S(\rho_{A}) + S(\rho_{B}) -S(\rho_{AB})~,
\end{equation}
where, $\rho_{A}\equiv Tr_{B}(\rho_{AB})$ and $\rho_{B}\equiv Tr_{A}(\rho_{AB})$ are the reduced 
density matrices corresponding to the detectors $A$ and $B$, and $S(\rho)\equiv -Tr(\rho\ln{\rho})$ 
signifies the von Neumann entropy corresponding to the density matrix $\rho$. With the density 
matrix (\ref{eq:detector-density-matrix}), and considering equal couplings between the  the two 
detectors and field, one can express the mutual information \cite{Simidzija:2018ddw} of 
(\ref{eq:MI-general}) as
\begin{eqnarray}\label{eq:MI-explicit}
 \mathcal{M}(\rho_{AB}) &=& c^2\big[P_{+}\ln{P_{+}} + P_{-}\ln{P_{-}} - 
P_{A}\ln{P_{A}}\nonumber\\
~&&~~~~~~~~ -~ P_{B}\ln{P_{B}}\big] + \mathcal{O}(c^4)~,
\end{eqnarray}
where, the quantities $P_{\pm}$ are given by
\begin{equation}\label{eq:P-pm}
 P_{\pm} = \frac{1}{2} \Big[ P_{A}+P_{B}\pm \sqrt{(P_{A}-P_{B})^2+4|P_{AB}|^2} 
\Big]~.
\end{equation}
In a system, if the mutual information is non-zero but the concurrence vanishes, then the 
correlation is considered classical. Consequently, we shall look into both concurrence and mutual 
information to understand the nature of the correlation between the two detectors.

\section{Conformal vacuum for two dimensional Schwarzschild black hole and FLRW universe} 
\label{Sec:conformal-vacuum}

In this section, we are going to discuss the conformal vacuum associated with a  $(1+1)$ 
dimensional 
Schwarzschild and \emph{Friedman-Lema\^itre-Robertson-Walker} (FLRW) spacetime. Although the 
concept 
is elaborated in literature \cite{book:Birrell, Das:2019aii} in great detail, here we shall provide 
a brief outline of the essential elements necessary for our study. Introducing a conformally 
invariant field in conformally flat spacetime allows one to realize flat-space-like quantum field 
theory in a curved background without moving to its asymptotic regions. The metric tensor 
corresponding to a conformally flat spacetime can be expressed as
\begin{eqnarray}\label{eq:conformal-metric}
 g_{\mu\nu} (x) = \Omega^2\,\eta_{\mu\nu} (x)~,
\end{eqnarray}
where $\Omega^2$ is the conformal factor, and $\eta_{\mu\nu} (x)$ denotes the metric tensor in a 
Minkowski spacetime. In a conformally flat spacetime it is necessary to consider a massless $(m=0)$ 
scalar field $\Phi(x)$ to obtain a conformally invariant wave equation. In particular, this wave 
equation is expressed as
\begin{eqnarray}\label{eq:conformal-EOM}
\big(\Box-\xi R\big)\,\Phi(x)=0~,
\end{eqnarray}
where the operator $\Box$ in the background spacetime is given by 
\scalebox{0.9}{$\Box\Phi=(\sqrt{-g})^{-1/2}\partial_{\mu}[\sqrt{-g}g^{\mu\nu}\partial_{\nu}\Phi]$}, 
$R$ represents the Ricci scalar, and $\xi$ the conformal coupling. In an $n-$dimensional spacetime 
this conformal coupling is of the form $\xi=(n-2)/4(n-1)$. Therefore, $\xi$ vanishes in 
$(1+1)-$dimensional spacetime and $\xi = 1/6$ for $(1+3)-$dimensions. It is to be noted that with a 
further decomposition of the conformally invariant field $\Phi=\Omega^{(2-n)/2}\bar{\Phi}$, the 
above wave equation reduces to that of the flat spacetime, 
$\eta^{\mu\nu}\partial_{\nu}\partial_{\mu}(\bar{\Phi})=0$. Then one could readily find out the mode 
solutions $\bar{u}_{k}$ of the field $\bar{\Phi}$, which are plane wave like. In terms of these mode 
functions one may express the scalar field $\Phi(x)$ as
\begin{eqnarray}\label{eq:conformal-field-decomposition}
 \Phi(x) = 
\Omega^{(2-n)/2}(x)\sum_{k}\big[\hat{a}_{k}\,\bar{u}_{k}(x)+\hat{a}^{\dagger}_{k}\,\bar{u}^{*}_{k}
(x)\big]~.
\end{eqnarray}
Here the annihilation operator $\hat{a}_{k}$ annihilates the conformal vacuum $|0\rangle$, i.e., 
$\hat{a}_{k}|0\rangle=0$. This brief summary highlights the advantages of considering a conformally 
coupled scalar field in a conformally flat spacetime to observe the effects of quantum field 
theory. 
It should be noted that all $(1+1)$ dimensional spacetimes and the spatially flat FLRW spacetime 
are conformally flat. In our subsequent analysis we shall briefly discuss the conformal nature of 
Schwarzschild and FLRW spacetimes.

\subsection{Two dimensional Schwarzschild spacetime}

In this part, we are going to elucidate on the Schwarzschild spacetime briefly. In particular, we 
shall concentrate on the $(1+1)$ dimensional representation of it. There are a few necessary 
reasons 
behind considering the $(1+1)$ dimensional representation. In our context, the most important 
reason 
is that all $(1+1)$ dimensional spacetimes are conformally flat, as we have previously mentioned. 
Then one can effortlessly utilize quantum field theory in this background. For semi-classical 
particle production, one usually needs the understanding of wave modes in ingoing and outgoing null 
paths in a black hole background. This picture is efficiently understandable in an $(1+1)$ 
dimensional representation of a generally higher-dimensional black hole spacetime to produce the 
thermal behavior of the Hawking effect. This outcome also becomes relevant in our scenario, as 
Entanglement harvesting is closely related to particle production. The $(1+1)$ dimensional 
representation of the Schwarzschild solution is considered to understand the entanglement 
harvesting 
conditions with static and free-falling detectors in literature \cite{Gallock-Yoshimura:2021yok}, 
which will also help us compare our results. 

In fact, the $(1+1)$ dimensional Schwarzschild black hole can be considered as the solution of a two dimensional Einstein-Dilaton theory \cite{Grumiller:2002nm, Das:2019aii} which is a dimensionally reduced form of higher dimensional usual Einstein's theory of gravity.
In particular, in $(1+1)$ dimensions the line-element in Schwarzschild background 
using the Schwarzschild coordinates $(\ts,r)$ is given by
\begin{eqnarray}\label{eq:metric-sch}
 ds^2 &=& -\Big(1-\frac{\rh}{r}\Big)d\ts^2+ 
\Big(1-\frac{\rh}{r}\Big)^{-1} dr^2 ~,
\end{eqnarray}
where, $\rh$ represents the Schwarzschild radius. It is observed that in terms of the tortoise 
coordinate $\rstar$, defined from
\begin{equation}\label{eq:tortoise-coord}
 d\rstar = \frac{dr}{1-\rh/r}~,
\end{equation}
the $(1+1)$ dimensional Schwarzschild metric becomes
\begin{equation}\label{eq:metric-sch-1p1}
 ds^2 = \Big(1-\frac{\rh}{r}\Big)\big[-d\ts^2+ 
d\rstar^2\big]~.
\end{equation}
The expression of this metric is analogous to the prescription of (\ref{eq:conformal-metric}) with 
$\Omega^2=(1-\rh/r)$, which ensures that the $(1+1)$ dimensional Schwarzschild black hole spacetime 
is conformally flat. One can easily obtain the scalar field decomposition like 
(\ref{eq:conformal-field-decomposition}) in this spacetime, with the wave modes now expressed in 
terms of $\ts$ and $\rstar$. In particular, the related conformal vacuum is known as the Boulware 
vacuum.

We mention that the conformal metric of (\ref{eq:metric-sch-1p1}) can also be represented in terms 
of the null coordinates $u=\ts-\rstar$ and $v=\ts+\rstar$, thus the Boulware modes are eligible to 
be represented in terms of these null coordinates. These null coordinates are related to the 
Kruskal coordinates as $V=2\rh e^{v/2\rh}$ and $U=-2\rh e^{-u/2\rh}$. One may move to these Kruskal 
coordinates also and find the metric of Eq. (\ref{eq:metric-sch-1p1}) again to be conformally flat, 
but with different conformal factors. We mention that considering different sets of these 
coordinates one obtains different conformal vacuums corresponding to the Schwarzschild black hole 
spacetime. Specifically with $u$ and $v$ the vacuum is Boulware, with $U$ and $v$ the vacuum is 
Unruh, and $U$ and $V$ the vacuum is Hartle-Hawking \cite{Juarez-Aubry:2018ofz, 
Gallock-Yoshimura:2021yok}.

A static observer in a Schwarzschild black hole spacetime does not observe the Boulware vacuum to 
be 
filled with particles. However, a freely in-falling observer perceives the Boulware vacuum to be 
thermal, as was pointed out by recent studies \cite{Scully:2017utk}. On the other hand, 
entanglement 
related observations are highly connected to the scenarios of particle creation in a curved 
spacetime \cite{Gallock-Yoshimura:2021yok}. Therefore, we shall consider these types of relevant 
paths, which can observe particles in these vacua, for the observation of entanglement harvesting.

\subsection{The FLRW spacetime}

After briefly elucidating on the static black hole spacetime we now proceed to the 
\emph{Friedman-Lema\^itre-Robertson-Walker} (FLRW) spacetime, which is an exact solution of the 
Einstein's field equations with positive cosmological constant. The FLRW metric was formulated to 
represent our universe and indicates spatial homogeneity, isotropy, and expansion. In 
(1+1)-dimensional FLRW spacetime the line element is given by
\begin{equation}\label{eq:metric-DS}
ds^2 =-dt^2+a^2(t)\,dx^2~,
\end{equation}
where $a(t)$ denotes the scale factor, which is different in different eras of the universe. In 
particular, in this work we shall consider the de Sitter era, where one can analytically pursue 
the calculations. In de Sitter era the scale factor is $a(t)=e^{t/\alpha_d}$, which describes the 
early expansion of the homogeneous, isotropic universe. In terms of the conformal time $\eta$, 
related to the coordinate time $t$ through $d\eta=dt/a(t)$, expressed in de Sitter background 
\begin{equation}\label{eta-tReln}
\eta=-\alpha_{d}\,e^{-t/\alpha_d}~,
\end{equation}
the line element becomes
\begin{equation}\label{eq:metric-DS-1p1}
 ds^2=a^2(\eta)(-d\eta^2+dx^2)~.
\end{equation}
Like the $(1+1)$ dimensional Schwarzschild spacetime here also one can observe that the background 
spacetime becomes conformally flat with $\Omega=a(\eta)$, and one can easily find out the scalar 
field decomposition like (\ref{eq:conformal-field-decomposition}) with respect to the coordinates 
$\eta$ and $x$.

On the other hand, in $(1+3)-$dimensional FLRW spacetime the line element is given by
\begin{eqnarray}\label{eq:metric-DS-1p3}
ds^2&=&-dt^2+a^2(t)|d\vec{x}|^2\nonumber\\&=&a^2(\eta)(-d\eta^2+|d\vec{x}|^2)~.
\end{eqnarray}
Here also the spacetime is conformally flat and one is eligible to decompose a conformally coupled 
scalar field in terms of plane wave modes. It is to be noted that, in both the $(1+1)$ and $(1+3)$ 
dimensional de Sitter spacetimes the conformal vacuums for observers with coordinates 
$(\eta,\vec{x})$ are perceived to be particle creating with respect to a co-moving observer 
with coordinates $(t,\vec{x})$.

In \cite{Chakraborty:2019ltu}, it is shown that an observer in a null like trajectory also observes 
the conformal vacuum particle generating. We have already stated semi-classical particle creation is 
closely related to non-trivial findings in the entanglement harvesting conditions. In our subsequent 
studies we shall be discussing about the null paths in these spacetimes which are related to 
particle creation. Furthermore, we shall investigate the entanglement harvesting conditions in these 
scenarios.

\section{Null paths related to particle creation from the conformal vacuum}

Here we discuss a class of trajectories, specifically the null paths, from which an observer 
perceives the conformal vacuum to be particle creating.
First, in a $(1+1)$ dimensional black hole Schwarzschild spacetime Eq. (\ref{eq:metric-sch-1p1}) 
signifies that a radially moving object is following an outgoing null path if $u=\ts-\rstar$ is 
constant along its trajectory. On the other hand, it is ingoing when $v=\ts+\rstar$ is constant. 
These two coordinates are often referred to as the retarded and advanced time coordinates or the 
outgoing and ingoing null coordinates. In terms of the Eddington-Finkelstein (EF) coordinates 
$(t,r)$, with $t+r=\ts+\rstar$ the metric (\ref{eq:metric-sch-1p1}) transforms to
\begin{equation}\label{eq:metric-sch-1p1-EF}
 ds^2 = -\Big(1-\frac{\rh}{r}\Big)dt^2 + \frac{2\rh}{r}dtdr + 
\Big(1+\frac{\rh}{r}\Big) dr^2~.
\end{equation}
%
%
\begin{figure}[h]
\centering
 \includegraphics[width=0.95\linewidth]{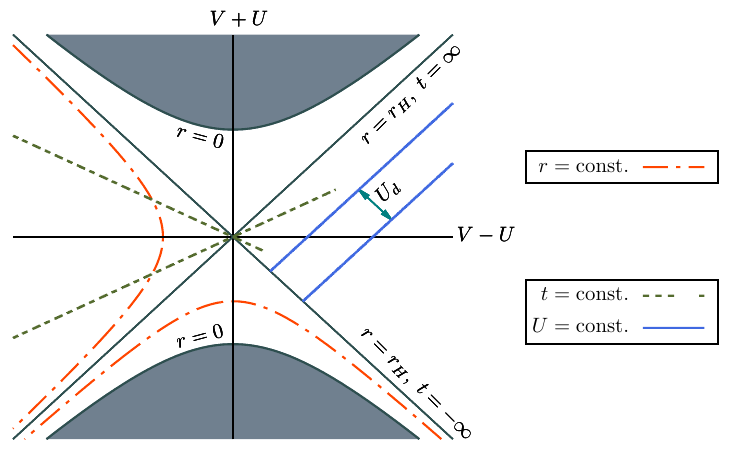}
 \caption{Schematic representation of two detectors in outgoing null paths depicted in a Kruskal 
diagram. In \emph{Eddington-Finkelstein} coordinates the two detectors are separated as 
$u_{A}-u_{b}=d$, while in Kruskal coordinates they are separated as 
$U_{d}=U_{B}-U_{A}=2\rh(1-e^{-d/2\rh})$, where $U_{j}=-2\rh~ e^{-u_{j}/2\rh}$.}
 \label{fig:Kruskal-null-outgoing}
\end{figure}
With these EF coordinates one can find out an outgoing null trajectory by making $ds^2=0$ in 
(\ref{eq:metric-sch-1p1-EF}) and considering the positive solution of $dr/dt$, see 
\cite{Dalui:2020qpt}, as
\begin{equation}\label{eq:outgoing-null-path}
 \frac{dt}{dr} = \frac{r/\rh+1}{r/\rh-1}~,
\end{equation}
which gives the path to be
\begin{equation}\label{eq:outgoing-null-coord}
 t = r+2\rh\ln{\Big[\frac{r}{\rh}-1\Big]}+d~.
\end{equation}
Here $d$ is a constant parameter arriving as an integration constant from Eq. 
(\ref{eq:outgoing-null-path}). In Fig. \ref{fig:Kruskal-null-outgoing} we have provided a Kruskal 
diagram depicting the null rays in a $(1+1)$ dimensional Schwarzschild black hole spacetime, and in 
this figure one also notices that $d$ distinguishes different outgoing null paths. We mention that, 
utilizing quantum field theory one can perceive particle production in conformal Boulware vacuum 
with respect to observers, moving along these null paths \cite{Chakraborty:2019ltu, Dalui:2020qpt}.

On the other hand, the outgoing and ingoing null coordinates in a general de Sitter background is 
given by $u=\eta- |\vec{x}|$ and $v=\eta+ |\vec{x}|$. One can simply understand that these 
expressions in $(1+1)$ dimensions become $u=\eta- x$ and $v=\eta+ x$.
With respect to the coordinate time $t$ an observer along these null paths perceives the conformal 
vacuum particle generating.
The scenarios of particle creation in curved spacetimes also influences the entanglement related 
observations. Therefore, we are going to consider these null trajectories in the Schwarzschild and 
FLRW spacetimes to understand the entanglement harvesting conditions from the relevant conformal 
vacuums.

\section{Green's function corresponding to outgoing null detectors}\label{Sec:Greens-fn-null-paths}

From Eq. (\ref{eq:cond-entanglement}) and (\ref{eq:all-integrals}) of Sec. \ref{Sec:model-set-up} 
we 
have seen that it is imperative to construct the Green's functions for a considered trajectory in a 
background spacetime to understand the entanglement harvesting conditions in this 
scenario. In this section, we will construct these necessary Green's functions along the null paths 
in the previously considered Schwarzschild and FLRW spacetimes. In this regard, we mention that 
rather than considering these Green's functions in their position space representations, we will 
take them in their momentum space representations. This consideration allows one to evaluate 
detector transition probabilities corresponding to a specific field mode frequency even with linear 
field-detector interaction in $(1+1)$ dimensions. With this particular consideration, we 
also observed that one could circumvent the issues related to the infrared cutoff inherent to the 
$(1+1)$ dimensional massless scalar field theory.

\subsection{Schwarzschild background}
\subsubsection{Boulware vacuum}

The positive frequency Boulware modes in terms of the null coordinates $u$ and 
$v$ are given by $e^{-i\omega u}$ and $e^{-i\omega v}$. One can then decompose 
a massless minimally coupled scalar field $\Phi$ in terms of these Boulware 
modes and suitably choosing the sets of creation and annihilation operators 
$\{\hat{a}_{k}^{B\dagger}, \hat{a}_{k}^{B}\}$ and $\{\hat{b}_{k}^{B\dagger}, 
\hat{b}_{k}^{B}\}$ as \cite{Hodgkinson:2013tsa}
\begin{eqnarray}
 \Phi &=& \int_{0}^{\infty} \frac{d\omega_{k}}{\sqrt{4\pi\omega_{k}}} 
\Big[\hat{a}_{k}^{B} e^{-i\omega_{k} u} + \hat{a}_{k}^{B\dagger} 
e^{i\omega_{k} u} \nonumber\\
~&& ~~~~~~~~+~~ \hat{b}_{k}^{B} e^{-i\omega_{k} v} + 
\hat{b}_{k}^{B\dagger} 
e^{i\omega_{k} v} \Big]~.
\end{eqnarray}
The ladder operators satisfy the commutation relation $\big[ \hat{a}_{k}^{B}, 
\hat{a}_{k'}^{B\dagger}\big]=\delta_{k,k'}$ and $\big[ \hat{b}_{k}^{B}, 
\hat{b}_{k'}^{B\dagger}\big]=\delta_{k,k'}$, where all other choices in the 
commutator vanishes. Also the Boulware vacuum $|0\rangle_{B}$ is now defined by 
the one annihilated by these annihilation operators $\hat{a}_{k}^{B} 
|0\rangle_{B} = 0 = \hat{b}_{k}^{B} |0\rangle_{B}$. Then using the above field 
decomposition one can get the positive frequency Wightman function with respect 
to the Boulware vacuum to be given by
\begin{eqnarray}\label{eq:Wightman-Boulware}
 G^{+}_{B}(X_{j},X_{l}) &=& _{B}\langle 0|\Phi(X_{j})\Phi(X_{l}) 
|0\rangle_{B}\nonumber\\
~&=& \int_{0}^{\infty} \frac{d\omega_{k}}{4\pi\omega_{k}} \big[e^{-i\omega_{k} 
(u_{j}-u_{l})} + e^{-i\omega_{k} 
(v_{j}-v_{l})}\big]~, \nonumber\\
\end{eqnarray}
where the subscript $j$ and $l$ correspond to the events $X_{j}$ and $X_{l}$ 
respectively. 

Now we shall be considering one detector, say detector $A$ with a non zero $d$, 
and detector $B$ with $d=0$. Then we shall be using $t_{A} = r_{A} + 
2\rh\ln{[r_{A}/\rh-1]} + d$ and $t_{B} = r_{B} + 2\rh\ln{[r_{B}/\rh-1]}$. Using 
these EF coordinates, for two detectors following outgoing null trajectories 
(\ref{eq:outgoing-null-coord}), one has the quantities
\begin{eqnarray}\label{eq:outgoing-null-BV-u}
 v'_{j}-v_{l} &=& \tsj'+\rstarj'-(\tsl+\rstarl)\nonumber\\
 ~&=& 2(r'_{j}-r_{l}) + 2\rh\ln{\bigg[\frac{r'_{j}-\rh}{r_{l}-\rh}\bigg]}\nonumber\\
 ~&& ~~~~~~~~~+ d\,(\delta_{jA}\, \delta_{lB} - \delta_{jB}\, \delta_{lA})~,
\end{eqnarray}
and
\begin{eqnarray}\label{eq:outgoing-null-BV-v}
 u'_{j}-u_{l} &=& \tsj'-\rstarj'-(\tsl-\rstarl) \nonumber\\
 ~&=& d\,(\delta_{jA}\, \delta_{lB} - 
\delta_{jB}\, \delta_{lA})~,
\end{eqnarray}
where $j$ and $l$ can represent either detector $A$ or $B$, with $\delta_{jl}$ denoting the 
\emph{Kronecker delta} defined as
\begin{eqnarray}
 \delta_{jl} &=& 0~,~~~~\textup{if}~j\neq l\nonumber\\
 ~&=& 1~,~~~~\textup{if}~j= l~.
\end{eqnarray}
Substitution of (\ref{eq:outgoing-null-BV-u}) and (\ref{eq:outgoing-null-BV-v}) in 
(\ref{eq:Wightman-Boulware}) provides us the required Green's function $G^{+}_{B}$ corresponding to 
the Boulware vacuum with respect to our observers.

\subsubsection{Unruh vacuum}

To discuss about the Unruh modes and the corresponding Unruh vacuum one needs an understanding of 
the Kruskal coordinates $V=2\rh e^{v/2\rh}$, and $U=-2\rh e^{-u/2\rh}$. Then the positive frequency 
Unruh modes in terms of the null coordinates $v$ and $U$ are given by $e^{-i\omega U}$ and 
$e^{-i\omega v}$. In terms of these Unruh modes a massless minimally coupled scalar field $\Phi$ is 
decomposed, choosing the sets of creation and annihilation operators $\{\hat{a}_{k}^{U\dagger}, 
\hat{a}_{k}^{U}\}$ and $\{\hat{b}_{k}^{U\dagger}, \hat{b}_{k}^{U}\}$, as
\begin{eqnarray}
 \Phi &=& \int_{0}^{\infty} \frac{d\omega_{k}}{\sqrt{4\pi\omega_{k}}} 
\Big[\hat{a}_{k}^{U} e^{-i\omega_{k} v} + \hat{a}_{k}^{U\dagger} 
e^{i\omega_{k} v} \nonumber\\
~&& ~~~~~~~~+~~ \hat{b}_{k}^{U} e^{-i\omega_{k} U} + 
\hat{b}_{k}^{U\dagger} 
e^{i\omega_{k} U} \Big]~.
\end{eqnarray}
The ladder operators satisfy the commutation relation $\big[ \hat{a}_{k}^{U}, 
\hat{a}_{k'}^{U\dagger}\big]=\delta_{k,k'}$ and $\big[ \hat{b}_{k}^{U}, 
\hat{b}_{k'}^{U\dagger}\big]=\delta_{k,k'}$, where all other choices in the 
commutator vanishes. Also the Unruh vacuum $|0\rangle_{U}$ is now defined by 
the one annihilated by these annihilation operators $\hat{a}_{k}^{U} 
|0\rangle_{U} = 0 = \hat{b}_{k}^{U} |0\rangle_{U}$. Then using the above field 
decomposition one can get the positive frequency Wightman function with respect 
to the Unruh vacuum to be given by
\begin{eqnarray}\label{eq:Wightman-Unruh}
 G^{+}_{U}(X_{j},X_{l}) &=& _{U}\langle 0|\Phi(X_{j})\Phi(X_{l}) 
|0\rangle_{U}\nonumber\\
~&=& \int_{0}^{\infty} \frac{d\omega_{k}}{4\pi\omega_{k}} \big[e^{-i\omega_{k} 
(v_{j}-v_{l})} + e^{-i\omega_{k} 
(U_{j}-U_{l})}\big]~, \nonumber\\
\end{eqnarray}
where the subscript $j$ and $l$ correspond to the events $X_{j}$ and $X_{l}$ respectively.
To write this in terms of our chosen trajectories we need to use the following transformation 
relations.
Using the EF coordinates, for two detectors in outgoing null trajectories 
(\ref{eq:outgoing-null-coord}), one has these relations
\begin{eqnarray}\label{eq:v-Diff-Un}
 v'_{j}-v_{l} &=& \tsj'+\rstarj'-(\tsl+\rstarl)\nonumber\\
 ~&=& 2(r'_{j}-r_{l}) + 2\rh\ln{\bigg[\frac{r'_{j}-\rh}{r_{l}-\rh}\bigg]}\nonumber\\
 ~&& ~~~~~~~~~+ d\,(\delta_{jA}\, \delta_{lB} - \delta_{jB}\, \delta_{lA})~,
\end{eqnarray}
and
\begin{eqnarray}\label{eq:U-Diff-Un}
 U'_{j}-U_{l} &=& -2\rh e^{-\frac{u'_{j}}{2\rh}}-(-2\rh 
e^{-\frac{u_{l}}{2\rh}})\nonumber\\
 ~&=& 2\rh\Big(1-e^{-\frac{d}{2\rh}}\Big)\,(\delta_{jA}\, \delta_{lB} - 
\delta_{jB}\, \delta_{lA}).
\end{eqnarray}

\subsection{de Sitter spacetime}
\subsubsection{(1+1)-dimensions}
Let us consider a massless minimally coupled scalar field $\Phi$ in the $(1+1)$ dimensional de 
Sitter background denoted by (\ref{eq:metric-DS-1p1}). In particular, the equation of motion for 
the 
field $\Phi$ suggests field mode solutions of the form $u_{\nu}=e^{\mp i\omega_{k}(\eta\mp x)}$. 
Let 
us construct like the Schwarzschild case the outgoing and ingoing null coordinates $u=\eta- x$ and 
$v=\eta+ x$. Then with a suitable set of creation and annihilation operators and with these mode 
functions one can decompose the scalar field as
\begin{eqnarray}
 \Phi &=& \int_{0}^{\infty} \frac{d\omega_{k}}{\sqrt{4\pi\omega_{k}}} 
\Big[\hat{a}_{k}^{D} e^{-i\omega_{k} u} + \hat{a}_{k}^{D\dagger} 
e^{i\omega_{k} u} \nonumber\\
~&& ~~~~~~~~+~~ \hat{b}_{k}^{D} e^{-i\omega_{k} v} + 
\hat{b}_{k}^{D\dagger} 
e^{i\omega_{k} v} \Big]~,
\end{eqnarray}
where the annihilation operators $\hat{a}_{k}^{D}$ and $\hat{b}_{k}^{D}$ annihilate the de Sitter 
vacuum $|0\rangle_{D}$. Then with the help of this field decomposition one can express the Green's 
function as
\begin{eqnarray}\label{eq:GreensFn-DS}
 G_D^{+}(X_{j},X_{l}) &=& ~_D\langle 0|\Phi(X_{j})\Phi(X_{l}) 
|0\rangle_{D}\nonumber\\
~&=& \int_{0}^{\infty} \frac{d\omega_{k}}{4\pi\omega_{k}} \big[e^{-i\omega_{k} 
(u_{j}-u_{l})} + e^{-i\omega_{k} (v_{j}-v_{l})}\big]. \nonumber\\
\end{eqnarray}
We also mention that along outgoing null trajectory $u=\eta-x$ is constant and along an ingoing 
null trajectory $v=\eta+x$ is constant. We consider our two observers Alice and Bob moving in 
outgoing null trajectories, and for Alice $\eta=x+d$ while for Bob $\eta=x$. With the help of 
Kronecker delta $\delta_{jl}$ one can collectively express these differences as
\begin{eqnarray}\label{eq:outgoing-null-DS}
u_{j}-u_{l} &=& d\,(\delta_{jA}\, \delta_{lB} - \delta_{jB}\, \delta_{lA})\nonumber\\
v_{j}-v_{l} &=& 2\alpha_{d}\,(e^{-t_{l}/\alpha_d}-e^{-t_{j}/\alpha_d}) \nonumber\\
~&& ~~~~~~ - d\,(\delta_{jA}\, \delta_{lB} - \delta_{jB}\, \delta_{lA})~,
\end{eqnarray}
where, $j$ and $l$ can take values of either $A$ or $B$. Substituting (\ref{eq:outgoing-null-DS}) 
in (\ref{eq:GreensFn-DS}) we will find our required $G_D^{+}$ with respect to the outgoing 
trajectory.

\subsubsection{(1+3)-dimensions}

In $(1+3)$ dimensional de Sitter spacetime also the conformal factor is $\Omega=a(\eta)$ like 
$(1+1)$ dimensions. However, the difference in the spacetime dimensionality results in a factor of 
$a(\eta)^{(2-4)/2}$ in the scalar field decomposition of (\ref{eq:conformal-field-decomposition}). 
In particular, in a $(1+3)-$dimensional FLRW universe a conformally coupled scalar field $\Phi$ in 
can be decomposed into modes and ladder operators as,
\begin{eqnarray}
\Phi &=& \int\frac{d^3k}{\sqrt{(2\pi)^3 2\omega_k}}\,\frac{1}{a(\eta)}\, 
\Big(e^{-i\omega_k\eta+i\vec{k}\cdot\vec{x}}\,\hat{b}_{\vec{k},\omega_k}\nonumber\\
~&& ~~~~~~~~~~~~~~~~~~~~~~~~~~~+~e^{
i\omega_k\eta-i\vec { k }
\cdot\vec{x}}\,\hat{b}^{\dagger}_{\vec{k},\omega_k}\Big)\,.
\end{eqnarray}
For simplicity, we choose the detectors to be outgoing along the $x$ axis. Then along this path 
$\Delta y=0=\Delta z$, and only the $k_{x}$ component from the factor $\vec{k}\cdot\vec{x}$ will 
survive in the Green's function evaluated with respect to the conformal vacuum. The Wightman 
functions for these outgoing null paths are given by
\begin{eqnarray}\label{eq:GreensFn-DS-1p3}
G_W(X_j,X_l)&=&\int\frac{d^3k}{(2\pi)^32\omega_k} 
\frac{e^{i{k_x}\Delta x_{jl}-i\omega_k\Delta\eta_{jl}
}}{a(\eta_j)a(\eta_l)}~.
\end{eqnarray}
For the case when the detectors are moving along the $x$ axis, i.e., the motion of the detectors 
are 
actually confined to a one dimensional line, the outgoing null paths corresponding to Alice and Bob 
are again $\eta=x+d$ and $\eta=x$. One can use the information of these paths to obtain the 
appropriate $\Delta x_{jl}$ and $\Delta\eta_{jl}$ in a straightforward manner. Then one can 
substitute these expressions in (\ref{eq:GreensFn-DS-1p3}) to obtain the necessary Green's 
functions 
in the null trajectories.

\section{Entanglement harvesting}\label{Sec:entanglement-harvesting}

\subsection{Schwarzschild background}

\subsubsection{Boulware vacuum}

As we have already discussed we are considering one detector, say detector $A$ with a non zero $d$, 
and detector $B$ with $d=0$. First we shall use the expression of the Wightman function estimated 
in the Boulware vacuum from Eq. (\ref{eq:Wightman-Boulware}). With the coordinate transformations 
of Eq. (\ref{eq:outgoing-null-BV-u}) and (\ref{eq:outgoing-null-BV-v}) suitable to an observer in 
an outgoing null path in a Schwarzschild black hole spacetime, one can compute the necessary 
integrals of (\ref{eq:cond-entanglement}) to investigate the entanglement harvesting condition. In 
particular, one can find out the individual detector transition probabilities $\mathcal{I}_{j}$ as
\begin{eqnarray}\label{eq:Ij-Sch-BV-1}
 \mathcal{I}_{j} &=& \int_{-\infty}^{\infty} dt'_{j}\int_{-\infty}^{\infty} 
dt_{j}~ e^{-i\Delta E^{j}(t'_{j}-t_{j})} G^{+}_{B}(X'_{j},X_{j})\nonumber\\
~&=& \int_{0}^{\infty} \frac{d\omega_{k}}{4\pi\omega_{k}}~ 
\mathcal{I}_{j_{\omega_{k}}}~.
\end{eqnarray}
Here the integrals $\mathcal{I}_{j_{\omega_{k}}}$ are represented as 
\begin{eqnarray}\label{eq:Ij-Sch-BV-2}
 \mathcal{I}_{j_{\omega_{k}}} &=& \int_{-\infty}^{\infty} 
dt'_{j}\int_{-\infty}^{\infty} 
dt_{j}\, e^{-i\Delta E^{j}(t'_{j}-t_{j})}\nonumber\\
~&&~~ \times~\big[e^{-i\omega_{k} 
(u_{j'}-u_{j})} + e^{-i\omega_{k} 
(v_{j'}-v_{j})}\big]~\nonumber\\
~&=& \int_{-\infty}^{\infty} 
dt'_{j}\int_{-\infty}^{\infty} 
dt_{j}\, e^{-i\Delta E^{j}(t'_{j}-t_{j})}\nonumber\\
~&\times& \bigg[1 + e^{-2i\omega_{k} 
(r_{j'}-r_{j})} \bigg(\tfrac{r_{j'}-\rh}{r_{j}-\rh}\bigg)^{-2i\rh\omega_{k}}\bigg]~,
\end{eqnarray}
where we have used the relations from Eq. (\ref{eq:outgoing-null-BV-u}) and 
(\ref{eq:outgoing-null-BV-v}). Here we mention that the integrations over the first additive unity 
provides multiplicative factors of the Dirac delta distributions $\delta(\Delta E^j)$, which makes 
the contribution of that part of the integration to vanish as the detector transition energy 
$\Delta E^j>0$. Then we consider the expression of (\ref{eq:outgoing-null-path}) for the 
realization 
of outgoing null paths in this calculation and the expression (\ref{eq:Ij-Sch-BV-2}) transforms 
into
\begin{eqnarray}\label{eq:Ij-Sch-BV-3}
 \mathcal{I}_{j_{\omega_{k}}} &=& \int_{\rh}^{\infty} 
dr_{j'}~ \frac{r_{j'}+\rh}{r_{j'}-\rh}\int_{\rh}^{\infty} 
dr_{j}~ \frac{r_{j}+\rh}{r_{j}-\rh}~ \nonumber\\
~&\times& e^{-i(\Delta 
E^{j} + 2\omega_{k})(r_{j'}-r_{j})}~\bigg(\tfrac{r_{j'}-\rh}{r_{j}-\rh}\bigg)^{-2i\rh(\Delta 
E^{j} + \omega_{k})}.
\end{eqnarray}\vspace{0.1cm}
%
\begin{figure}[h]
\centering
 \includegraphics[width=0.95\linewidth]{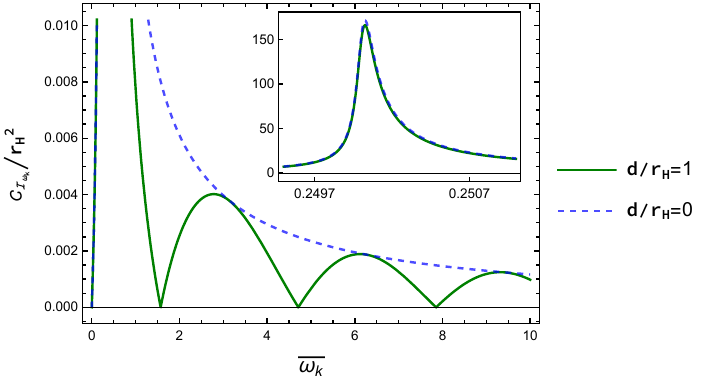}
 \caption{The quantity $\mathcal{C}_{\mathcal{I}_{\omega_{k}}}/\rh^2 = 
(|\mathcal{I}_{\varepsilon_{\omega_{k}}}| -\mathcal{I}_{j_{\omega_{k}}})/\rh^2$, signifying the 
concurrence, is plotted for two outgoing null detectors in a $(1+1)$ dimensional Schwarzschild black 
hole spacetime with respect to the dimensionless frequency of the field $\overline{\omega}_{k} = 
\rh\omega_{k}$ for fixed detector transition energies $\overline{\Delta E^{A}}=\rh\Delta E^{A}=0.5$, 
$\overline{\Delta E^{B}}=\rh\Delta E^{B}=0.5$. The other parameter is fixed at $d/\rh=0$ and $d/\rh=1$.}
 \label{fig:Concurrence-Schwarzschild-d0-dEsame}
\end{figure}

%
%
\begin{figure}[h]
\centering
 \includegraphics[width=0.85\linewidth]{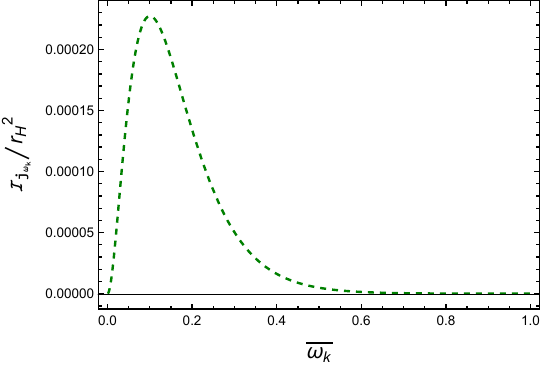}
 \caption{The quantity $\mathcal{I}_{j_{\omega_{k}}}/\rh^2$ is plotted for two outgoing 
null detectors in a $(1+1)$ dimensional Schwarzschild black hole spacetime with respect to the 
frequency of the field $\overline{\omega}_{k} = \rh\omega_{k}$ for fixed detector transition 
energies $\overline{\Delta E^{A}}=\rh\Delta E^{A}=0.5$, $\overline{\Delta E^{B}}=\rh\Delta 
E^{B}=0.5$. One should note that $\mathcal{I}_{j_{\omega_{k}}}$ is independent of the parameter $d$.}
 \label{fig:Ij-Schwarzschild-d0-dEsame}
\end{figure}
%
\begin{figure}[h]
\centering
 \includegraphics[width=0.95\linewidth]{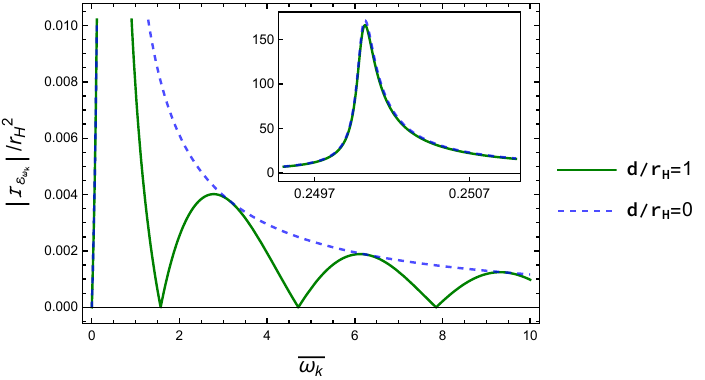}
 \caption{The quantity $|\mathcal{I}_{\varepsilon_{\omega_{k}}}|/\rh^2$ is plotted for two outgoing 
null detectors in a $(1+1)$ dimensional Schwarzschild black hole spacetime with respect to the 
frequency of the field $\overline{\omega}_{k} = \rh\omega_{k}$ for fixed detector transition 
energies $\overline{\Delta E^{A}}=\rh\Delta E^{A}=0.5$, $\overline{\Delta E^{B}}=\rh\Delta E^{B}=0.5$, 
and fixed $d/\rh=0$ and $d/\rh=1$.}
 \label{fig:Ie-Schwarzschild-d0-dEsame}
\end{figure}
\begin{figure}[h]
\centering
 \includegraphics[width=0.95\linewidth]{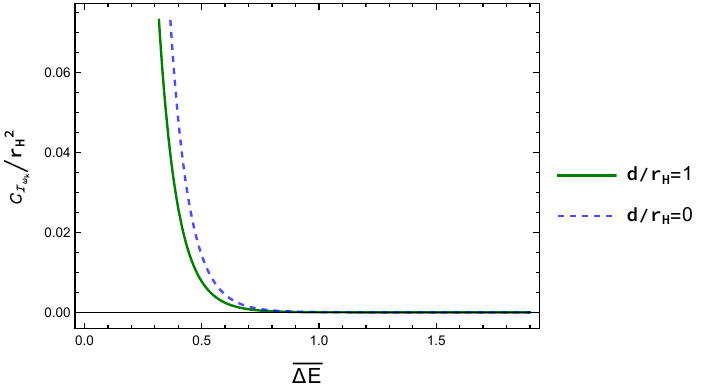}
 \caption{The quantity $\mathcal{C}_{\mathcal{I}_{\omega_{k}}}/\rh^2$ is plotted for two outgoing 
null detectors in a $(1+1)$ dimensional Schwarzschild black hole spacetime with respect to the 
dimensionless transition energy $\overline{\Delta E} = \rh\Delta E$ of the detectors, where $\Delta 
E = \Delta E^{A} = \Delta E^{B}$. The dimensionless field mode frequency is fixed at 
$\overline{\omega}_{k} = 1$ and the dimensionless distance between the null paths are $d/\rh=0$ and $d/\rh=1$.}
 \label{fig:Concurrence-Schwarzschild-d0-VdE}
\end{figure}

%
%
%
\begin{figure}[h]
\centering
 \includegraphics[width=0.95\linewidth]{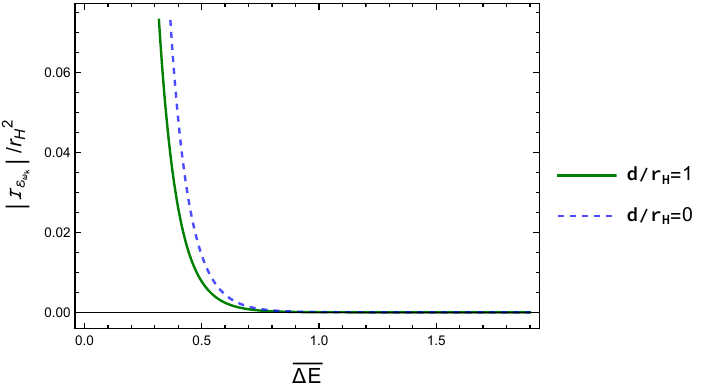}
 \caption{The quantity $|\mathcal{I}_{\varepsilon_{\omega_{k}}} |/\rh^2$ is plotted for two outgoing 
null detectors in a $(1+1)$ dimensional Schwarzschild black hole spacetime with respect to the 
dimensionless transition energy $\overline{\Delta E} = \rh\Delta E$ of the detectors, where $\Delta 
E = \Delta E^{A} = \Delta E^{B}$. The dimensionless field mode frequency is fixed at 
$\overline{\omega}_{k} = 1$ and the other fixed parameter are $d/\rh=0$ and  $d/\rh=1$.}
 \label{fig:Ie-Schwarzschild-d0-VdE}
\end{figure}

\begin{figure}[h]
\centering
 \includegraphics[width=0.85\linewidth]{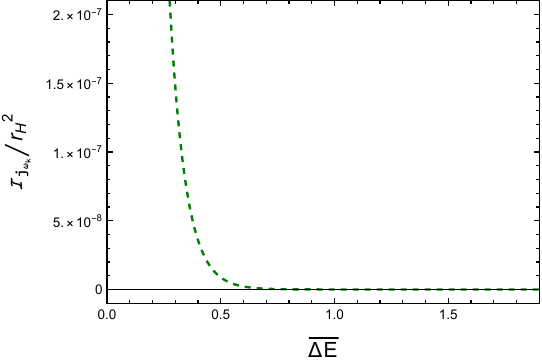}
 \caption{The quantity $\mathcal{I}_{j_{\omega_{k}}}/\rh^2$ is plotted for two outgoing 
null detectors in a $(1+1)$ dimensional Schwarzschild black hole spacetime with respect to the 
dimensionless transition energy $\overline{\Delta E} = \rh\Delta E$ of the detectors, where $\Delta 
E = \Delta E^{A} = \Delta E^{B}$. The dimensionless field mode frequency is fixed at 
$\overline{\omega}_{k} = 1$. One should note that $\mathcal{I}_{j_{\omega_{k}}}$ is independent of $d$.}
 \label{fig:Ij-Schwarzschild-d0-VdE}
\end{figure}
\begin{figure}[h]
\centering
\includegraphics[width=0.85\linewidth]{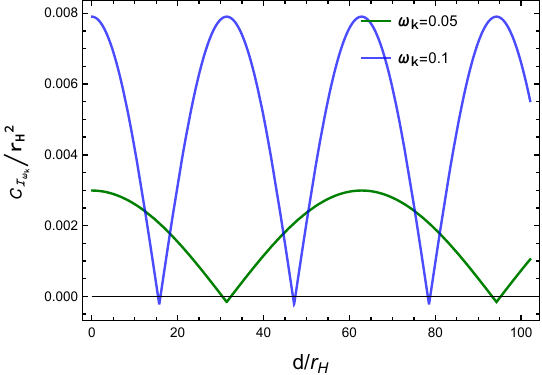}\\
 \includegraphics[width=0.85\linewidth]{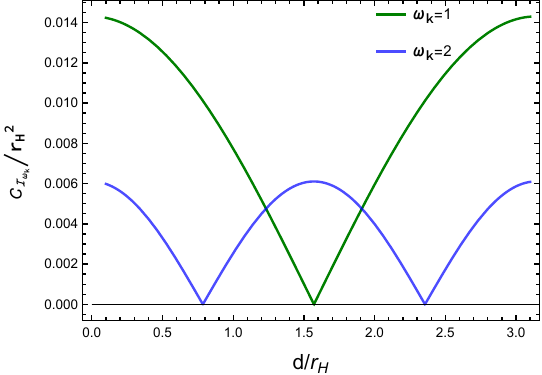}
 \caption{The quantity $\mathcal{C}_{\mathcal{I}_{\omega_{k}}} /\rh^2$ is plotted for two outgoing null detectors in different parallel paths in a $(1+1)$ dimensional Schwarzschild black hole spacetime with respect to the separation between the two paths $d/\rh$. The dimensionless frequency of the field are fixed at $\overline{\omega}_{k}=0.05$ and $\overline{\omega}_{k}=0.1$, respectively in the upper plot and shows entanglement shadow regions.  Whereas in the lower plot, the dimensionless frequency of the field are fixed at $\overline{\omega}_{k}=1$ and $\overline{\omega}_{k}=2$, respectively and shows entanglement shadow points instead of the shadow regions.The detector transition energy is fixed at $\overline{\Delta E} = 0.5$.}
 \label{fig:Concurrence-Schwarzschild-wk0p4-dd}
\end{figure}
%

Now one may consider a change of variables $y'_{j}=r_{j'}/\rh-1$ and $y_{j}=r_{j}/\rh-1$. Then this 
integral simplifies to
\begin{eqnarray}\label{eq:Ij-Sch-BV-4}
 \mathcal{I}_{j_{\omega_{k}}} &=& \rh^2 \Bigg|\int_{0}^{\infty} dy_{j}\, 
\frac{y_{j}+2}{y_{j}}\,\frac{e^{i\rh(\Delta E^{j}+2\omega_{k})y_{j}}}{y_{j}^{-2i\rh(\Delta 
E^{j}+\omega_{k})}}\Bigg|^2.
\end{eqnarray}
To evaluate this integral we introduce regulator of the form $y_{j}^{\epsilon}\,e^{-\epsilon 
y_{j}}$, where $\epsilon$ is a real positive parameter. Then the actual value of the integral is 
obtained by taking the limit $\epsilon\to 0$ after evaluating the regulated integral as
\begin{eqnarray}\label{eq:Ij-Sch-BV-5}
\lim_{\epsilon\to0} &\bigg[& \int_{0}^{\infty} dy_{j}\, \frac{y_{j}+2}{y_{j}}\,\frac{e^{i\rh(\Delta 
E^{j}+2\omega_{k})y_{j}-\epsilon y_{j}}}{y_{j}^{-2i\rh(\Delta 
E^{j}+\omega_{k})-\epsilon}}\bigg]\nonumber\\
~&=& e^{-\pi\rh(\Delta E^{j}+\omega_{k})}\,\Gamma(2i\rh(\Delta E^{j}+\omega_{k}))\,\nonumber\\
~&& \times\frac{2\omega_{k}\big(\rh(\Delta E^{j}+2\omega_{k})\big)^{-2i\rh(\Delta 
E^{j}+\omega_{k})}}{\Delta E^{j}+2\omega_{k}}\, .
\end{eqnarray}
The entire integral $\mathcal{I}_{j_{\omega_k}}$ from (\ref{eq:Ij-Sch-BV-5}) becomes 
\begin{eqnarray}\label{eq:Ij-Sch-BV-6}
 \mathcal{I}_{j_{\omega_{k}}} &=& \frac{4\pi\rh\omega_{k}^2}{(\Delta 
E^{j}+\omega_{k})(\Delta 
E^{j}+2\omega_{k})^2}~\nonumber\\
~&&~~~~~~\times~~\frac{1}{e^{4\pi\rh(\Delta 
E^{j}+\omega_{k})}-1}~,
\end{eqnarray}
where we have used the \emph{Gamma function} identity $\Gamma(iz)\Gamma(-iz) = \pi/(z\sinh{\pi 
z})$. 
This signifies a somewhat Planckian distribution with respect to the detector transition energy 
$\Delta E^{j}$ plus energy of each Boulware mode $\omega_{k}$. On the other hand, for the 
evaluation 
of the integral $\mathcal{I}_{\varepsilon}$ we express it with the help of Eq. 
(\ref{eq:Ie-integral}) as
\begin{equation}\label{eq:Ie-Sch-BV-1}
 \mathcal{I}_{\varepsilon} = -\mathcal{I}^{W}_{\varepsilon} - 
\mathcal{I}^{R}_{\varepsilon}~,
\end{equation}
where one has 
\begin{eqnarray}\label{eq:IeW-Sch-BV-1}
 \mathcal{I}^{W}_{\varepsilon} &=& \int_{-\infty}^{\infty}d\tau_{B} 
\int_{-\infty}^{\infty}d\tau_{A}~e^{i(\Delta 
E^{B}\tau_{B}+\Delta E^{A}\tau_{A})} G_{W}(X_{B},X_{A})\nonumber\\
~&=& \int_{0}^{\infty} \frac{d\omega_{k}}{4\pi\omega_{k}}~ 
\mathcal{I}^{W}_{\varepsilon_{\omega_{k}}}~,
\end{eqnarray}
and
\begin{eqnarray}\label{eq:IeR-Sch-BV-1}
~\mathcal{I}^{R}_{\varepsilon} &=& \int_{-\infty}^{\infty}d\tau_{B} 
\int_{-\infty}^{\infty}d\tau_{A}~e^{i(\Delta 
E^{B}\tau_{B}+\Delta E^{A}\tau_{A})}  \theta(T_{A}-T_{B})\nonumber\\
~&& ~~~~~~~~~\times \big[G_{W}\left(X_{A},X_{B}\right)-G_{W}\left(X_{B},X_{A}
\right) \big]~\nonumber\\
~&=& \int_{0}^{\infty} \frac{d\omega_{k}}{4\pi\omega_{k}}~ 
\mathcal{I}^{R}_{\varepsilon_{\omega_{k}}}~.
\end{eqnarray}
One should note here the detector times $\tau_{j}$ are denoted by the EF times, i.e., $\tau_{j} = 
t_{j}$. On the other hand, the times $T_{j}$ appearing in the Heaviside step function due to field 
decomposition are the Schwarzschild times $t_{s_{j}}$ as the field decomposition has been done with 
respect to the Boulware modes. With the help of Eq. (\ref{eq:outgoing-null-BV-u}) and 
(\ref{eq:outgoing-null-BV-v}) one can express the integral 
$\mathcal{I}^{W}_{\varepsilon_{\omega_{k}}}$ as
\begin{eqnarray}\label{eq:IeW-Sch-BV-2}
 \mathcal{I}^{W}_{\varepsilon_{\omega_{k}}} &=& \int_{-\infty}^{\infty} 
dt_{B}\int_{-\infty}^{\infty} dt_{A}~ e^{i(\Delta E^{B}t_{B}+\Delta E^{A}t_{A})}~\nonumber\\
~&& \times~\big[e^{-i\omega_{k} (u_{B}-u_{A})} + e^{-i\omega_{k} (v_{B}-v_{A})}\big]\nonumber\\
~&=& \int_{-\infty}^{\infty} dt_{B}\int_{-\infty}^{\infty} dt_{A}~ e^{i(\Delta E^{B}t_{B}+\Delta 
E^{A}t_{A})}~ \bigg[e^{i\omega_{k}d} \nonumber\\
~&& + e^{-i\omega_{k} (2r_{B}-2r_{A}-d)} 
\bigg(\frac{r_{B}-\rh}{r_{A}-\rh}\bigg)^{-2i\rh\omega_{k}}\bigg]~.
\end{eqnarray}
Here also one can observe that the integration over the first quantity with $e^{i\omega_{k}d}$ as 
multiplicative factor will provide the multiplication of Dirac delta distributions $\delta(\Delta 
E^A)$ and $\delta(\Delta E^B)$. Therefore that part of the integral will vanish as the detector 
transition energy $\Delta E^j>0$. Now like the evaluation of $\mathcal{I}_{j_{\omega_{k}}}$ we 
utilize Eq. (\ref{eq:outgoing-null-coord}) and consider a change of variables $y_{B}=r_{B}/\rh-1$ 
and $y_{A}=r_{A}/\rh-1$, which will simplify the above integral to
\begin{eqnarray}\label{eq:IeW-Sch-BV-3}
 \mathcal{I}^{W}_{\varepsilon_{\omega_{k}}} &=& \rh^2~e^{i d(\Delta E^{A}+\omega_{k})}
\nonumber\\
 ~&\times & \Bigg[\int_{0}^{\infty} dy_{B}~ \frac{y_{B}+2}{y_{B}}~\frac{e^{i\rh(\Delta 
E^{B}-2\omega_{k})(y_{B}+1)}}{y_{B}^{-2i\rh(\Delta E^{B}-\omega_{k})}}~\nonumber\\
~&\times& \int_{0}^{\infty} dy_{A}~ 
\frac{y_{A}+2}{y_{A}}~\frac{e^{i\rh(\Delta E^{A}+2\omega_{k})(y_{A}+1)}}{y_{A}^{-2i\rh(\Delta 
E^{A}+\omega_{k})}}\Bigg].
\end{eqnarray}
The analytic expression of this integral (\ref{eq:IeW-Sch-BV-3}) is given in Eq. 
(\ref{eq:IeW-Sch-BV-4}) of Appendix \ref{Appn:IW-IR-BV}.
On the other hand, one can also express $\mathcal{I}^{R}_{\varepsilon}$, which 
contains the contribution from a retarded Green's function, from Eq. (\ref{eq:IeR-Sch-BV-1}) as
\begin{eqnarray}\label{eq:IeR-Sch-BV-2}
 \mathcal{I}^{R}_{\varepsilon_{\omega_{k}}} &=& \int_{-\infty}^{\infty} 
dt_{A}\int_{-\infty}^{\infty} dt_{B}\, e^{i(\Delta E^{B}t_{B}+\Delta 
E^{A}t_{A})}\,\theta(t_{s_A}-t_{s_B})\nonumber\\
~&& \times\big[e^{-i\omega_{k} (u_{A}-u_{B})} + 
e^{-i\omega_{k} 
(v_{A}-v_{B})} \nonumber\\
~&& - e^{-i\omega_{k} (u_{B}-u_{A})} - e^{-i\omega_{k} 
(v_{B}-v_{A})}\big]~.
\end{eqnarray}
Here we mention that the Schwarzschild time $\ts$ and the EF time $t$ are related among themselves 
as $t+r=\ts+\rstar$. Along an outgoing null trajectory (\ref{eq:outgoing-null-coord}) one readily 
gets \scalebox{0.9}{$\theta(t_{s_A}-t_{s_B})=\theta(r_{\star_{A}}+d-r_{\star_{B}})$}. Then for any 
$d>0$ as one takes $r_{A}>r_{B}$ the outcome $t_{s_A}>t_{s_B}$ is guaranteed, i.e., 
\scalebox{0.9}{$\theta(r_{A}-r_{B})\Rightarrow\theta(t_{s_A}-t_{s_B})$} for $d>0$. In our analysis 
we have considered $d>0$, transformed the coordinate $t$ to $r$ using relation 
(\ref{eq:outgoing-null-coord}), and utilized the $\theta(r_{A}-r_{B})$ expression to change the 
limit of $r_{B}$ to $[\rh,r_{A}]$ from $[\rh,\infty)$. Then one can proceeds to evaluate the 
integral of (\ref{eq:IeR-Sch-BV-2}) as
\begin{eqnarray}\label{eq:IeR-Sch-BV-3}
 \mathcal{I}^{R}_{\varepsilon_{\omega_{k}}} 
&=& \int_{\rh}^{\infty} 
dr_{A}\, \frac{r_{A}+\rh}{r_{A}-\rh}\int_{\rh}^{r_{A}} 
dr_{B}\, \frac{r_{B}+\rh}{r_{B}-\rh}\nonumber\\ 
~&& \times~ e^{i\{\Delta 
E^{B} r_{B}+\Delta 
E^{A} (r_{A}+d)\}}~ \bigg(\frac{r_{B}}{\rh}-1\bigg)^{2i\rh\Delta 
E^{B}}\nonumber\\
~&& \times~ \bigg(\frac{r_{A}}{\rh}-1\bigg)^{2i\rh\Delta 
E^{A}} \bigg[e^{-i\omega_{k} (2r_{A}+d-2r_{B})}\nonumber\\
~&& \times~
\bigg(\frac{r_{A}-\rh}{r_{B}-\rh}\bigg)^{-2i\rh\omega_{k}} - e^{-i\omega_{k} 
(2r_{B}-2r_{A}-d)}\nonumber\\
~&& \times~\bigg(\frac{r_{B}-\rh}{r_{A}-\rh}\bigg)^{-2i\rh\omega_{k}}
- 2i\sin{(\omega_{k}d)}\bigg]~.
\end{eqnarray}
Here we notice that the contribution from the quantity $e^{-i\omega_{k} (u_{A}-u_{B})}- 
e^{-i\omega_{k} (u_{B}-u_{A})} = - 2i\sin{(\omega_{k}d)}$ is independent of $t_{j}$. With the 
change of variables $y_{B}=r_{B}/\rh-1$ and $y_{A}=r_{A}/\rh-1$, the 
above integral simplifies to
\begin{eqnarray}\label{eq:IeR-Sch-BV-4}
\mathcal{I}^{R}_{\varepsilon_{\omega_{k}}} &=& \rh^2 \Bigg[e^{-i\omega_{k} 
\,d}\int_{0}^{\infty} 
dy_{A}~ \frac{y_{A}+2}{y_{A}}~\frac{e^{i\rh(\Delta 
E^{A}-2\omega_{k})(y_{A}+1)}}{y_{A}^{-2i\rh(\Delta 
E^{A}-\omega_{k})}} \nonumber\\
~&& \int_{0}^{y_{A}} 
dy_{B}~ \frac{y_{B}+2}{y_{B}}~\frac{e^{i\rh(\Delta 
E^{B}+2\omega_{k})(y_{B}+1)}}{y_{B}^{-2i\rh(\Delta 
E^{B}+\omega_{k})}}\nonumber\\
~&& - e^{i\omega_{k} 
\,d}\int_{0}^{\infty} 
dy_{A}~ \frac{y_{A}+2}{y_{A}}~\frac{e^{i\rh(\Delta 
E^{A}+2\omega_{k})(y_{A}+1)}}{y_{A}^{-2i\rh(\Delta 
E^{A}+\omega_{k})}} \nonumber\\
~&& \int_{0}^{y_{A}} 
dy_{B}~ \frac{y_{B}+2}{y_{B}}~\frac{e^{i\rh(\Delta 
E^{B}-2\omega_{k})(y_{B}+1)}}{y_{B}^{-2i\rh(\Delta 
E^{B}-\omega_{k})}}\nonumber\\
~&& 
- 2i\sin{(\omega_{k}d)} \int_{0}^{\infty} 
dy_{A}~ \frac{y_{A}+2}{y_{A}}~\frac{e^{i\rh\Delta 
E^{A}(y_{A}+1)}}{y_{A}^{-2i\rh\Delta 
E^{A}}} \nonumber\\
~&& \int_{0}^{y_{A}} 
dy_{B}~ \frac{y_{B}+2}{y_{B}}~\frac{e^{i\rh\Delta 
E^{B}(y_{B}+1)}}{y_{B}^{-2i\rh\Delta 
E^{B}}}\Bigg]\,e^{i\Delta 
E^{A} d}\,.
\end{eqnarray}
One may go through Appendix \ref{Appn:IW-IR-BV}, Eq. (\ref{eq:IeR-Sch-BV-5}), for an analytical 
evaluation of this integral, which we have estimated introducing regulators. These regulators make 
the otherwise divergent integrals convergent.
One gets the analytical expression of this integral in terms of the \emph{Gamma functions} 
$\Gamma(x)$ and \emph{Hypergeometric functions} $~_2F_1(a,b;c;x)$. We also mention that the quantity with 
a multiplicative $2i\sin{(\omega_{k}d)}$ term here has negligible contribution compared to the 
other terms, see Appendix \ref{Appn:IW-IR-Unruh}.

Like the concurrence defined in Eq. (\ref{eq:concurrence-I}) one perceives that in the symmetric 
case the quantity \scalebox{0.9}{$\mathcal{C}_{\mathcal{I}_{\omega_{k}}} =  
(|\mathcal{I}_{\varepsilon_{\omega_{k}}}| -\mathcal{I}_{j_{\omega_{k}}})$} represents the 
concurrence corresponding to a specific field mode frequency $\omega_{k}$, where 
\scalebox{0.9}{$\mathcal{I}_{\varepsilon_{\omega_{k}}} = \mathcal{I}_{\varepsilon_{\omega_{k}}}^{W}+ 
\mathcal{I}_{\varepsilon_{\omega_{k}}}^{R}$}.
%
In Fig. \ref{fig:Concurrence-Schwarzschild-d0-dEsame} we have plotted the dimensionless quantity \scalebox{0.9}{$\mathcal{C}_{\mathcal{I}_{\omega_{k}}} /\rh^2$}, which represents the concurrence, as a function of the dimensionless frequency \scalebox{0.9}{$\overline{\omega}_{k} = \rh\omega_{k}$} for fixed \scalebox{0.9}{$\overline{\Delta E^{A}}=\rh\Delta E^{A}=0.5$}, \scalebox{0.9}{$\overline{\Delta E^{B}}=\rh\Delta E^{B}=0.5$}, and for $d/\rh=0$ and $d/\rh=1$ respectively. For $d/\rh=0$ the entanglement harvesting takes a peak at a certain $\overline{\omega}_{k}$. Whereas, for $d/\rh=1$ there are some periodic shadow points. Moreover, the maximum amount of \scalebox{0.9}{$\mathcal{C}_{\mathcal{I}_{\omega_{k}}}$} dampens as \scalebox{0.9}{$\overline{\omega}_{k}$} increases in successive periods. Although, in this case we are getting shadow points but it is possible to get shadow regions for large values of $d/\rh$. This will be clarified in a short while. Interestingly, this periodicity is not due to the \scalebox{0.9}{$\mathcal{I}_{j_{\omega_{k}}}$} (see Fig. \ref{fig:Ij-Schwarzschild-d0-dEsame}). Whereas, it can be observed from Fig. \ref{fig:Ie-Schwarzschild-d0-dEsame} that \scalebox{0.9}{$|\mathcal{I}_{\varepsilon_{\omega_{k}}}|$} provides the major contribution in the concurrence and also has similar periodicity. Therefore, this periodic nature of \scalebox{0.9}{$\mathcal{C}_{\mathcal{I}_{\omega_{k}}}$} for $d\neq 0$ as a function of \scalebox{0.9}{$\overline{\omega}_{k}$} is due to the non-local term.

On the other hand, in Fig. \ref{fig:Concurrence-Schwarzschild-d0-VdE} we have plotted \scalebox{0.9}{$\mathcal{C}_{\mathcal{I}_{\omega_{k}}} /\rh^2$} as a function of the dimensionless transition energy \scalebox{0.9}{$\overline{\Delta E} = \rh\Delta E$} of the detectors for fixed \scalebox{0.9}{$\overline{\omega}_{k} = 1$} and for $d/\rh=0$ and $d/\rh=1$ respectively, where \scalebox{0.9}{$\Delta E = \Delta E^{A} = \Delta E^{B}$}. These plots also proclaim the possibility of entanglement harvesting for low values of \scalebox{0.9}{$\Delta E$}.
We also see that the amount of harvested entanglement decreases with increasing detector transition energy.
Fig. \ref{fig:Ie-Schwarzschild-d0-VdE} and Fig. \ref{fig:Ij-Schwarzschild-d0-VdE} confirm that major contribution in concurrence comes from \scalebox{0.9}{$|\mathcal{I}_{\varepsilon_{\omega_{k}}}|$} rather than \scalebox{0.9}{$\mathcal{I}_{j_{\omega_{k}}}$}. 
%

Next in Fig. \ref{fig:Concurrence-Schwarzschild-wk0p4-dd} we have plotted \scalebox{0.9}{$\mathcal{C}_{\mathcal{I}_{\omega_{k}}} /\rh^2$} with respect to the dimensionless distance \scalebox{0.9}{$d/\rh$} between the two detectors' null trajectories for fixed \scalebox{0.9}{$\overline{\Delta E} = 0.5$} and different \scalebox{0.9}{$\overline{\omega}_{k}$} respectively. These plots and corresponding numerical values confirm that entanglement harvesting is happening in a periodic manner with respect to \scalebox{0.9}{$d/\rh$}, with period depending on \scalebox{0.9}{$\overline{\omega}_{k}$}. 
We observe that this period decreases with increasing values of $\overline{\omega_{k}}$.
They also confirm that there are periodic \scalebox{0.9}{$d/\rh$} values where harvesting stops, i.e., \scalebox{0.9}{$\mathcal{C}_{\mathcal{I}_{\omega_{k}}} /\rh^2$} becomes zero. We further observed that for small $\overline{\omega_{k}}~(=0.05,\,0.1)$ there are entanglement harvesting shadow regions while for $\overline{\omega_{k}}~(=1,\,2)$ these regions become point-like. This figure also confirms that for large $d$ values \scalebox{0.9}{$\mathcal{C}_{\mathcal{I}_{\omega_{k}}} /\rh^2$} as a function of  $\overline{\omega_{k}}$ shows shadow regions rather than points, which we was promised to be shown earlier.
The entanglement harvesting shadow regions have also been observed earlier in the case of black 
holes \cite{Henderson:2017yuv, Robbins:2020jca}. However, in \cite{Henderson:2017yuv} and 
\cite{Robbins:2020jca}, the entanglement harvesting shadow regions are observed in different 
spacetimes, related to the analysis near the event horizons of static and rotating BTZ black holes, 
respectively.

\subsubsection{Unruh vacuum}

Here we consider the Wightman function estimated with the Unruh vacuum from Eq. 
(\ref{eq:Wightman-Unruh}) to estimate integrals of (\ref{eq:cond-entanglement}) to investigate the 
entanglement harvesting condition. In particular, one can find out the individual detector 
transition probabilities $\mathcal{I}_{j}$ as
\begin{eqnarray}\label{eq:Ij-Sch-UV-1}
 \mathcal{I}_{j} &=& \int_{-\infty}^{\infty} dt'_{j}\int_{-\infty}^{\infty} 
dt_{j}~ e^{-i\Delta E^{j}(t'_{j}-t_{j})} G^{+}_{B}(X'_{j},X_{j})\nonumber\\
~&=& \int_{0}^{\infty} \frac{d\omega_{k}}{4\pi\omega_{k}}~ 
\mathcal{I}_{j_{\omega_{k}}}~,
\end{eqnarray}
where $\mathcal{I}_{j_{\omega_{k}}}$ are now given by 
\begin{eqnarray}\label{eq:Ij-Sch-UV-2}
 \mathcal{I}_{j_{\omega_{k}}} &=& \int_{-\infty}^{\infty} 
dt'_{j}\int_{-\infty}^{\infty} 
dt_{j} e^{-i\Delta E^{j}(t'_{j}-t_{j})}\nonumber\\
~&&~~ \times~\big[e^{-i\omega_{k} 
(U_{j'}-U_{j})} + e^{-i\omega_{k} 
(v_{j'}-v_{j})}\big]~.
\end{eqnarray}
With the help of Eq. (\ref{eq:v-Diff-Un}) and (\ref{eq:U-Diff-Un}) one can observe that this 
integral is same as the one from Eq. (\ref{eq:Ij-Sch-BV-3}) of the Boulware vacuum case. Then this 
should also provide the same result of Eq. (\ref{eq:Ij-Sch-BV-6}).
%
%
Let us now evaluate the integrals $\mathcal{I}^{W}_{\varepsilon_{\omega_{k}}}$ and 
$\mathcal{I}^{R}_{\varepsilon_{\omega_{k}}}$ from (\ref{eq:IeW-Sch-BV-1}) and 
(\ref{eq:IeR-Sch-BV-1}) with the Green's functions (\ref{eq:Wightman-Unruh}) evaluated from the 
Unruh vacuum. First let us proceed to calculate $\mathcal{I}^{W}_{\varepsilon_{\omega_{k}}}$, which 
in this case turns out to be
\begin{widetext}
\begin{eqnarray}\label{eq:IeW-Sch-UV-1}
 \mathcal{I}^{W}_{\varepsilon_{\omega_{k}}} &=& \int_{-\infty}^{\infty} 
dt_{B}\int_{-\infty}^{\infty} 
dt_{A}~ e^{i(\Delta E^{B}t_{B}+\Delta E^{A}t_{A})}~\big[e^{-i\omega_{k} 
(U_{B}-U_{A})} + e^{-i\omega_{k} 
(v_{B}-v_{A})}\big]\nonumber\\
~&=& \int_{-\infty}^{\infty} 
dt_{B}\int_{-\infty}^{\infty} 
dt_{A}\, e^{i(\Delta E^{B}t_{B}+\Delta E^{A}t_{A})} 
\bigg[\exp\Big\{i\omega_{k}2\rh\big(1-e^{-\frac{d}{2\rh}}\big)\Big\}  + e^{-i\omega_{k} 
(2r_{B}-2r_{A}-d)} \bigg(\tfrac{r_{B}-\rh}{r_{A}-\rh}\bigg)^{-2i\rh\omega_{k}}\bigg].
\end{eqnarray}
\end{widetext}
The integration with the first term inside the square bracket will vanish due to the Dirac delta 
distributions $\delta(\Delta E^{j})$, with $\Delta E^{j}>0$. Then this integral becomes exactly same 
with the one for the Boulware vacuum from Eq. (\ref{eq:IeW-Sch-BV-3}). On the other hand, in a 
similar manner one can evaluate the quantity $\mathcal{I}^{R}_{ \varepsilon_{\omega_{k}}}$ as
\begin{widetext}
\begin{eqnarray}\label{eq:IeR-Sch-UV-1}
 \mathcal{I}^{R}_{\varepsilon_{\omega_{k}}} &=& \int_{-\infty}^{\infty} 
dt_{A}\int_{-\infty}^{\infty} dt_{B}~ e^{i(\Delta E^{B}t_{B}+\Delta 
E^{A}t_{A})}\Big[\theta(t_{s_A}-t_{s_B})~\big\{e^{-i\omega_{k} 
(v_{A}-v_{B})} 
- e^{-i\omega_{k} (v_{B}-v_{A})}\big\} \nonumber\\
~&& ~~~~~~~~~~~~~~~~~~~~~~~~~~~~~~~~~~~~~~~~~~~~~~~~~~~~~~~~~~~~~~~~+~ 
\theta(T_{K_A}-T_{K_B})~ \big\{e^{-i\omega_{k} (U_{A}-U_{B})} - e^{-i\omega_{k} 
(U_{B}-U_{A})}\big\}\Big]\nonumber\\
~&=& \int_{\rh}^{\infty} 
dr_{A}\, \frac{r_{A}+\rh}{r_{A}-\rh}\int_{\rh}^{r_{A}} 
dr_{B}\, \frac{r_{B}+\rh}{r_{B}-\rh}~ e^{i\{\Delta 
E^{B} r_{B}+\Delta 
E^{A} (r_{A}+d)\}}~ \bigg(\frac{r_{B}}{\rh}-1\bigg)^{2i\rh\Delta 
E^{B}} \bigg(\frac{r_{A}}{\rh}-1\bigg)^{2i\rh\Delta 
E^{A}} \nonumber\\
~&&\times \bigg[e^{-i\omega_{k} (2r_{A}+d-2r_{B})} 
\bigg(\tfrac{r_{A}-\rh}{r_{B}-\rh}\bigg)^{-2i\rh\omega_{k}} - e^{-i\omega_{k} (2r_{B}-2r_{A}-d)} 
\bigg(\tfrac{r_{B}-\rh}{r_{A}-\rh}\bigg)^{-2i\rh\omega_{k}} - 
2i\sin\Big\{\omega_{k}2\rh\big(1-e^{-\frac{d}{2\rh}}\big)\Big\} \bigg].\nonumber\\
\end{eqnarray}
\end{widetext}
Here we note that while dealing with modes which are represented in terms of the EF null coordinates 
the Heaviside step function arising from the Feynman propagator should take the time to be the 
Schwarzschild time $\ts$. While for modes which are represented in terms of the Kruskal null 
coordinates $(U,\,V)$, one should take the relevant time to be the Kruskal time $T_{K} = (U+V)/2$. 
This Kruskal time is expressed in terms of the Schwarzschild time and the tortoise coordinate as 
$T_{K} = 2\rh\,e^{\rstar/2\rh}\,\sinh{(\ts/2\rh)}$. Using this expression and the null paths followed by Alice and Bob as $t_{s_{A}}-r_{\star_{A}} = d>0$ and 
$t_{s_{B}}-r_{\star_{B}} = 0$, respectively one finds that $T_{K_A}>T_{K_B}$ implies $r_A> r_B$ (see Appendix \ref{Appn:Stepfn-simplification-Unruh}). Therefore the Heaviside function $\theta(T_{K_A}-T_{K_B})$ can be replaced by $\theta(r_{A}-r_{B})$. Similarly, as explained earlier, $t_{s_A}>t_{s_B}$ implies $r_A>r_B$ and hence we replace $\theta(t_{s_A}-t_{s_B})$ by $\theta(r_{A}-r_{B})$. Using these the last result of (\ref{eq:IeR-Sch-UV-1}) has been obtained. 
We observe that the expression of $\mathcal{I}^{R}_{ \varepsilon_{\omega_{k}}}$ from Eq. 
(\ref{eq:IeR-Sch-UV-1}) and (\ref{eq:IeR-Sch-BV-4}) are mathematically same except for the last 
term. In Appendix \ref{Appn:IW-IR-Unruh} we have shown that this last term 
$2i\sin\big\{\omega_{k}2\rh\big( 1-e^{-d/2\rh}\big)\big\}$ is many order smaller than the rest of 
the expressions for the same set of parameter values. Therefore, the nature of entanglement 
harvesting with respect to the Unruh vacuum is similar to that in the Boulware vacuum.

\subsection{de Sitter universe}

\subsubsection{$(1+1)-$dimensions}

Let us now evaluate the integrals of Eq. (\ref{eq:all-integrals}) for detectors in null 
trajectories 
in a de Sitter spacetime, so that one can check the entanglement condition 
(\ref{eq:cond-entanglement}) and also quantify the harvested entanglement using 
(\ref{eq:concurrence-I}). In this regard, we first consider the integral $\mathcal{I}_j$, which 
using the Green's function of Eq. (\ref{eq:GreensFn-DS}) is represented as
%

%
\begin{figure}[h]
\centering
 \includegraphics[width=0.85\linewidth]{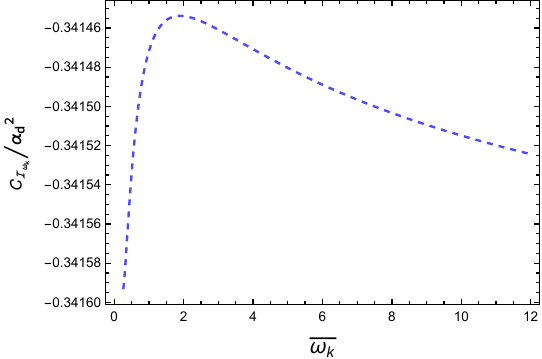}
 \caption{The quantity $\mathcal{C}_{\mathcal{I}_{\omega_{k}}}/\alpha_{d}^2$, signifying the concurrence, is plotted for two outgoing null detectors in a $(1+1)$ dimensional de Sitter spacetime with respect to the dimensionless frequency of the field $\overline{\omega}_{k} = \omega_{k}\,\alpha_{d}$ for fixed dimensionless detector transition energy $\overline{\Delta E}=\Delta E\,\alpha_{d}=0.5$. The other fixed parameter is $d/\alpha_{d}=0$.}
 \label{fig:Concurrence-DS-d0-dwk}
\end{figure}
%

%
%
%
\begin{figure}[h]
\centering
 \includegraphics[width=0.85\linewidth]{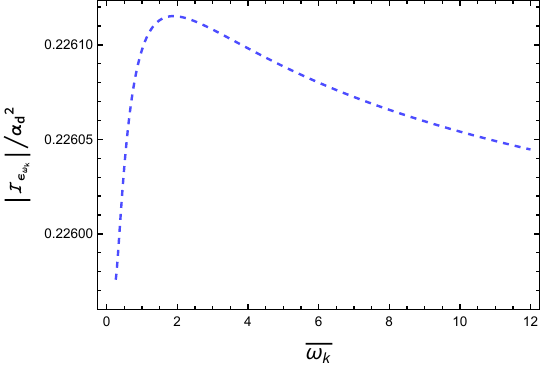}
 \caption{The quantity $|\mathcal{I}_{\varepsilon_{\omega_{k}}} |/\alpha_{d}^2$ is plotted for two outgoing null detectors in a $(1+1)$ dimensional de Sitter spacetime with respect to the dimensionless frequency of the field $\overline{\omega}_{k}$ for fixed $\overline{\Delta E}=0.5$ and $d/\alpha_{d}=0$.}
 \label{fig:Ie-DS-d0-dwk}
\end{figure}
%

%
%
%
\begin{figure}[h]
\centering
 \includegraphics[width=0.85\linewidth]{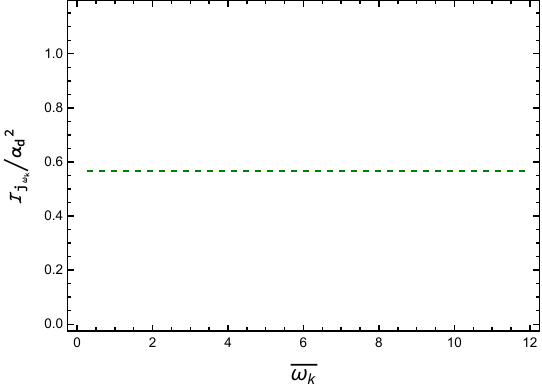}
 \caption{The quantity $\mathcal{I}_{j_{\omega_{k}}}/\alpha_{d}^2$, signifying the 
individual detector transition probability, is plotted as a function of the dimensionless frequency 
of the field $\overline{\omega}_{k}$ for a detector in an outgoing null path in a $(1 + 1)$ 
dimensional de Sitter spacetime. The dimensionless transition energy of the detector is fixed at 
$\overline{\Delta E} = 0.5$. From this figure, one can notice that the individual detector transition 
probability, in this case, is independent of the frequency of the field $\overline{\omega}_{k}$. 
From (\ref{eq:Ij-DS-1p1-5}) one should note that they are also independent of $d/\alpha_{d}$.}
 \label{fig:Ij-DS-d0-dwk}
\end{figure}
%

%
\begin{figure}[h]
\centering
 \includegraphics[width=0.85\linewidth]{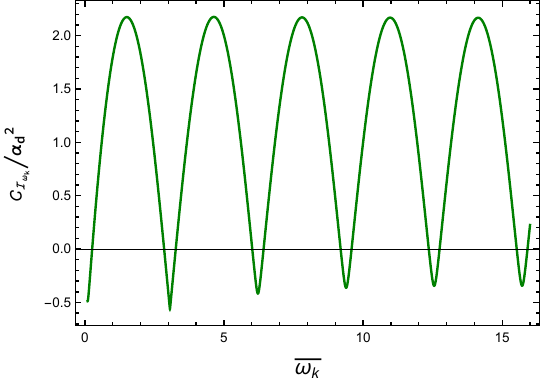}
 \caption{The quantity $\mathcal{C}_{\mathcal{I}_{\omega_{k}}}/\alpha_{d}^2$, signifying the 
concurrence, is plotted for two outgoing null detectors in a $(1+1)$ dimensional de Sitter spacetime 
with respect to the dimensionless frequency of the field $\overline{\omega}_{k}$ for fixed 
dimensionless detector transition energy $\overline{\Delta E}=0.5$, and fixed $d/\alpha_{d}=1$.}
 \label{fig:Concurrence-DS-d1-dwk}
\end{figure}
%

%
%
%
\begin{figure}[h]
\centering
 \includegraphics[width=0.85\linewidth]{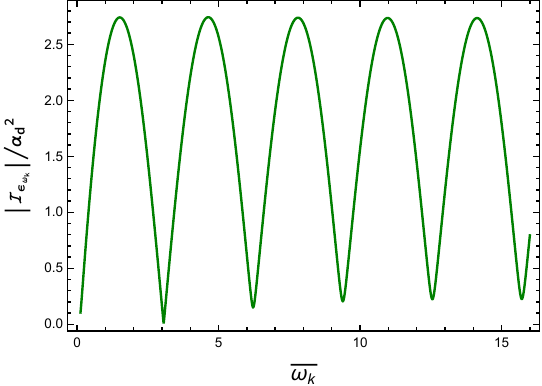}
 \caption{The quantity $|\mathcal{I}_{\varepsilon_{\omega_{k}}}|/\alpha_{d}^2$ is plotted for two 
outgoing null detectors in a $(1+1)$ dimensional de Sitter spacetime with respect to the 
dimensionless frequency of the field $\overline{\omega}_{k}$ for fixed $\overline{\Delta E}=0.5$, and 
$d/\alpha_{d}=1$.}
 \label{fig:Ie-DS-d1-dwk}
\end{figure}
\begin{figure}[h]
\centering
 \includegraphics[width=0.85\linewidth]{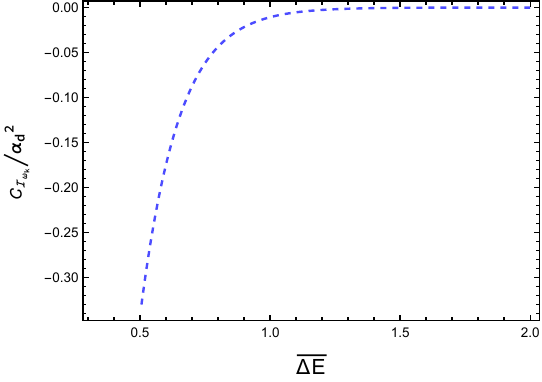}
 \caption{The quantity $\mathcal{C}_{\mathcal{I}_{\omega_{k}}}/\alpha_{d}^2$, signifying the 
concurrence, is plotted for two outgoing null detectors in a $(1+1)$ dimensional de Sitter spacetime 
with respect to the dimensionless transition energy $\overline{\Delta E}$ of the detectors for fixed 
dimensionless frequency of the field $\overline{\omega}_{k}=0.2$, and fixed $d/\alpha_{d}=0$.}
 \label{fig:Concurrence-DS-d0-dEsame}
\end{figure}
%

%
%
%
\begin{figure}[h]
\centering
 \includegraphics[width=0.85\linewidth]{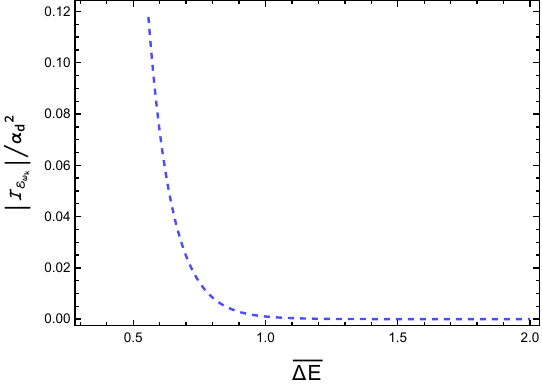}
 \caption{The quantity $|\mathcal{I}_{\varepsilon_{\omega_{k}}}|/\alpha_{d}^2$ is plotted for two 
outgoing null detectors in a $(1+1)$ dimensional de Sitter spacetime with respect to the 
dimensionless transition energy $\overline{\Delta E}$ of the detectors for fixed dimensionless 
frequency of the field $\overline{\omega}_{k}=0.2$, and fixed $d/\alpha_{d}=0$.}
 \label{fig:Ie-DS-d0-dEsame}
\end{figure}
%

%
%
\begin{figure}[h]
\centering
 \includegraphics[width=0.85\linewidth]{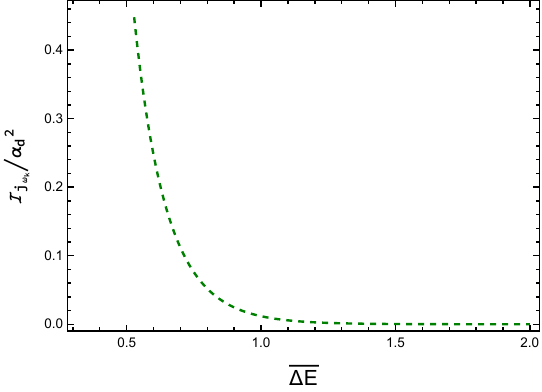}
 \caption{The quantity $\mathcal{I}_{j_{\omega_{k}}}/\alpha_{d}^2$ is plotted for two outgoing null 
detectors in a $(1+1)$ dimensional de Sitter spacetime with respect to the dimensionless transition 
energy $\overline{\Delta E}$ of the detectors for fixed dimensionless frequency of the field 
$\overline{\omega}_{k}=0.2$. This nature will be the same for all $d/\alpha_{d}$ values as the quantity is independent of $d$.}
 \label{fig:Ij-DS-d0-dEsame}
\end{figure}

\begin{figure}[h]
\centering
 \includegraphics[width=0.85\linewidth]{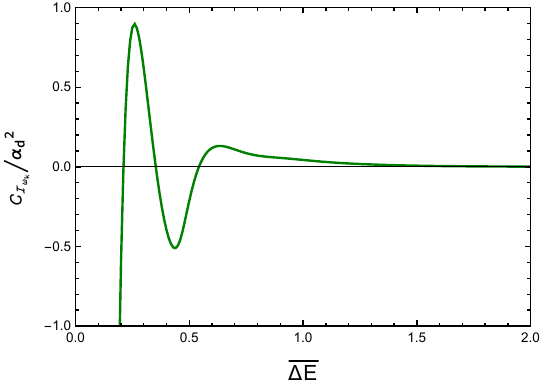}
 \caption{The quantity $\mathcal{C}_{\mathcal{I}_{\omega_{k}}}/\alpha_{d}^2$, signifying the 
concurrence, is plotted for two outgoing null detectors in a $(1+1)$ dimensional de Sitter spacetime 
with respect to the dimensionless transition energy $\overline{\Delta E}$ of the detectors for fixed 
dimensionless frequency of the field $\overline{\omega}_{k}=0.2$, and fixed $d/\alpha_{d}=1$.}
 \label{fig:Concurrence-DS-d1-dEsame}
\end{figure}

%
%
\begin{figure}[h]
\centering
 \includegraphics[width=0.85\linewidth]{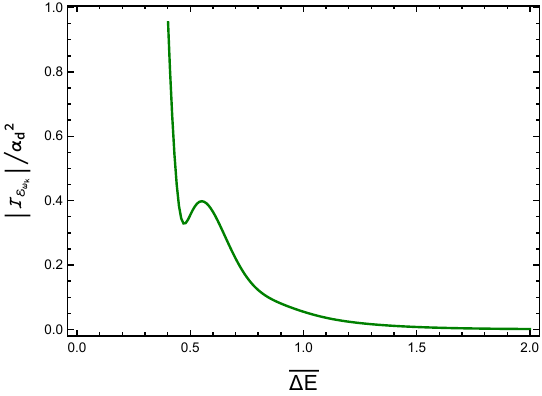}
 \caption{The quantity $|\mathcal{I}_{\varepsilon_{\omega_{k}}}|/\alpha_{d}^2$ is plotted for two 
outgoing null detectors in a $(1+1)$ dimensional de Sitter spacetime with respect to the 
dimensionless transition energy $\overline{\Delta E}$ of the detectors for fixed dimensionless 
frequency of the field $\overline{\omega}_{k}=0.2$, and fixed $d/\alpha_{d}=1$.}
 \label{fig:Ie-DS-d1-dEsame}
\end{figure}

%
%
%
%
\begin{figure}[h]
\centering
 \includegraphics[width=0.85\linewidth]{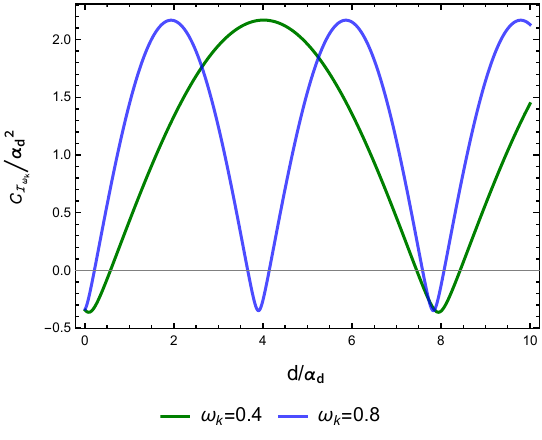}\\
 \caption{The quantity $\mathcal{C}_{\mathcal{I}_{\omega_{k}}} /\alpha_{d}^2$ is plotted for two 
outgoing null detectors in different parallel paths in a $(1+1)$ dimensional de-Sitter spacetime 
with respect to the separation between the two paths $d/\alpha_{d}$ for different dimensionless frequencies of 
the field. The detector transition energies are fixed at $\overline{\Delta E}=0.5$.}
 \label{fig:Concurrence-DS-dd}
\end{figure}


\begin{eqnarray}\label{eq:Ij-DS-1p1-1}
\mathcal{I}_j &=& \int_{-\infty}^{\infty}dt_j^{\prime} \int_{-\infty}^{\infty}dt_j e^{-i\Delta 
E^j(t_j^{\prime}-t_j)}\, G_D^{+}(X'_{j},X_{j}) \nonumber\\
~&=&\int_{0}^{\infty}\frac{d\omega_k}{4\pi\omega_{k}}\,\mathcal{I}_{j_{\omega_k}}~.
\end{eqnarray}
Here the detector times $\tau_{j}$ are represented by the de Sitter coordinate times $t_{j}$.
Like the previous Schwarzschild case here also we shall be evaluating $\mathcal{I}_{j_{\omega_k}}$, 
which correspond to a certain frequency of the field mode, to arrive at the entanglement harvesting 
condition. With the consideration of outgoing null paths for both Alice and Bob one can express 
the integrals $\mathcal{I}_{j_{\omega_k}}$ as
\begin{eqnarray}\label{eq:Ij-DS-1p1-2}
\mathcal{I}_{j_{\omega_k}} &=& \int_{-\infty}^{\infty}dt_j^{\prime} \int_{-\infty}^{\infty}dt_j 
e^{-i\Delta E^j(t_j^{\prime}-t_j)}\nonumber\\
&& \times~\big[e^{2i\omega_{k} \alpha_{d}(e^{-t^{\prime}_j/\alpha_d}-e^{-t_j/\alpha_d}) }  + 
1\big]~.
\end{eqnarray}
Here one can observe that the integration over the second additive unity is simple and provides 
multiplicative factors of the Dirac delta distribution $\delta(\Delta E^j)$. For non zero detector 
transition energy $\Delta E^j>0$ these quantities are bound to make the concerned part of the 
integral vanish. Then the previous integral with the change of variables $e^{-t/\alpha_d}=z$ can be 
evaluated as
\begin{eqnarray}\label{eq:Ij-DS-1p1-3}
\mathcal{I}_{j_{\omega_k}} = \alpha_d^2\,\bigg|\int^{\infty}_{0}
dz\,z^{i\Delta E^j\alpha_d-1}\,e^{2i\omega_k\alpha_dz}\bigg|^2~.
\end{eqnarray}
To evaluate this integral we introduce regulators of the form $(z^{\epsilon}\,e^{-\epsilon z})$, 
where $\epsilon$ is a positive real parameter with $\epsilon\ll1$. One can get the actual value of 
the integral by taking the limit $\epsilon\to 0$ after evaluating the regulated integral as
\begin{eqnarray}\label{eq:Ij-DS-1p1-4}
&&\lim_{\epsilon\to0}\,\bigg[\int_{0}^{\infty}dz\,z^{i\Delta 
E^j\alpha_d-1+\epsilon}\,e^{(2i\omega_k\alpha_d-\epsilon)z}\bigg]\nonumber\\
~&=& e^{-\pi\alpha_d\Delta E^j/2}\, (2\omega_k\alpha_d)^{-i\alpha_d\Delta 
E^j}\,\Gamma(i\alpha_d\Delta E^j)~.
\end{eqnarray}
Then one can promptly express $\mathcal{I}_{j_{\omega_k}}$ as
\begin{eqnarray}\label{eq:Ij-DS-1p1-5}
&&\mathcal{I}_{j_{\omega_k}}=\frac{2\pi\alpha_d}{\Delta 
E^j}\,\frac{1}{e^{2\pi\alpha_d\Delta E^j}-1}~,
\end{eqnarray}
where we have used the \emph{Gamma function} identity $\Gamma(iz)\Gamma(-iz) = \pi/(z\sinh{\pi 
z})$. Let us now evaluate the integral $\mathcal{I}_{\epsilon}$, which can again be expressed as
\begin{eqnarray}\label{eq:Ie-DS-1p1-1}
\mathcal{I}_{\epsilon} &=& -\int_0^{\infty} 
\frac{d\omega_k}{4\pi\omega_k}\,\Big[\mathcal{I}^{W}_{\epsilon_{\omega_k}} + 
\mathcal{I}^{R}_{\epsilon_{\omega_k}}\Big]~,
\end{eqnarray}
which is in same way that we have considered in the Schwarzschild case. Here, one can evaluate the 
quantity $\mathcal{I}^{W}_{\epsilon_{\omega_k}}$ as
\begin{eqnarray}\label{eq:IeW-DS-1p1-1}
~&& \mathcal{I}^{W}_{\epsilon_{\omega_k}} = 
\int_{-\infty}^{\infty}dt_B\int_{-\infty}^{\infty}dt_A\, 
e^{i(\Delta E^{A} t_A+\Delta E^{B} t_B)}\, \nonumber\\ 
~&& ~~~~~~\times~\Big\{e^{-i\omega_{k}(u_{B}-u_{A})} +   
e^{-i\omega_{k}(v_{B}-v_{A})}\Big\}\nonumber\\
&=& \int_{-\infty}^{\infty}dt_B\int_{-\infty}^{\infty}dt_A\,e^{i(\Delta E^{A} t_A+\Delta E^{B} 
t_B)}\, \nonumber\\ 
~&\times& \Big[ e^{i\omega_{k}d} ~+~ e^{-i\omega_{k}d} \, e^{2i\omega_{k} 
\alpha_{d}(e^{-t_B/\alpha_d}-e^{-t_A/{\alpha_d}})}  \Big].
\end{eqnarray}
Here also one can observe that the integration over the first quantity with $e^{i\omega_{k}d}$ as 
multiplicative factor will provide the multiplication of Dirac delta distributions $\delta(\Delta 
E^A)$ and $\delta(\Delta E^B)$. Therefore that part of the integral will vanish as the detector 
transition energy $\Delta E^j>0$. Now with the change of variables $z_{j} = e^{-t_{j}/\alpha_{d}}$ 
one simplifies the previous integral as
\begin{eqnarray}\label{eq:IeW-DS-1p1-2}
\mathcal{I}^{W}_{\epsilon_{\omega_k}} &=& \alpha_d^2\,e^{-i\omega_{k}d}\,\int^{\infty}_{0} 
dz_{A}\,\int^{\infty}_{0} dz_{B}\,e^{2i\omega_k\alpha_d(z_{B}-z_{A})}\nonumber\\
~&& ~~~~~~~~~~~~~\times\,z_{A}^{-i\Delta E^A\alpha_d-1}\,z_{B}^{-i\Delta 
E^B\alpha_d-1}~.
\end{eqnarray}
By introducing regulators of the form $(z_{A}z_{B})^{\epsilon}\,e^{-\epsilon (z_{A} + z_{B})}$, 
with 
a positive real parameter $\epsilon$, one can evaluate this integral. The explicit expression after 
the integration is carried out, is provided in the Appendix. \ref{Appn:IW-IR-DS-1p1}. We now 
proceed 
to evaluate the integral $\mathcal{I}^{R}_{\epsilon_{\omega_k}}$, which can be expressed as
\begin{eqnarray}\label{eq:IeR-DS-1p1-1}
\mathcal{I}^{R}_{\epsilon_{\omega_k}} &=& 
\int_{-\infty}^{\infty}dt_B\int_{-\infty}^{\infty}dt_A\, 
e^{i(\Delta E^{A} t_A+\Delta E^{B} t_B)}\,\theta(\eta_{A}-\eta_{B}) \nonumber\\ 
~&& \times~\Big\{e^{-i\omega_{k}(u_{A}-u_{B})} +   
e^{-i\omega_{k}(v_{A}-v_{B})} \nonumber\\
~&&~~~~~~~ - e^{-i\omega_{k}(u_{B}-u_{A})} -   
e^{-i\omega_{k}(v_{B}-v_{A})}\Big\}\nonumber\\
&=& \alpha_d^2\,\int^{\infty}_{0} 
dz_{A}\,\int^{\infty}_{z_{A}} dz_{B}\,z_{A}^{-i\Delta E^A\alpha_d-1}\,z_{B}^{-i\Delta 
E^B\alpha_d-1}\nonumber\\
~&& \times~\Big\{e^{i\omega_{k}d}\,e^{2i\omega_k\alpha_d(z_{A}-z_{B})} - 
e^{-i\omega_{k}d}\,e^{2i\omega_k\alpha_d(z_{B}-z_{A})}\nonumber\\
~&& ~~~~~~~~~~ - 2i\sin{\omega_{k}d}\Big\}.
\end{eqnarray}
In de Sitter background, the real scalar field is decomposed with respect to the positive and 
negative frequency modes, represented in the conformal time $\eta$. Therefore, the time $T_{j}$ 
inside the Heaviside step function here is denoted by the conformal time.
Here for positive $\alpha_{d}$ one obtains $\theta(\eta_{A}-\eta_{B}) = \theta(t_{A}-t_{B}) = 
\theta(z_{B}-z_{A})$ using the relation (\ref{eta-tReln}), and we have used this fact to realize the 
previous  expression. We also mention that utilization of the function $\theta(z_{B}-z_{A})$ 
transformed the $z_{B}$ integration range from $[0,\infty)$ to $[z_{A},\infty)$ in the 
representation of Eq. (\ref{eq:IeR-DS-1p1-1}).
This integral can be evaluated numerically with the introduction of the regulator of the form 
$(z_{A}z_{B})^{\epsilon}\,e^{-\epsilon (z_{A} + z_{B})}$, with positive real $\epsilon$.

In Fig. \ref{fig:Concurrence-DS-d0-dwk} we have plotted the dimensionless quantity \scalebox{0.9}{$\mathcal{C}_{\mathcal{I}_{\omega_{k}}} /\alpha_{d}^2$}, representing the concurrence, as a function of the dimensionless frequency \scalebox{0.9}{$\overline{\omega}_{k} = \omega_{k}\,\alpha_{d}$} for fixed \scalebox{0.9}{$\overline{\Delta E^{A}}=\alpha_{d}\Delta E^{A}=0.5$, $\overline{\Delta E^{B}}=\alpha_{d}\Delta E^{B}=0.5$}, and $d=0$. This plot asserts that entanglement harvesting is not possible when the detectors move along the same path. We observe from Fig. \ref{fig:Ie-DS-d0-dwk} and \ref{fig:Ij-DS-d0-dwk} that here \scalebox{0.9}{$\mathcal{I}_{j_{\omega_{k}}}$} is larger than \scalebox{0.9}{$|\mathcal{I}_{\varepsilon_{\omega_{k}}}|$} and as a result \scalebox{0.9}{$\left(|\mathcal{I}_{\varepsilon} |- \sqrt{\mathcal{I}_{A} \mathcal{I}_{B}}\right)/\alpha_{d}^2$} is negative. However, Fig. \ref{fig:Concurrence-DS-d1-dwk}, obtained for the same parameters but \scalebox{0.9}{$d/\alpha_{d}=1$}, shows that \scalebox{0.9}{$\mathcal{C}_{\mathcal{I}_{\omega_{k}}} /\rh^2$} is now positive making entanglement harvesting possible in this scenario. 
%
Moreover, one observes that there are periodic entanglement harvesting regions with respect to frequency $\overline{\omega}_{k}$. Like earlier case here also this periodicity is only due to \scalebox{0.9}{$|\mathcal{I}_{\varepsilon_{\omega_{k}}}|$}, see Fig. \ref{fig:Ij-DS-d0-dwk} and \ref{fig:Ie-DS-d1-dwk}.
%

We now plot \scalebox{0.9}{$\mathcal{C}_{\mathcal{I}_{\omega_{k}}} /\alpha_{d}^2$} with respect to the dimensionless transition energy $\overline{\Delta E}$ of the detectors for fixed \scalebox{0.9}{$\overline{\omega}_{k} = 0.2$} and $d=0$ in Fig. \ref{fig:Concurrence-DS-d0-dEsame}. This plot also states that for $d=0$ the entanglement harvesting is not possible. This claim is supported by Fig. \ref{fig:Ie-DS-d0-dEsame} and \ref{fig:Ij-DS-d0-dEsame}. However, from Fig. \ref{fig:Concurrence-DS-d1-dEsame} with \ref{fig:Ie-DS-d1-dEsame}, where the similar plots are obtained for $d/\alpha_{d}=1$, we observe that entanglement harvesting is possible in certain discrete ranges of \scalebox{0.9}{$\overline{\Delta E}$}.

Finally in Fig. \ref{fig:Concurrence-DS-dd} we plot the concurrence \scalebox{0.9}{$\mathcal{C}_{\mathcal{I}_{\omega_{k}}} /\alpha_{d}^2$} with respect to $d/\alpha_{d}$ for fixed \scalebox{0.9}{$\overline{\Delta E} = 0.5$} and different $\overline{\omega}_{k}$. This plot is consistent with the findings of Fig. \ref{fig:Concurrence-DS-d0-dwk} and Fig. \ref{fig:Concurrence-DS-d0-dEsame}, reconfirming that for $d/\alpha_{d}=0$ harvesting is not possible. Fig. \ref{fig:Concurrence-DS-dd} indicates that like the Schwarzschild case here also harvesting is periodic with respect to $d/\alpha_{d}$. Here also one perceives the occurrence of entanglement harvesting shadow regions, the length of which decreases with increasing frequency $\overline{\omega}_{k}$. However, here we have seen that with increasing $\overline{\omega}_{k}$ (in the range $[10^{-4},10^{4}]$) the shadow regions do not become shadow points.

%
\vspace{0.2cm}

\subsubsection{(1+3)-dimensions}

In $(1+3)-$dimensional de Sitter spacetime one can express the integrals from Eq. 
(\ref{eq:all-integrals}), essential for perceiving the entanglement harvesting 
(\ref{eq:cond-entanglement}), as 
\begin{eqnarray}\label{eq:Ij-DS-1p3-1}
\mathcal{I}_j&=&\int_{-\infty}^{\infty}dt_j^{\prime}
 \int_{-\infty}^{\infty}dt_j\, e^{-i\Delta E^j(t_j^{\prime}-t_j)} G_D^{+}(X'_{j},X_{j})\nonumber\\
&&~~~ \nonumber\\
~&=&\int \frac{d^2k_{\perp}}{(2\pi)^3 
2\omega_{k}}\int_{0}^{\infty}dk_{x}~\mathcal{I}_{j_{\omega_k}}~,
\end{eqnarray}
where, $\omega_{k}^2 = k_{\perp}^2 + k_{x}^2$, and $k_{\perp}^2 = k_{y}^2 +k_{z}^2$. These
integrations over $t_j^{\prime}$, and $t_j$ can be solved for detectors in outgoing null 
paths using the Green's functions (\ref{eq:GreensFn-DS-1p3}) as
\begin{eqnarray}\label{eq:Ij-DS-1p3-2}
\mathcal{I}_{j_{\omega_k}} &=& \int_{-\infty}^{\infty}dt_j^{\prime}\int_{-\infty}^{\infty}dt_j\, 
\,\frac{e^{-i\Delta E^j(t_j^{\prime}-t_j)}}{a(\eta_j')a(\eta_j)}\nonumber\\
~&\times&~ \Big[e^{i{k_x}\Delta{x}_{j'j}-i\omega_k\Delta\eta_{j'j}
} + e^{-i{k_x}\Delta{x}_{j'j}-i\omega_k\Delta\eta_{j'j}
}\Big]\nonumber\\
&=& \int_{-\infty}^{\infty}dt_j^{\prime}\int_{-\infty}^{\infty}dt_j\, 
 e^{-i\Delta 
E^j(t_j^{\prime}-t_j)}\nonumber\\
&\times& 
\frac{1}{e^{t'_{j}/\alpha_d + t_j/\alpha_d}}\, 
\Big[e^{i\alpha_d(\omega_k-k_{x})(e^{-t'_{j}/\alpha_d}-e^{
-t_j/\alpha_d})} \nonumber\\
~&& ~~~+ e^{i\alpha_d(\omega_k + k_x)(e^{-t'_{j}/\alpha_d}-e^{
-t_j/\alpha_d})}\Big]~.
\end{eqnarray}
With the change of variables $z_{j}=e^{-t_{j}/\alpha_d}$ this expression changes into
\begin{eqnarray}\label{eq:Ij-DS-1p3-3}
\mathcal{I}_{j_{\omega_k}} &=& \alpha_d^2 \Bigg[\bigg| \int_{0}^{\infty}dz_j\, 
z_j^{-i\alpha_d\Delta 
E^j}e^{-i\alpha_d(\omega_k - k_x)z_j}\bigg|^2\nonumber\\
&&~ + \bigg| \int_{0}^{\infty}dz_j\, z_j^{-i\alpha_d\Delta 
E^j}e^{-i\alpha_d(\omega_k + k_x)z_j}\bigg|^2\Bigg].
\end{eqnarray}
%

%
\begin{figure}[h]
\centering
 \includegraphics[width=0.85\linewidth]{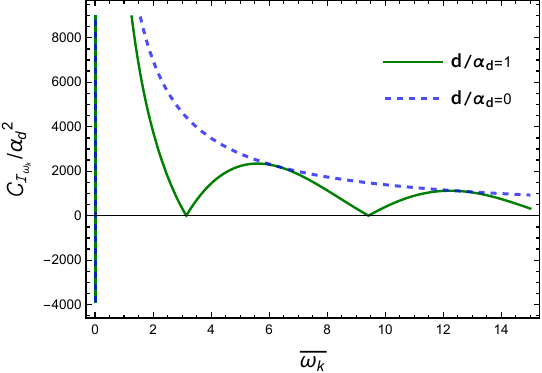}
 \caption{The quantity $\mathcal{C}_{\mathcal{I}_{\omega_{k}}}/\alpha_{d}^2$, signifying the concurrence, is plotted for two outgoing null detectors in a $(1+3)$ dimensional de Sitter spacetime with respect to the dimensionless frequency of the field $\overline{\omega}_{k} = \omega_{k}\,\alpha_{d}$ for fixed dimensionless detector transition energy $\overline{\Delta E}=\Delta E\,\alpha_{d}=0.5$, and fixed $d/\alpha_{d}=0$ and $d/\alpha_{d}=1$ respectively. We considered $\overline{k}_{x} = \overline{\omega}_{k}/2$. }
 \label{fig:Concurrence-DS-d0-dwk-1p3D}
\end{figure}
%
%
%
\begin{figure}[h]
\centering
 \includegraphics[width=0.85\linewidth]{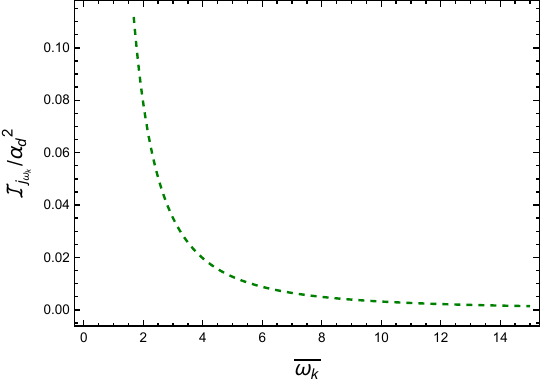}
 \caption{The quantity $\mathcal{I}_{j_{\omega_{k}}}/\alpha_{d}^2$, signifying the concurrence, is plotted for two outgoing null detectors in a $(1+3)$ dimensional de Sitter spacetime with respect to the dimensionless frequency of the field $\overline{\omega}_{k} = \omega_{k}\,\alpha_{d}$ for fixed $\overline{\Delta E}=0.5$. We considered $\overline{k}_{x} = \overline{\omega}_{k}/2$. This plot is valid for all the $d/\alpha_{d}$ values.}
 \label{fig:Ij-DS-d0-dwk-1p3D}
\end{figure}

%
%

%
\begin{figure}[h]
\centering
 \includegraphics[width=0.85\linewidth]{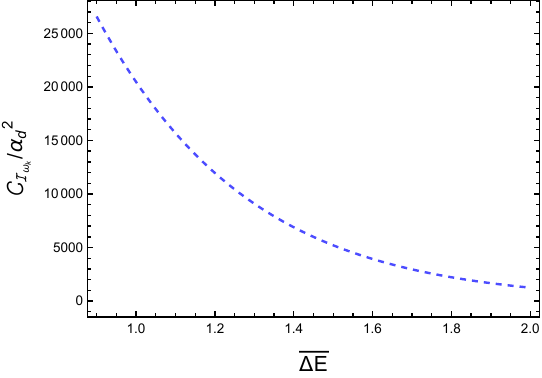}
 \caption{The quantity $\mathcal{C}_{\mathcal{I}_{\omega_{k}}}/\alpha_{d}^2$, signifying the concurrence, is plotted for two outgoing null detectors in a $(1+3)$ dimensional de Sitter spacetime with respect to the dimensionless transition energy $\overline{\Delta E}$ of the detectors for fixed dimensionless frequency of the field $\overline{\omega}_{k}=0.2$, $\overline{k}_{x} = \overline{\omega}_{k}/2$, and $d/\alpha_{d}=0$.}
 \label{fig:Concurrence-DS-d0-dEsame-1p3D}
\end{figure}
%
%
%
%
%
\begin{figure}[h]
\centering
 \includegraphics[width=0.85\linewidth]{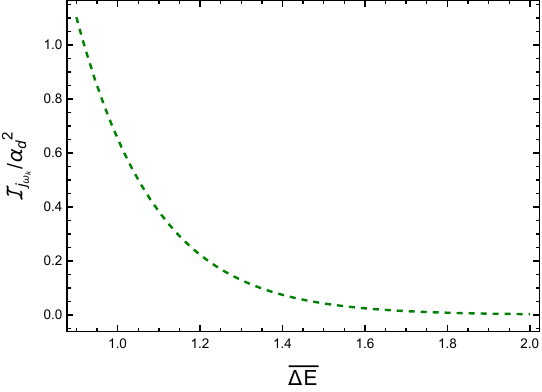}
 \caption{The quantity $\mathcal{I}_{j_{\omega_{k}}}/\alpha_{d}^2$ is plotted for two outgoing null 
detectors in a $(1+3)$ dimensional de Sitter spacetime with respect to the dimensionless transition 
energy $\overline{\Delta E}$ of the detectors for fixed $\overline{\omega}_{k}=0.2$, and
$\overline{k}_{x} = \overline{\omega}_{k}/2$. This plot is valid for all the $d/\alpha_{d}$ values.}
 \label{fig:Ij-DS-d0-dEsame-1p3D}
\end{figure}
\begin{figure}[h]
\centering
 \includegraphics[width=0.85\linewidth]{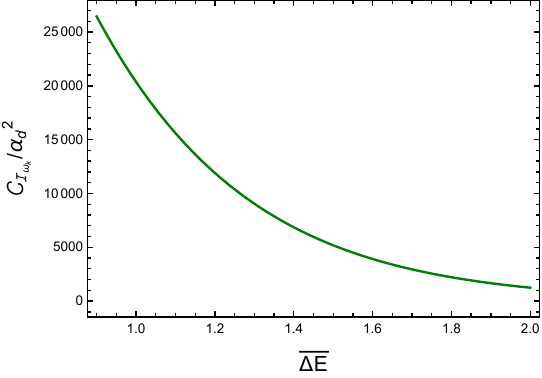}
 \caption{The quantity $\mathcal{C}_{\mathcal{I}_{\omega_{k}}}/\alpha_{d}^2$, signifying the concurrence, is plotted for two outgoing null detectors in a $(1+3)$ dimensional de Sitter spacetime with respect to the dimensionless transition energy $\overline{\Delta E}$ of the detectors for fixed dimensionless frequency of the field $\overline{\omega}_{k}=0.2$, $\overline{k}_{x} = \overline{\omega}_{k}/2$, and $d/\alpha_{d}=1$.}
 \label{fig:Concurrence-DS-d1-dEsame-1p3D}
\end{figure}
\begin{figure}[h]
\centering
 \includegraphics[width=0.85\linewidth]{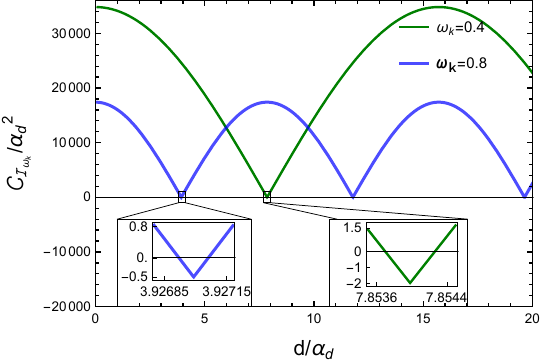} 
  \includegraphics[width=0.85\linewidth]{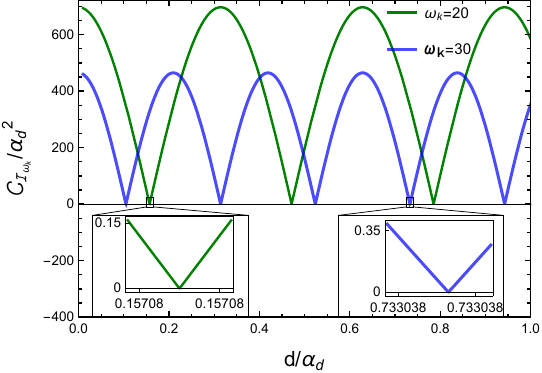}
 \caption{The quantity $\mathcal{C}_{\mathcal{I}_{\omega_{k}}} /\alpha_{d}^2$ is plotted for two outgoing null detectors in different parallel paths in a $(1+3)$ dimensional de-Sitter spacetime with respect to the separation between the two paths $d/\alpha_{d}$. The frequency of the field are fixed at $\overline{\omega}_{k}=0.4$ and $\overline{\omega}_{k}=0.8$, respectively in the upper plot (showing very small shadow regions) and $\overline{\omega}_{k}=20$ and $\overline{\omega}_{k}=30$, respectively in the lower plot (showing shadow points). Here we fixed $\overline{k}_{x} = \overline{\omega}_{k}/2$, and $\overline{\Delta E} = 0.5$.}
 \label{fig:Concurrence-DS-dd-1p3D}
\end{figure}


These integrals can performed using regulators of the form $(z^{\epsilon}\,e^{-\epsilon z})$, and 
in the limit $\epsilon\to0$ results in
\begin{eqnarray}\label{eq:Ij-DS-1p3-4}
\mathcal{I}_{j_{\omega_k}} &=&
\bigg[\frac{1}{(\omega_k - k_x)^2} + \frac{1}{(\omega_k + 
k_x)^2}\bigg]\, \pi\alpha_d\Delta{E}^{j}\,\nonumber\\
~&&~~~~~~~~\times~\frac{1}{e^{2\pi\alpha_{d}\Delta{E}^{j}}-1}~.
\end{eqnarray}
Now one can proceed to evaluate the value of $\mathcal{I}_{\epsilon}$ in a similar manner. In 
particular $\mathcal{I}_{\epsilon}$ can be expressed as
\begin{eqnarray}\label{eq:Ie-DS-1p3-1}
\mathcal{I}_{\epsilon} &=& -\int \frac{d^2k_{\perp}}{(2\pi)^3 
2\omega_{k}}\int_{0}^{\infty}dk_{x}\,\Big[\mathcal{I}^{W}_{\epsilon_{\omega_k}} + 
\mathcal{I}^{R}_{\epsilon_{\omega_k}}\Big]\,.
\end{eqnarray}
Here the first integral $\mathcal{I}^{W}_{\epsilon_{\omega_k}}$, which has emerged due to the 
Wightman function, is
\begin{eqnarray}\label{eq:IeW-DS-1p3-1}
\mathcal{I}^{W}_{\epsilon_{\omega_k}} &=& 
\int_{-\infty}^{\infty}dt_B \int_{-\infty}^{\infty}dt_A\, 
e^{i(\Delta E^A t_A+\Delta E^B t_B)}\,\frac{1}{a(\eta_A)a(\eta_B)}\nonumber\\
~&\times&  \Big[e^{i{k_x}\Delta{x}_{BA}-i\omega_k\Delta\eta_{BA}
} + e^{-i{k_x}\Delta{x}_{BA}-i\omega_k\Delta\eta_{BA}
}\Big].
\end{eqnarray}
With a change of variables $z_{j}=e^{-t_{j}/\alpha_d}$ this integral turns into
\begin{eqnarray}\label{eq:IeW-DS-1p3-2}
\mathcal{I}^{W}_{\epsilon_{\omega_k}} &=& \alpha_d^2
\int_{0}^{\infty}dz_B\,\int_{0}^{\infty}dz_A\, z_B^{-i\alpha_d\Delta 
E^B}\, z_A^{-i\alpha_d\Delta 
E^A}\,\nonumber\\
~&& \times\Big[e^{i\alpha_d(\omega_k - k_x)(z_B-z_{A})} e^{ik_{x}d} \nonumber\\
~&& ~~~~+~ 
e^{i\alpha_d(\omega_k + k_x)(z_B-z_{A})} e^{-ik_{x}d}\Big]~.
\end{eqnarray}
This integral can be straightforwardly evaluated with the introduction of the regulator 
$(z_{A}z_{B})^{\epsilon}\,e^{-\epsilon (z_{A}+z_{B})}$. We mention that the other integral 
$\mathcal{I}^{R}_{\epsilon_{\omega_k}}$, arriving from the retarded Green's function, can also be 
provided a final form in a similar manner after the change of variables as
\begin{eqnarray}\label{eq:IeR-DS-1p3-1}
\mathcal{I}^{R}_{\epsilon_{\omega_k}} &=& 
\alpha_d^2
\int_{0}^{\infty}dz_A\,\int_{z_A}^{\infty}dz_B\, z_B^{-i\alpha_d\Delta 
E^B}\, z_A^{-i\alpha_d\Delta 
E^A}\,\nonumber\\
~&& \times\Big[e^{i\alpha_d(\omega_k - k_x)(z_A-z_{B})} e^{-ik_{x}d} \nonumber\\
~&& ~~~~+~ 
e^{i\alpha_d(\omega_k + k_x)(z_A-z_{B})} e^{ik_{x}d} \nonumber\\
~&& ~~~~-~e^{i\alpha_d(\omega_k - k_x)(z_B-z_{A})} e^{ik_{x}d} \nonumber\\
~&& ~~~~-~ 
e^{i\alpha_d(\omega_k + k_x)(z_B-z_{A})} e^{-ik_{x}d} \Big]~,
\end{eqnarray}
which can also be evaluated in a similar manner introducing a regulator same as the one in the 
precious case, see Appendix \ref{Appn:IW-IR-DS-1p3}.

In Fig. \ref{fig:Concurrence-DS-d0-dwk-1p3D} we first plot \scalebox{0.9}{$\mathcal{C}_{\mathcal{I}_{\omega_{k}}} /\alpha_{d}^2$} as a function of \scalebox{0.9}{$\overline{\omega}_{k}$} for fixed \scalebox{0.9}{$\overline{\Delta E^{A}}=0.5=\overline{\Delta E^{B}}$} and for $d/\alpha_{d}=0$ and $d/\alpha_{d}=1$ respectively. While in Fig. \ref{fig:Ij-DS-d0-dwk-1p3D} we have represented the individual detector transition probability with respect to \scalebox{0.9}{$\overline{\omega}_{k}$} for the same set of parameters.
%
%
These plots suggest the possibility of entanglement harvesting in the considered parameter range. For non-zero $d/\alpha_{d}$ there are periodic entanglement harvesting shadow points with respect to $\overline{\omega}_{k}$. However, the amplitude of the oscillations keeps decreasing. Like the Schwarzschild case, one can also have shadow regions instead of shadow points for high $d$ values. For $d/\alpha_{d}=0$ the plot suggests that entanglement harvesting decreases with increasing field mode frequency. On the other hand, in Fig. \ref{fig:Concurrence-DS-d0-dEsame-1p3D} we have depicted the concurrence with respect to the dimensionless transition energy \scalebox{0.9}{$\overline{\Delta E}$} for $d/\alpha_{d}=0$ and fixed \scalebox{0.9}{$\overline{\omega}_{k} = 0.2$}. While the individual detector transition probability is depicted in Fig. \ref{fig:Ij-DS-d0-dEsame-1p3D}.
Moreover, for $d/\alpha_{d}=1$ and for the same other parameters we have depicted the concurrence 
in Fig. \ref{fig:Concurrence-DS-d1-dEsame-1p3D}.
These figures also predict the possibility of entanglement harvesting for the considered parameter ranges, and state that entanglement harvesting decreases with increasing detector transition energy. It should be noted that unlike the $(1+1)$ dimensional case in $(1+3)$ dimensions entanglement harvesting is possible for $d/\alpha_{d}=0$ in similar parameter ranges.

In Fig. \ref{fig:Concurrence-DS-dd-1p3D} we portray the concurrence with respect to the dimensionless distance $d/\alpha_{d}$ between the two detectors' null trajectories for fixed \scalebox{0.9}{$\overline{\Delta E} = 0.5$} and different \scalebox{0.9}{$\overline{\omega}_{k}$}. Like the previous Schwarzschild and $(1+1)$ dimensional de Sitter cases these plots show a periodicity of obtained concurrence with respect to $d/\alpha_{d}$. However, the plots are more like the Schwarzschild case than the $(1+1)$ dimensional de Sitter case. In the present case one periodically perceives  entanglement harvesting shadow regions and points in low ($\overline{\omega}_{k}=0.4,\,0.8$) and high ($\overline{\omega}_{k}=20,\,30$) frequency regimes respectively. We provide a discussion on the possible reason for these aforesaid similarities between $(1+1)$ Schwarzschild and $(1+3)$ de Sitter in the discussion section.

%
\vspace{0.2cm}

\section{Quantification of ``true harvesting"}\label{Sec:Entanglement-estimator}
So far, we have studied the entanglement harvesting between the two detectors moving in null trajectories in different backgrounds through concurrence. Although the fields are considered in the vacuum state, there is a possibility that the harvested entanglement can have different origins. The reason is the following. We have seen that the concurrence depends on both the local $\mathcal{I}_{j}$ and non-local $\mathcal{I}_{\varepsilon}$ terms, and both of them depend on the respective Wightman functions. These Wightman functions can be expressed as a sum of contributions from the commutator and anti-commutator of the field operators. Now since the commutator is proportional to the identity operator, its expectation value is independent of the chosen field state. Hence this part in the non-local term does not confirm whether such contribution to the concurrence is due to the vacuum fluctuations of the field. On the other hand, the expectation value of the anti-commutator is state-dependent. Therefore, this part of the non-local term carries the information about the contribution to the entanglement due to the vacuum fluctuation of the field. Under these circumstances, the former part can be interpreted as harvesting via the communication between the detectors. While the latter part accounts for the entanglement through the vacuum fluctuation of the field even if the detectors are causally disconnected. Therefore, the anti-commutator depending part measures the ``true harvesting" \cite{Martin-Martinez:2015psa, Tjoa:2020eqh, Gallock-Yoshimura:2021yok, Tjoa:2021roz}. In this scenario, in order to quantify the true harvesting, one should investigate the individual contributions of the commutator and anti-commutator in the $\mathcal{I}_{\varepsilon}$ term of concurrence rather than the same in the $\mathcal{I}_{j}$ terms, as the former one contains the information about the communication between the two detectors. This idea was first initiated in \cite{Martin-Martinez:2015psa} and later has been used for the black hole case in \cite{Tjoa:2020eqh, Gallock-Yoshimura:2021yok}. A quantitative estimator of true harvesting has been proposed recently in \cite{Tjoa:2021roz}.

Inspired by these investigations here we will draw a comparison between the commutator and anti-commutator contributions of the non-local term to understand the role of vacuum fluctuations in the entanglement harvesting.  To continue our discussion in this direction we point out that the expression of the integral $\mathcal{I}_{\varepsilon}$ from Eq. (\ref{eq:Ie-integral}) can also be cast into the form 
\begin{eqnarray}
 \mathcal{I}_{\varepsilon} =~ -\mathcal{I}^{+}_{\varepsilon} - 
\mathcal{I}^{-}_{\varepsilon}~.\label{eq:Ie-integral-PM}
\end{eqnarray}
The quantities $\mathcal{I}^{+}_{\varepsilon}$ and $\mathcal{I}^{-}_{\varepsilon}$ correspond to integrals which contain the vacuum expectations of the field anti-commutator and commutator exclusively. The 
explicit expressions of these quantities are
\begin{eqnarray}
\mathcal{I}^{+}_{\varepsilon} &=& \int_{-\infty}^{\infty}d\tau_{B} 
\int_{-\infty}^{\infty}d\tau_{A}~\scalebox{0.91}{$e^{i(\Delta 
E^{B}\tau_{B}+\Delta E^{A}\tau_{A})} $}\nonumber\\
~&& 
\times\scalebox{0.87}{$\big\{G_{W}(X_{B},X_{A})+G_{W}(X_{A},X_{B})\big\}/2$}\,;
\label{eq:Ie-integral-IP}\\
\mathcal{I}^{-}_{\varepsilon} &=& \int_{-\infty}^{\infty}d\tau_{B} 
\int_{-\infty}^{\infty}d\tau_{A}~\scalebox{0.91}{$e^{i(\Delta 
E^{B}\tau_{B}+\Delta E^{A}\tau_{A})} $}\nonumber\\
~&\times& 
\scalebox{0.87}{$\Big[\big\{G_{W}(X_{B},X_{A})-G_{W}(X_{A},X_{B})\big\}/2$}
\nonumber\\
~&+& \scalebox{0.87}{$ \theta(T_{A}-T_{B}) 
\left\{G_{W}\left(X_{A},X_{B}\right)-G_{W}\left(X_{B},X_{A}\right)\right\}
\Big]$}.\label{eq:Ie-integral-IM}
\end{eqnarray}
Let us now examine the contribution of which of the terms between $\mathcal{I}^{+}_{\varepsilon}$ and $\mathcal{I}^{-}_{\varepsilon}$ dominates in the concurrence for different parameter spaces.

\subsection{$(1+1)$ dimensional Schwarzschild spacetime}
 The integrals from Eq. (\ref{eq:Ie-integral-IP}) and (\ref{eq:Ie-integral-IM}) can be simplified noting that $G_{W}(X_{A},X_{B}) = G^{*}_{W}(X_{B},X_{A})$. One may also expresses the integrals in a form $\mathcal{I}^{\pm}_{\varepsilon} = \int_{0}^{\infty} (d\omega_{k}/4\pi\omega_{k})~ \mathcal{I}^{\pm}_{\varepsilon_{\omega_{k}}}$, which are similar to the ones provided in Eq. (\ref{eq:IeW-Sch-BV-1}) and (\ref{eq:IeR-Sch-BV-1}). Then the expressions of $\mathcal{I}^{\pm}_{\varepsilon_{\omega_{k}}}$ are
\begin{eqnarray}
\mathcal{I}^{+}_{\varepsilon_{\omega_{k}}} &=& \big[\mathcal{I}^{W}_{\varepsilon_{\omega_{k}}}(\Delta E)+\mathcal{I}^{W^{*}}_{\varepsilon_{\omega_{k}}}(-\Delta E)\big]/2~,
\label{eq:Ie-integral-IP-2}\\
\mathcal{I}^{-}_{\varepsilon_{\omega_{k}}} &=& \big[\mathcal{I}^{W}_{\varepsilon_{\omega_{k}}}(\Delta E)-\mathcal{I}^{W^{*}}_{\varepsilon_{\omega_{k}}}(-\Delta E)\big]/2+\mathcal{I}^{R}_{\varepsilon_{\omega_{k}}}(\Delta E)\,;\nonumber\\
\label{eq:Ie-integral-IM-2}
\end{eqnarray}
where the the integrals $\mathcal{I}^{W}_{\varepsilon_{\omega_{k}}}(\Delta E)$ and $\mathcal{I}^{R}_{\varepsilon_{\omega_{k}}}(\Delta E)$ are obtained from Eqs. (\ref{eq:IeW-Sch-BV-2}) and (\ref{eq:IeR-Sch-BV-2}). Now one should note that the quantity $\mathcal{I}^{+}_{\varepsilon_{\omega_{k}}}$ is reminiscent of the field anti-commutator and $\mathcal{I}^{-}_{\varepsilon_{\omega_{k}}}$ corresponds to the expectation of the field commutator. Following the first discussion of this section one can then assign true harvesting to the contribution of $\mathcal{I}^{+}_{\varepsilon_{\omega_{k}}}$. While $\mathcal{I}^{-}_{\varepsilon_{\omega_{k}}}$ contributes to the entanglement harvesting through communication channel.
%
%
\begin{figure}[h]
\centering
\includegraphics[width=0.95\linewidth]{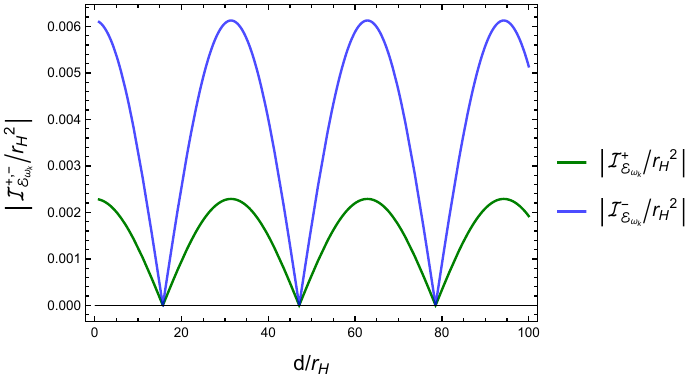}\\
 \caption{The quantities $|\mathcal{I}_{{\mathcal{E}}_{\omega_{k}}}^{+}| /\rh^2$  and  $|\mathcal{I}_{\mathcal{E}_{\omega_{k}}}^{-}| /\rh^2$ are plotted in green and blue lines, respectively for two outgoing null detectors in different parallel paths in a $(1+1)$ dimensional Schwarzschild black hole spacetime with respect to the separation between the two paths $d/\rh$. The dimensionless frequency of the field are fixed at $\overline{\omega}_{k}=0.1$.The detector transition energy is fixed at $\overline{\Delta E} = 0.5$.}
 \label{fig:Estimator-Schwarzschild}
\end{figure}
%
%
\begin{figure}[h]
\centering
\includegraphics[width=0.95\linewidth]{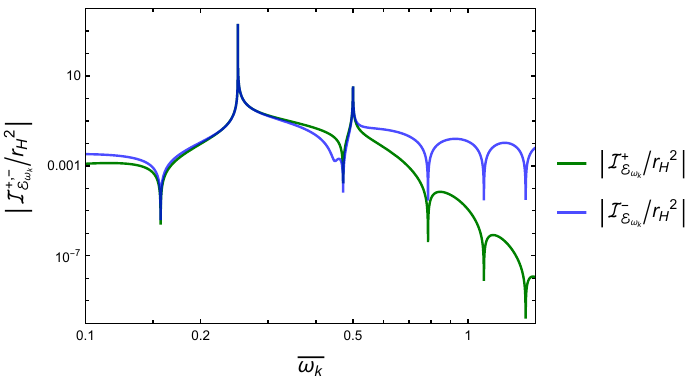}\\
 \caption{The quantities $|\mathcal{I}_{\mathcal{E}_{\omega_{k}}}^{+}| /\rh^2$  and  $|\mathcal{I}_{\mathcal{E}_{\omega_{k}}}^{-}| /\rh^2$ are plotted in green and blue lines, respectively for two outgoing null detectors in different parallel paths in a $(1+1)$ dimensional Schwarzschild black hole spacetime with respect to the separation between the two paths $d/\rh$. The dimensionless frequency of the field are fixed at $d/r_{H}=10$. The detector transition energy is fixed at $\overline{\Delta E} = 0.5$. It should be mentioned that these plots are expressed in $Log-Log$ manner for the convenience of representation.}
 \label{fig:Estimator-Schwarzschild-wk}
\end{figure}

Let us now discuss the features of these harvesting characterizing quantities. In Fig. \ref{fig:Estimator-Schwarzschild} and \ref{fig:Estimator-Schwarzschild-wk} we have plotted the the absolute values of the quantities $\mathcal{I}^{\pm}_{\varepsilon_{\omega_{k}}}$ respectively as functions of the the distance $d/\rh$ and the dimensionless frequency $\overline{\omega_{k}}$. From Fig. \ref{fig:Estimator-Schwarzschild} one observes that for considered fixed frequency and detector transition energy, the entanglement harvesting is greater through the communication channel. There is a lower contribution from the anti-commutator, which corresponds to true harvesting. However, an interesting thing to note here is that these quantities also vary periodically with the distance $d/\rh$, and the dips in these quantities are at the same places. From Fig. \ref{fig:Estimator-Schwarzschild-wk} we see that either of $|\mathcal{I}^{+}_{\varepsilon_{\omega_{k}}}|$ or $|\mathcal{I}^{-}_{\varepsilon_{\omega_{k}}}|$ may dominate in low $\overline{\omega_{k}}$ region and their minima are at the same values of $\overline{\omega_{k}}$. However, for large $\overline{\omega_{k}}$ the contribution from $\mathcal{I}^{+}_{\varepsilon_{\omega_{k}}}$ becomes negligible compared to the other term. Therefore the vacuum fluctuation almost does not play any role for large $\overline{\omega_{k}}$.

\subsection{$(1+1)$ dimensional de Sitter spacetime}
In $(1+1)$ de Sitter spacetime also one can express the integrals $\mathcal{I}^{\pm}_{\varepsilon} = \int_{0}^{\infty} (d\omega_{k}/4\pi\omega_{k})~ \mathcal{I}^{\pm}_{\varepsilon_{\omega_{k}}}$. The expressions of $\mathcal{I}^{\pm}_{\varepsilon_{\omega_{k}}}$ are given by Eq. (\ref{eq:Ie-integral-IP-2}) and (\ref{eq:Ie-integral-IM-2}). However, here the integrals $\mathcal{I}^{W}_{\varepsilon_{\omega_{k}}}(\Delta E)$ and $\mathcal{I}^{R}_{\varepsilon_{\omega_{k}}}(\Delta E)$ are realized from Eq. (\ref{eq:IeW-DS-1p1-1}) and (\ref{eq:IeR-DS-1p1-1}) for the $(1+1)$ de Sitter case.

\begin{figure}[h]
\centering
\includegraphics[width=0.95\linewidth]{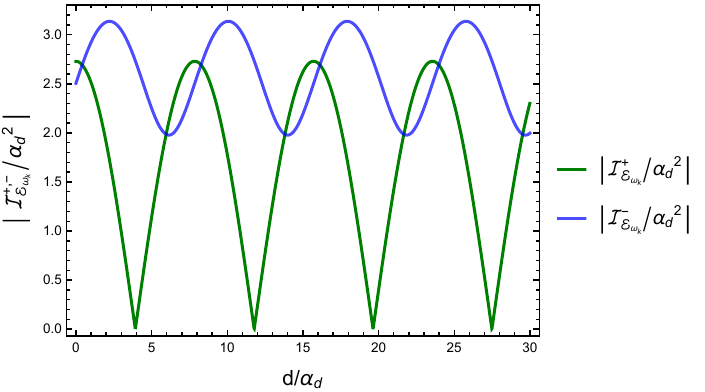}\\
 \caption{The quantities $|\mathcal{I}_{\mathcal{E}_{\omega_{k}}}^{+}| /\rh^2$  and  $|\mathcal{I}_{\mathcal{E}_{\omega_{k}}}^{-}| /\rh^2$ are plotted in green and blue lines, respectively for two outgoing null detectors in different parallel paths in a $(1+1)$ dimensional de Sitter spacetime with respect to the separation between the two paths $d/\rh$. The dimensionless frequency of the field are fixed at $\overline{\omega}_{k}=0.4$. The detector transition energy is fixed at $\overline{\Delta E} = 0.5$.}
 \label{fig:Estimator-deSitter1p1}
\end{figure}

\begin{figure}[h]
\centering
\includegraphics[width=0.95\linewidth]{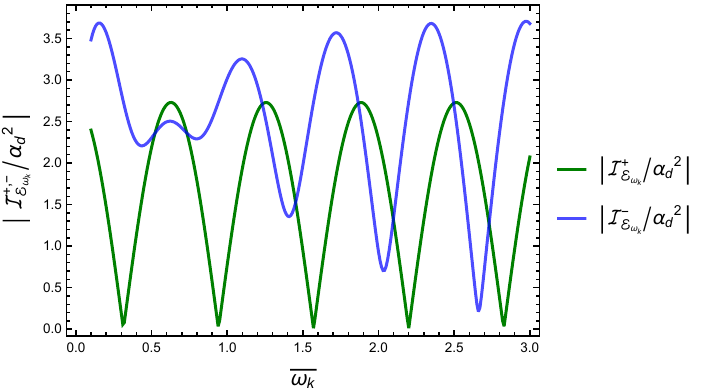}\\
 \caption{The quantities $|\mathcal{I}_{\mathcal{E}_{\omega_{k}}}^{+}| /\rh^2$  and  $|\mathcal{I}_{\mathcal{E}_{\omega_{k}}}^{-}| /\rh^2$ are plotted in green and blue lines, respectively for two outgoing null detectors in different parallel paths in a $(1+1)$ dimensional de Sitter spacetime with respect to the separation between the two paths $d/\rh$. The dimensionless frequency of the field are fixed at $d/\alpha_{d}=5$.The detector transition energy is fixed at $\overline{\Delta E} = 0.5$.}
 \label{fig:Estimator-deSitter1p1-wk}
\end{figure}

In Fig. \ref{fig:Estimator-deSitter1p1} and \ref{fig:Estimator-deSitter1p1-wk} we have plotted the integrals $|\mathcal{I}^{+}_{\varepsilon_{\omega_{k}}}|$ and $|\mathcal{I}^{-}_{\varepsilon_{\omega_{k}}}|$ as functions of the dimensionless distance and field frequency respectively. Here the dip in these quantities with respect to $d/\alpha_{d}$ do not exactly match with the dip in the concurrence, see Fig. \ref{fig:Concurrence-DS-dd}. This may be due to the fact that unlike the Schwarzschild case, here (see Fig. \ref{fig:Estimator-deSitter1p1}) the $|\mathcal{I}^{+}_{\varepsilon_{\omega_{k}}}|$ and $|\mathcal{I}^{-}_{\varepsilon_{\omega_{k}}}|$ do not have the kinks at the same positions. These contributions periodically dominate each other in the total harvesting. From the second Fig. \ref{fig:Estimator-deSitter1p1-wk} one can observe that the amplitude of the $|\mathcal{I}^{-}_{\varepsilon_{\omega_{k}}}|$ term is frequency dependent. Here also one observes that the $|\mathcal{I}^{+}_{\varepsilon_{\omega_{k}}}|$ and $|\mathcal{I}^{-}_{\varepsilon_{\omega_{k}}}|$ terms dominate each other in different $\overline{\omega}_{k}$.

\subsection{$(1+3)$ dimensional de Sitter spacetime}
In $(1+3)$ de Sitter spacetime we express the integrals $\mathcal{I}^{\pm}_{\varepsilon} = \int ~d^2k_{\perp}/(16\pi^3 \omega_{k})\int_{0}^{\infty}dk_{x}~ \mathcal{I}^{\pm}_{\varepsilon_{\omega_{k}}}$. The expressions of $\mathcal{I}^{\pm}_{\varepsilon_{\omega_{k}}}$ are again given by Eq. (\ref{eq:Ie-integral-IP-2}) and (\ref{eq:Ie-integral-IM-2}). Here the integrals $\mathcal{I}^{W}_{\varepsilon_{\omega_{k}}}(\Delta E)$ and $\mathcal{I}^{R}_{\varepsilon_{\omega_{k}}}(\Delta E)$ are realized from Eqs. (\ref{eq:IeW-DS-1p3-1}) and (\ref{eq:IeR-DS-1p3-1}) corresponding to the $(1+3)$ dimensional de Sitter case.

\begin{figure}[h]
\centering
\includegraphics[width=0.95\linewidth]{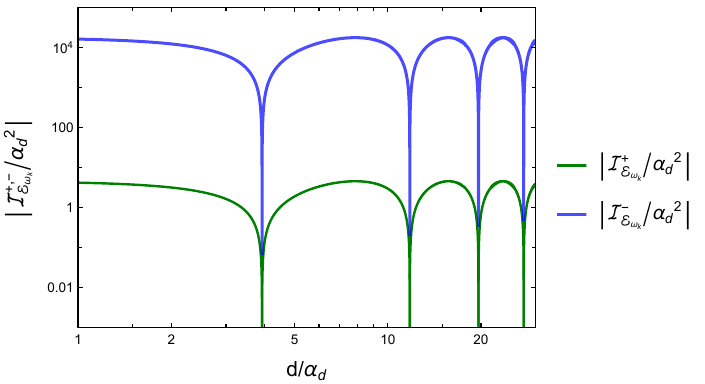}\\
 \caption{The quantities $|\mathcal{I}_{\mathcal{E}_{\omega_{k}}}^{+}| /\rh^2$  and  $|\mathcal{I}_{\mathcal{E}_{\omega_{k}}}^{-}| /\rh^2$ are plotted in green and blue lines, respectively for two outgoing null detectors in different parallel paths in a $(1+3)$ dimensional de Sitter spacetime with respect to the separation between the two paths $d/\rh$. The dimensionless frequency of the field are fixed at $\overline{\omega}_{k}=0.8$. The detector transition energy is fixed at $\overline{\Delta E} = 0.5$.}
 \label{fig:Estimator-deSitter1p3}
\end{figure}
\begin{figure}[h]
\centering
\includegraphics[width=0.95\linewidth]{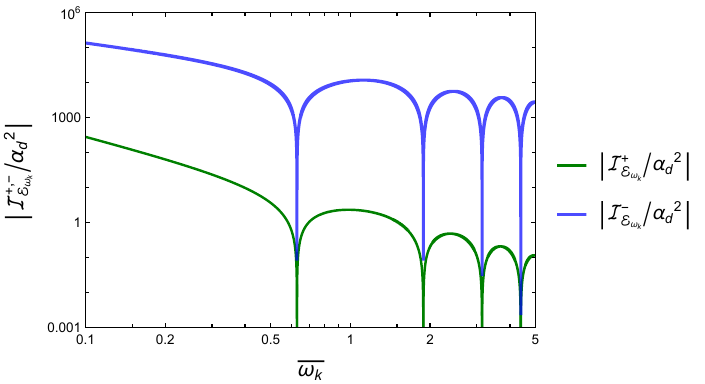}\\
 \caption{The quantities $|\mathcal{I}_{\mathcal{E}_{\omega_{k}}}^{+}| /\rh^2$  and  $|\mathcal{I}_{\mathcal{E}_{\omega_{k}}}^{-}| /\rh^2$ are plotted in green and blue lines, respectively for two outgoing null detectors in different parallel paths in a $(1+3)$ dimensional de Sitter spacetime with respect to the separation between the two paths $d/\rh$. The dimensionless frequency of the field are fixed at $d/\alpha_{d}=5$. The detector transition energy is fixed at $\overline{\Delta E} = 0.5$.}
 \label{fig:Estimator-deSitter1p3-wk}
\end{figure}

In Fig. \ref{fig:Estimator-deSitter1p3} and \ref{fig:Estimator-deSitter1p3-wk} we have plotted the integrals $|\mathcal{I}^{\pm}_{\varepsilon_{\omega_{k}}}|$ as functions of the dimensionless distance and field frequency respectively. Like the $(1+1)$ dimensional Schwarzschild case here also we are getting the dips at the same positions for both of these contributions. This is a notable difference from the $(1+1)$ dimensional de Sitter case. However, the current scenario is a bit different from the $(1+1)$ dimensional Schwarzschild case. Here, unlike the Schwarzschild case, $|\mathcal{I}^{+}_{\varepsilon_{\omega_{k}}}|$ is always many orders lower than $|\mathcal{I}^{-}_{\varepsilon_{\omega_{k}}}|$. Therefore, most of the entanglement is expected to be harvested through the communication channel.

\section{Mutual information}\label{Sec:Mutual-information}

\subsection{In Schwarzschild spacetime with respect to the Boulware and 
Unruh vacuum}

To talk about mutual information of two outgoing null detectors in a Schwarzschild black hole 
spacetime one needs to evaluate the value of the quantity $P_{AB}$, thus the integral 
$\mathcal{I}_{AB}$. One can express this integral as
\begin{eqnarray}
  \mathcal{I}_{AB} &=& \int_{-\infty}^{\infty}d\tau_{B} 
\int_{-\infty}^{\infty}d\tau_{A}~e^{-i(\Delta E^{B}\tau_{B}-\Delta E^{A}\tau_{A})} 
G_{W}(X_{B},X_{A})\nonumber\\
~&=& ~~~~\int_{0}^{\infty} \frac{d\omega_{k}}{4\pi\omega_{k}}~ \mathcal{I}_{AB_{\omega_{k}}}~.
\end{eqnarray}
Now we shall be evaluating $\mathcal{I}_{AB_{\omega_{k}}}$ corresponding to a certain field mode 
frequency $\omega_{k}$. In particular considering field mode decomposition corresponding to the 
Boulware vacuum one can get
\begin{widetext}
\begin{eqnarray}\label{eq:IAB-wk-Boulware}
 \mathcal{I}_{AB_{\omega_{k}}} &=& \int_{-\infty}^{\infty}d\tau_{B} 
\int_{-\infty}^{\infty}d\tau_{A}~e^{-i(\Delta E^{B}\tau_{B}-\Delta E^{A}\tau_{A})} 
~ \Bigg[ e^{i\omega_{k}d} + e^{-i\omega_{k}(2 r_{B}-2 
r_{A}-d)}\bigg(\frac{r_{B}-\rh}{r_{A}-\rh}\bigg)^{-2i\omega_{k}\rh} \Bigg]\nonumber\\
~&=& \rh^2~ e^{i(\Delta E^{A}+\omega_{k})d+i(\Delta E^{A}-\Delta E^{B})\rh} \int_{0}^{\infty} 
dy_{B}~\frac{y_{B}+2}{y_{B}} \int_{0}^{\infty} dy_{A}~\frac{y_{A}+2}{y_{A}}\nonumber\\
~&& e^{i(\Delta 
E^{A}y_{A}-\Delta E^{B}y_{B})\rh} y_{A}^{2i\Delta E^{A}\rh} y_{B}^{-2i\Delta E^{B}\rh} 
\Bigg[1+e^{-2i\omega_{k}\rh(y_{B}-y_{A})} 
\bigg(\frac{y_{B}}{y_{A}}\bigg)^{-2i\omega_{k}\rh}\Bigg]~.
\end{eqnarray}
\end{widetext}
\begin{widetext}
On the other hand, considering field mode decomposition corresponding to the Unruh vacuum 
one gets
%
\begin{eqnarray}\label{eq:IAB-wk-Unruh}
 \mathcal{I}_{AB_{\omega_{k}}} &=& \int_{-\infty}^{\infty}d\tau_{B} 
\int_{-\infty}^{\infty}d\tau_{A}~e^{-i(\Delta E^{B}\tau_{B}-\Delta E^{A}\tau_{A})} \Big[ 
\exp\Big\{i\omega_{k}2\rh\big(1-e^{-\frac{d}{2\rh}}\big)\Big\} + e^{-i\omega_{k}(2 r_{B}-2 
r_{A}-d)}\big(\frac{r_{B}-\rh}{r_{A}-\rh}\big)^{-2i\omega_{k}\rh} \Big]
\nonumber
\\
&=& \rh^2~ e^{i\Delta E^{A}d+i(\Delta E^{A}-\Delta E^{B})\rh} \int_{0}^{\infty} 
dy_{B}~\frac{y_{B}+2}{y_{B}} \int_{0}^{\infty} dy_{A}~\frac{y_{A}+2}{y_{A}}~ e^{i(\Delta 
E^{A}y_{A}-\Delta E^{B}y_{B})\rh}
\nonumber
\\
&& y_{A}^{2i\Delta E^{A}\rh} y_{B}^{-2i\Delta E^{B}\rh} 
\Big[\exp\Big\{i\omega_{k}2\rh\big(1-e^{-\frac{d}{2\rh}}\big)\Big\}  + e^{i\omega_{k}d} 
e^{-2i\omega_{k}\rh(y_{B}-y_{A} )} \Big(\frac{y_{B}}{y_{A}}\Big)^{-2i\omega_{k}\rh}\Big]~.
\end{eqnarray}
\end{widetext}

\begin{figure}[h]
\centering
 \includegraphics[width=0.85\linewidth]{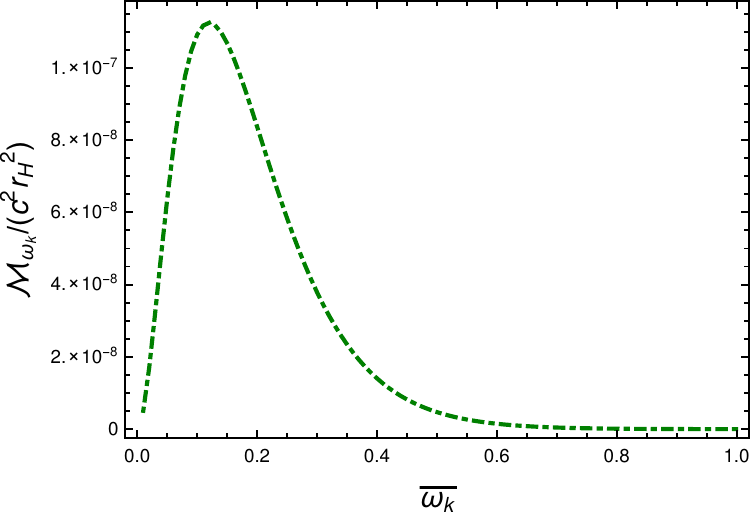}
 \caption{The mutual information $\mathcal{M}_{\omega_{k}}(\rho_{AB})/(c^2\rh^2)$ perceived by two 
out going null detectors corresponding to the Boulware and Unruh vacuum in plotted with respect to 
the dimensionless frequency of the field modes $\overline{\omega}_{k} = \rh\omega_{k}$ for fixed 
transition frequency $\overline{\Delta E} = \rh\Delta E = 1$.}
 \label{fig:Mutual-inf-Dwk-Schchld}
\end{figure}
%

\begin{figure}[h]
\centering
 \includegraphics[width=0.85\linewidth]{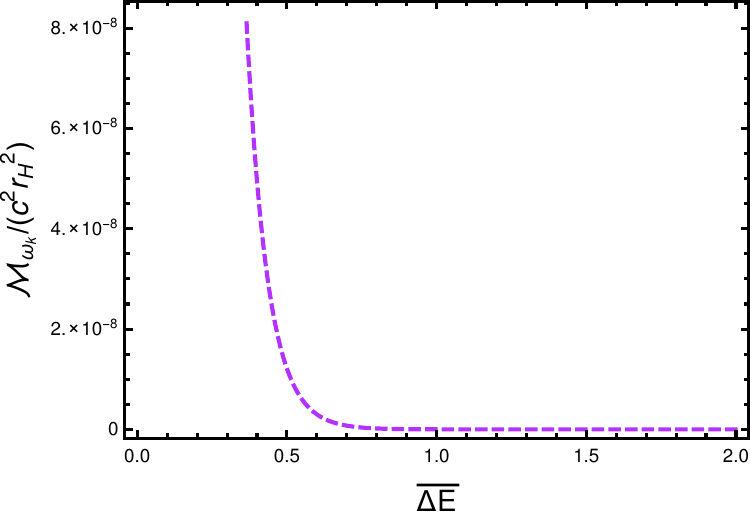}
 \caption{The mutual information $\mathcal{M}_{\omega_{k}}(\rho_{AB})/(c^2 \rh^2)$ perceived by two 
out going null detectors corresponding to the Boulware and Unruh vacuum in plotted with respect to 
the dimensionless transition frequency $\overline{\Delta E} = \rh\Delta E$  for fixed dimensionless 
field mode frequency $\overline{\omega}_{k} = \rh\omega_{k} = 1$.}
 \label{fig:Mutual-inf-DdE-Schchld}
\end{figure}
%
One should notice that the first terms in the square brackets on the right hand side of Eq. 
(\ref{eq:IAB-wk-Boulware}) and (\ref{eq:IAB-wk-Unruh}) vanishes and when $\Delta E^{A}=\Delta 
E^{A}=\Delta E$ both of these integrals get significantly simplified and can be evaluated to be
\begin{eqnarray}
  \mathcal{I}_{AB_{\omega_{k}}} &=& e^{i(\Delta E^{A}+\omega_{k})d} 
\frac{4\pi\rh\omega_{k}^2}{(\Delta E+\omega_{k})(\Delta E+2\omega_{k})^2}~\nonumber\\ 
~&&~~~~~~\times~~\frac{1}{e^{4\pi\rh(\Delta E+\omega_{k})}-1}~,
\end{eqnarray}
which is same as $\mathcal{I}_{j_{\omega_{k}}}$ up-to a phase factor for equal detector transition 
energies. Now it should be mentioned that unlike the expression of the concurrence, the mutual 
information has multiple disparate multiplicative expectation values of the monopole moment 
operators $m_{j}(0)$. In that case one cannot take a common factor of them out from the expression 
of the mutual information. In particular, from the operator form of $m_{j}(0) = 
|E_{1}^{j}\rangle\langle E_{0}^{j}| + |E_{0}^{j}\rangle\langle E_{1}^{j}|$ one can get the 
expectation values $\langle E_{1}^{j}| m_{j}(0) |E_{0}^{j}\rangle = 1$. In that case 
$P_{A_{\omega_{k}}} = \mathcal{I}_{A_{\omega_{k}}}$, $P_{B_{\omega_{k}}} = 
\mathcal{I}_{B_{\omega_{k}}}$, and $P_{AB_{\omega_{k}}} = \mathcal{I}_{AB_{\omega_{k}}}$, where 
$P_{j} = \int_{-\infty}^{\infty} dk/(4\pi\omega_{k})~~ P_{j_{\omega_{k}}}$. For $\Delta E^{A}=\Delta 
E^{A}=\Delta E$, let us consider $\mathcal{I}_{A_{\omega_{k}}} = \mathcal{I}_{B_{\omega_{k}}} = 
|\mathcal{I}_{AB_{\omega_{k}}}| = \mathcal{I}_{{\omega_{k}}}$. Then $P_{+_{\omega_{k}}} = 
\mathcal{I}_{{\omega_{k}}}$ and $P_{-_{\omega_{k}}} = 0$, and one can get the mutual information 
for fixed field frequency $\omega_{k}$ as
\begin{eqnarray}
 \mathcal{M}_{\omega_{k}}(\rho_{AB}) &=& c^2~ 2~ \mathcal{I}_{{\omega_{k}}}\ln{2} + 
\mathcal{O}(c^4)~.
\end{eqnarray}

Therefore we have observed that in both the Boulware and Unruh vacuum cases the mutual information 
corresponding to a certain field mode frequency up to $\mathcal{O}(c^2)$ are the same and independent 
of the distance $d$, between different outgoing null paths. Since mutual information is independent 
of $d$ and non-vanishing, the correlation is classical at the values of $d$ where entanglement 
harvesting does not occur. In Fig. \ref{fig:Mutual-inf-Dwk-Schchld} and 
\ref{fig:Mutual-inf-DdE-Schchld} we have plotted the dimensionless mutual information 
\scalebox{0.9}{$\mathcal{M}_{\omega_{k}}(\rho_{AB})/(c^2 \rh^2)$} with respect to the dimensionless 
parameters \scalebox{0.9}{$\overline{\omega}_{k}$} and \scalebox{0.9}{$\overline{\Delta E}$} 
respectively. It is observed that the mutual information decreases with increasing detector 
transition energy.

\subsection{de Sitter universe}

\subsubsection{(1+1)-dimensions}

Now we proceed to talk about the mutual information of two outgoing null detectors in a de Sitter 
background. Again we express the integral $\mathcal{I}_{AB} = \int_{0}^{\infty} 
d\omega_{k}/(4\pi\omega_{k})~ \mathcal{I}_{AB_{\omega_{k}}}$ to evaluate $P_{AB}$. 
%
%
Furthermore, this integral $\mathcal{I}_{AB_{\omega_{k}}}$ corresponding to the de Sitter vacuum 
is
\begin{eqnarray}\label{eq:IAB-wk-DS}
 \mathcal{I}_{AB_{\omega_{k}}} &=& \int_{-\infty}^{\infty}d\tau_{B} 
\int_{-\infty}^{\infty}d\tau_{A}~e^{-i(\Delta E^{B}\tau_{B}-\Delta E^{A}\tau_{A})} 
~\nonumber\\
~&& \Big[ e^{i\omega_{k}d} + e^{-i\omega_{k}d} e^{-2i\omega_{k} 
\alpha_{d}(e^{-t_A/\alpha_d}-e^{-t_B/\alpha_d}) } \Big].\nonumber\\
\end{eqnarray}
Like the previous cases here also the integration over $e^{i\omega_{k}d}$ vanishes using the 
properties of the Dirac delta distribution for $\Delta E^{j}>0$. Then with the change of variables 
$e^{-t_{j}/\alpha_d}=z_{j}$ the integral $\mathcal{I}_{AB_{\omega_{k}}}$ becomes
\begin{eqnarray}\label{eq:IAB-wk-DS}
 \mathcal{I}_{AB_{\omega_{k}}} &=& e^{-i\omega_{k}d}\,\alpha_d^2 \int_{0}^{\infty} 
dz_{B} \int_{0}^{\infty} dz_{A}\, e^{-2i\omega_{k} 
\alpha_{d}(z_A-z_B)}\nonumber\\
~&&~~~~\times~ z_{A}^{-i\alpha_{d}\Delta E^{A}-1}\,z_{B}^{i\alpha_{d}\Delta E^{B}-1}~.
\end{eqnarray}
%

\begin{figure}[h]
\centering
 \includegraphics[width=0.85\linewidth]{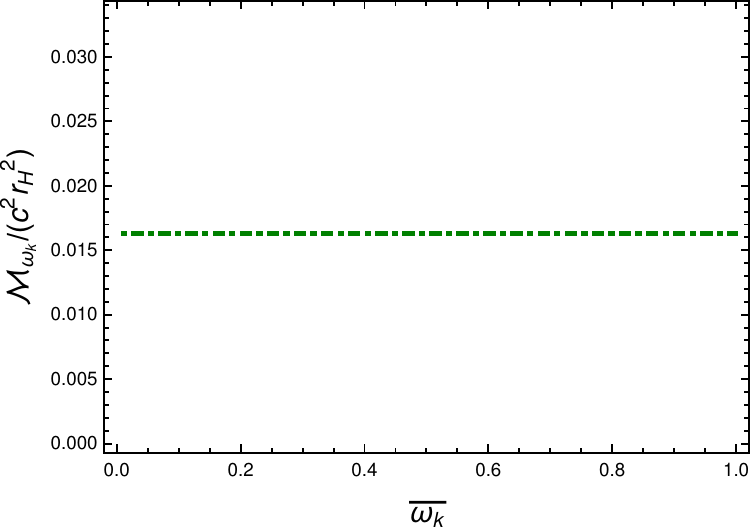}
 \caption{The mutual information $\mathcal{M}_{\omega_{k}}(\rho_{AB})/(c^2\alpha_{d}^2)$ in $(1+1)$ 
dimensional de Sitter spacetime as perceived by two out going null detectors is plotted with respect 
to the dimensionless frequency of the field modes $\overline{\omega}_{k} = \alpha_{d}\omega_{k}$ for 
fixed transition frequency $\overline{\Delta E} = \alpha_{d}\Delta E = 1$, and fixed 
$d/\alpha_{d}=0$.}
 \label{fig:Mutual-inf-Dwk-DS}
\end{figure}

%
\begin{figure}[h]
\centering
 \includegraphics[width=0.85\linewidth]{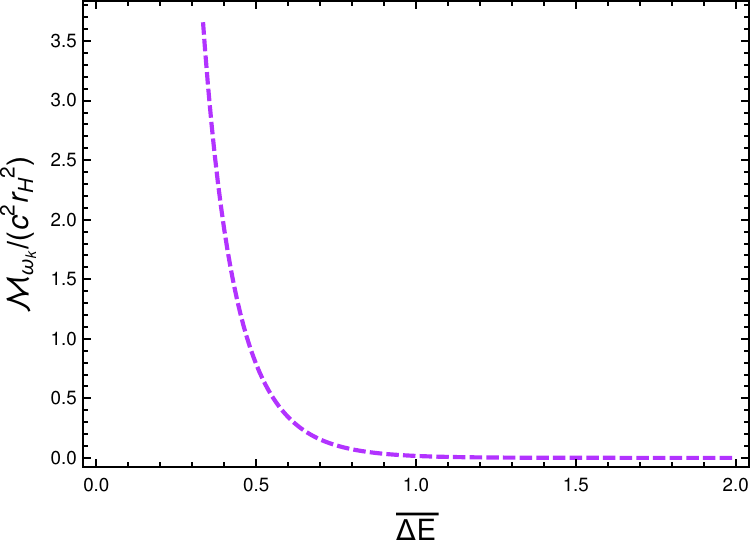}
 \caption{The mutual information $\mathcal{M}_{\omega_{k}}(\rho_{AB})/(c^2 \rh^2)$ in $(1+1)$ 
dimensional de Sitter spacetime as perceived by two out going null detectors is plotted with respect 
to the dimensionless transition frequency $\overline{\Delta E} = \rh\Delta E$  for fixed 
dimensionless field mode frequency $\overline{\omega}_{k} = \rh\omega_{k} = 1$, and 
fixed $d/\alpha_{d}=0$.}
 \label{fig:Mutual-inf-DdE-DS}
\end{figure}

%
When $\Delta E^{A}=\Delta E^{B}=\Delta E$ this integral gets significantly simplified and is 
evaluated to be
\begin{eqnarray}
  \mathcal{I}_{AB_{\omega_{k}}} &=& e^{-i\omega_{k}d}~ \frac{2\pi\alpha_d}{\Delta 
E}\,\frac{1}{e^{2\pi\alpha_d\Delta E}-1}~,
\end{eqnarray}
which is same as $\mathcal{I}_{j_{\omega_{k}}}$ up-to a phase factor. In that case one cannot take 
a 
common factor of them out from the expression of the mutual information. With the consideration of 
the form of the monopole moment operator $m_{j}(0) = |E_{1}^{j}\rangle\langle E_{0}^{j}| + 
|E_{0}^{j}\rangle\langle E_{1}^{j}|$ one gets the expectation values $\langle E_{1}^{j}| m_{j}(0) 
|E_{0}^{j}\rangle = 1$. In that case $P_{A_{\omega_{k}}} = \mathcal{I}_{A_{\omega_{k}}}$, 
$P_{B_{\omega_{k}}} = \mathcal{I}_{B_{\omega_{k}}}$, and $P_{AB_{\omega_{k}}} = 
\mathcal{I}_{AB_{\omega_{k}}}$, where $P_{j} = \int_{-\infty}^{\infty} dk/(4\pi\omega_{k})~~ 
P_{j_{\omega_{k}}}$. For $\Delta E^{A}=\Delta E^{A}=\Delta E$, we further consider 
$\mathcal{I}_{A_{\omega_{k}}} = \mathcal{I}_{B_{\omega_{k}}} = |\mathcal{I}_{AB_{\omega_{k}}}| = 
\mathcal{I}_{{\omega_{k}}}$. Then $P_{+_{\omega_{k}}} = \mathcal{I}_{{\omega_{k}}}$ and 
$P_{-_{\omega_{k}}} = 0$, and one can get the mutual information for a certain field frequency 
$\omega_{k}$ as
\begin{eqnarray}
 \mathcal{M}_{\omega_{k}}(\rho_{AB}) &=& c^2~ 2~ \mathcal{I}_{{\omega_{k}}}\ln{2} + 
\mathcal{O}(c^4)~.
\end{eqnarray}
We observe that this mutual information is independent of the distance $d$, between different 
outgoing null paths, and the field mode frequency $\omega_{k}$, see (\ref{eq:Ij-DS-1p1-5}) for 
$\mathcal{I}_{{\omega_{k}}}$. The Mutual information 
\scalebox{0.9}{$\mathcal{M}_{\omega_{k}}(\rho_{AB})/(c^2\rh^2)$} in this scenario is plotted as 
functions of $\overline{\omega}_{k}$ and \scalebox{0.9}{$\overline{\Delta E}$} respectively in Fig. 
\ref{fig:Mutual-inf-Dwk-DS} and \ref{fig:Mutual-inf-DdE-DS}. From these figures one asserts that 
mutual information decreases with increasing detector transition energy.

\subsubsection{(1+3)-dimensions}

We are going to evaluate the integral $\mathcal{I}_{AB}$ for the estimation of the quantity 
$P_{AB}$. This integral can be further expressed as
\begin{eqnarray}
  \mathcal{I}_{AB} &=& \int_{-\infty}^{\infty}d\tau_{B} 
\int_{-\infty}^{\infty}d\tau_{A}~e^{-i(\Delta E^{B}\tau_{B}-\Delta E^{A}\tau_{A})} 
G_{W}(X_{B},X_{A})\nonumber\\
~&=& \int \frac{d^2k_{\perp}}{(2\pi)^2 
2\omega_{k}}\int_{0}^{\infty}dk_{x}~ \mathcal{I}_{AB_{\omega_{k}}}~.
\end{eqnarray}
We are now going to evaluate the integral $\mathcal{I}_{AB_{\omega_{k}}}$ in $(1+3)$ 
dimensions corresponding to the conformal vacuum as
\begin{eqnarray}\label{eq:IAB-wk-DS-1p3}
 \mathcal{I}_{AB_{\omega_{k}}} &=& \int_{-\infty}^{\infty}d\tau_{B} 
\int_{-\infty}^{\infty}d\tau_{A}~e^{-i(\Delta E^{B}\tau_{B}-\Delta 
E^{A}\tau_{A})}\,e^{-\frac{t_{B}+t_{A}}{\alpha_{d}}}
~\nonumber\\
~&& \Big[ e^{ik_{x} d} e^{i\alpha_{d}(\omega_{k}-k_{x}) 
(e^{-t_B/\alpha_d}-e^{-t_A/\alpha_d}) } \nonumber\\
~&& ~+\, e^{-ik_{x} d} e^{i\alpha_{d}(\omega_{k}+k_{x}) 
(e^{-t_B/\alpha_d}-e^{-t_A/\alpha_d}) } \Big]~.
\end{eqnarray}
%

\begin{figure}[h]
\centering
 \includegraphics[width=0.85\linewidth]{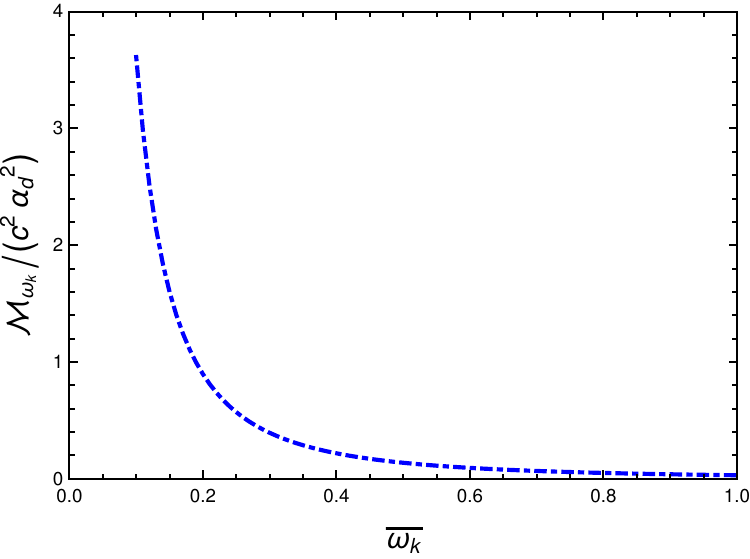}
 \caption{The mutual information $\mathcal{M}_{\omega_{k}}(\rho_{AB})/(c^2\rh^2)$ perceived by two 
out going null detectors in $(1+3)-$dimensional de Sitter spacetime corresponding to the conformal 
vacuum is plotted with respect to the dimensionless frequency of the field modes 
$\overline{\omega}_{k} = \alpha_{d}\omega_{k}$ for fixed transition frequency $\overline{\Delta E} = 
\alpha_{d}\Delta E = 1$, and fixed $d/\alpha_{d}=1$. Here we have also considered $\alpha_{d} k_{x} 
= \overline{\omega}_{k}/2$.}
 \label{fig:Mutual-inf-Dwk-DS-1p3}
\end{figure}
%
\begin{figure}[h]
\centering
 \includegraphics[width=0.85\linewidth]{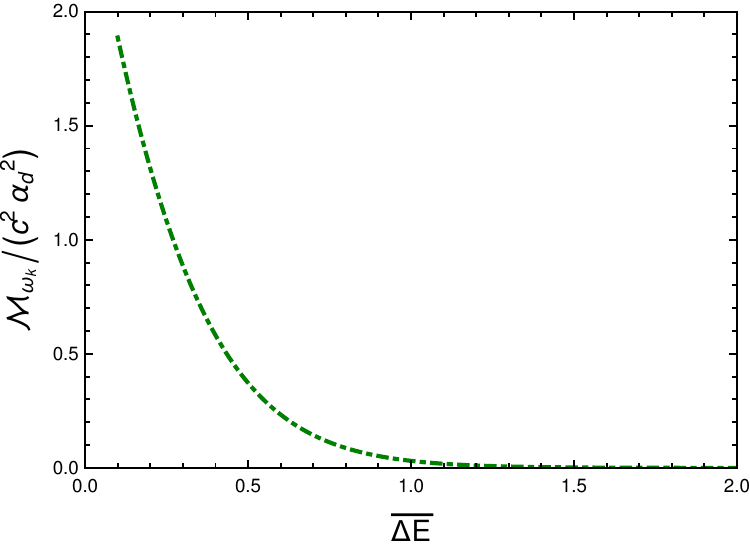}
 \caption{The mutual information $\mathcal{M}_{\omega_{k}}(\rho_{AB})/(c^2\rh^2)$ perceived by two 
out going null detectors in $(1+3)-$dimensional de Sitter spacetime corresponding to the conformal 
vacuum is plotted with respect to the dimensionless transition frequency $\overline{\Delta E} = 
\alpha_{d}\Delta E$ for fixed frequency of the field modes and other parameters 
$\overline{\omega}_{k} = \alpha_{d}\omega_{k} = 1$, $\alpha_{d} k_{x} = \overline{\omega}_{k}/2$ and 
$d/\alpha_{d} = 1$.}
 \label{fig:Mutual-inf-DdE-DS-1p3}
\end{figure}
%
\begin{figure}[h]
\centering
 \includegraphics[width=0.85\linewidth]{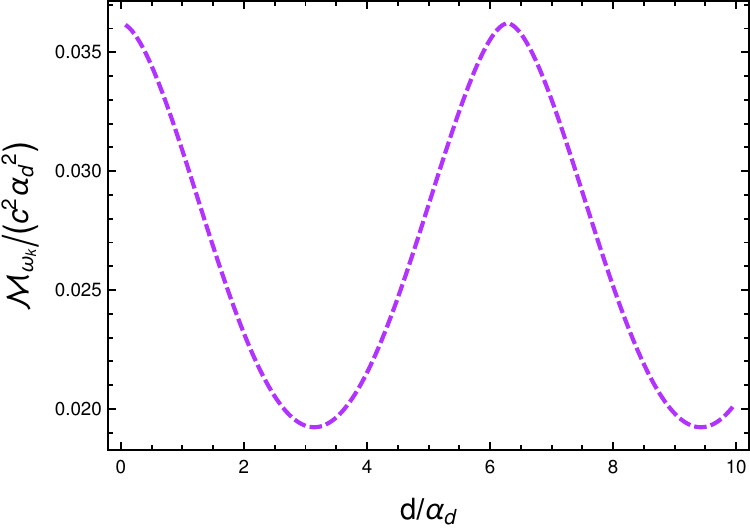}
 \caption{The mutual information $\mathcal{M}_{\omega_{k}}(\rho_{AB})/(c^2\rh^2)$ perceived by two 
out going null detectors in $(1+3)-$dimensional de Sitter spacetime corresponding to the conformal 
vacuum is plotted with respect to the dimensionless parameter $d/\alpha_{d}$, for fixed transition 
frequency $\overline{\Delta E} = \alpha_{d}\Delta E = 1$, frequency of the field modes 
$\overline{\omega}_{k} = \alpha_{d}\omega_{k} = 1$, and $\alpha_{d} k_{x} = 
\overline{\omega}_{k}/2$.}
 \label{fig:Mutual-inf-Dd-DS-1p3}
\end{figure}

Now one may consider a change of variables $z_{j}=e^{-t_{j}/\alpha_d}$. When $\Delta E^{A}=\Delta 
E^{B}=\Delta E$ this integral gets significantly simplified and is evaluated to be
\begin{eqnarray}
  \mathcal{I}_{AB_{\omega_{k}}} &=& \bigg[\frac{e^{ik_{x} d}}{(\omega_k - k_x)^2} + 
\frac{e^{-ik_{x} d}}{(\omega_k + 
k_x)^2}\bigg]\, \nonumber\\
~&&~~~~\times~\pi\alpha_d\Delta{E}\,\frac{1}{e^{2\pi\alpha_{d}\Delta{E}}-1}~.
\end{eqnarray}
Unlike the $(1+1)$ dimensional case here the integral $\mathcal{I}_{AB_{\omega_{k}}}$ is not 
equivalent to $\mathcal{I}_{j_{\omega_{k}}}$ up-to a phase factor. In $(1+3)$ dimensions one also 
observes that the mutual information will be dependent on the parameter $d$ separating two outgoing 
null paths. With the expression of the monopole moment operator to be $m_{j}(0) = 
|E_{1}^{j}\rangle\langle E_{0}^{j}| + |E_{0}^{j}\rangle\langle E_{1}^{j}|$ one gets the expectation 
values $\langle E_{1}^{j}| m_{j}(0) |E_{0}^{j}\rangle = 1$. In that case $P_{A_{\omega_{k}}} = 
\mathcal{I}_{A_{\omega_{k}}}$, $P_{B_{\omega_{k}}} = \mathcal{I}_{B_{\omega_{k}}}$, and 
$P_{AB_{\omega_{k}}} = \mathcal{I}_{AB_{\omega_{k}}}$, where
\begin{eqnarray}
P_{j} =  \int \frac{d^2k_{\perp}}{(2\pi)^2 
2\omega_{k}}\int_{0}^{\infty}dk_{x}~ P_{j_{\omega_{k}}}~.
\end{eqnarray}
When $d=0$ and $\Delta E^{A}=\Delta E^{A}=\Delta E$ the integral $\mathcal{I}_{AB_{\omega_{k}}}$ 
becomes same with the $\mathcal{I}_{j_{\omega_{k}}}$. In that scenario one can further consider 
$\mathcal{I}_{A_{\omega_{k}}} = \mathcal{I}_{B_{\omega_{k}}} = |\mathcal{I}_{AB_{\omega_{k}}}| = 
\mathcal{I}_{{\omega_{k}}}$. Then $P_{+_{\omega_{k}}} = \mathcal{I}_{{\omega_{k}}}$ and 
$P_{-_{\omega_{k}}} = 0$, and one can get the mutual information for a certain field frequency 
$\omega_{k}$ as
\begin{eqnarray}
 \mathcal{M}_{\omega_{k}}(\rho_{AB}) &=& c^2~ 2~ \mathcal{I}_{{\omega_{k}}}\ln{2} + 
\mathcal{O}(c^4)~.
\end{eqnarray}
Then the plots corresponding to $\mathcal{M}_{\omega_{k}}(\rho_{AB})$ for $d=0$ should be 
qualitatively same with the plots of $\mathcal{I}_{j_{\omega_{k}}}$ from Fig. 
\ref{fig:Ij-DS-d0-dwk-1p3D} and \ref{fig:Ij-DS-d0-dEsame-1p3D}.

On the other hand, when \scalebox{0.9}{$\Delta E^{A}=\Delta E^{A}=\Delta E$} and $d\neq0$, one gets 
\scalebox{0.9}{$P_{A_{\omega_{k}}} = \mathcal{I}_{A_{\omega_{k}}}$, $P_{B_{\omega_{k}}} = 
\mathcal{I}_{B_{\omega_{k}}}$, $P_{AB_{\omega_{k}}} = \mathcal{I}_{AB_{\omega_{k}}}$} and 
\scalebox{0.9}{$P_{\pm_{\omega_{k}}} = 
\mathcal{I}_{j_{\omega_{k}}}\pm|\mathcal{I}_{AB_{\omega_{k}}}|$}. Using these expressions we have 
obtained the mutual information for $d\neq0$. We included the plots \ref{fig:Mutual-inf-Dwk-DS-1p3} 
and \ref{fig:Mutual-inf-DdE-DS-1p3} which represent the change of the mutual information with 
respect to \scalebox{0.9}{$\overline{\omega}_{k}$} and \scalebox{0.9}{$\overline{\Delta E}$} when 
$d/\alpha_{d} = 1$. Qualitatively these plots are not different from the ones with $d/\alpha_{d} = 
0$. Furthermore, unlike the $(1+1)$ dimensional de Sitter case in $(1+3)$ dimensions, the mutual 
information is dependent, in fact periodically dependent, on $d/\alpha_{d}$ and this fact is 
graphically presented in Fig. \ref{fig:Mutual-inf-Dd-DS-1p3}.

\section{Discussion}\label{Sec:discussion}

This work investigates the entanglement harvesting with detectors in outgoing null trajectories from 
the conformal vacuums in de Sitter and $(1+1)$ dimensional Schwarzschild spacetimes. In particular, 
we considered the integral representation of Green's functions to estimate the Harvesting 
conditions corresponding to a specific field mode frequency. We observed that entanglement 
harvesting is possible, and it is maximum at a particular field mode frequency in the $(1+1)$ 
dimensional Schwarzschild black hole spacetime with detectors in the same outgoing null trajectory. 
For this specific trajectory of the same outgoing null path of the two detectors, one observes that 
$(1+1)$ de Sitter spacetime does not exhibit any entanglement extraction, for the same set of 
considered parameter values.
It signifies that though both detectors are moving along the same outgoing null path in $(1+1)$ de 
Sitter spacetime, there is no quantum correlation among them. We also observed that these two 
detectors can be classically correlated as the mutual information between them turned out to be non 
zero.
In contrast, $(1+3)$ dimensional de Sitter spacetime shows entanglement harvesting, for the two 
detectors in same outgoing null path. When the detectors are in different outgoing null paths, 
distance $d$ apart, we observed that the concurrence becomes periodic with respect to the distance 
$d$ with period and amplitude depending on the field mode frequency. 
%
%
We also observed that the concurrence vanishes at periodic regions and points in $d$ respectively for low and high field mode frequencies in the $(1+1)$ Schwarzschild and $(1+3)$ de Sitter spacetimes. In comparison, in $(1+1)$ dimensional de Sitter spacetime, there are only periodic regions of $d$ with no harvesting. Therefore, concerning the harvested entanglement in terms of the concurrence, one finds the $(1+1)$ dimensional Schwarzschild and $(1+3)$ dimensional de Sitter spacetimes to exhibit qualitatively the same features.

We also observed that in $(1+1)$ dimensional Schwarzschild and $(1+3)$ dimensional de Sitter 
spacetime, regardless of whether one takes the same or different outgoing null paths, the 
concurrence continuously decreases with increasing detector transition energy. In $(1+1)$ 
dimensional de Sitter spacetime and for a certain non-zero value of the distance $d$, one harvests 
entanglement in discrete ranges of the detector transition energy, see Fig. 
\ref{fig:Concurrence-DS-d1-dEsame}. In this scenario, we also investigated the role of the vacuum fluctuations of the field in the harvested entanglement.

We observed that the mutual information corresponding to a specific field mode frequency is 
generally independent of the distance $d$ in $(1+1)$ dimensional Schwarzschild and de Sitter 
spacetime. In contrast, this mutual information is periodically dependent on $d$ in $(1+3)$ 
dimensional de Sitter background, but it never becomes zero like the concurrence. This analysis 
asserts that in these spacetimes, one can obtain certain outgoing null paths for the two detectors 
where there is no quantum communication between the two detectors. However, classical communication 
is still possible as perceived through mutual information.\vspace{0.2cm}

In this analysis we intentionally refrain from making any comment on Hartle-Hawking vacuum for black hole case. 
This is because we could not properly analyse the related integrals since our choices of regulators could not make the integrals convergent, both at the analytical and numerical levels. 
One should note that we have 
specifically considered the outgoing null paths for the detectors. In this regard, one could have considered the 
ingoing null paths as well. In particular, we have checked the case with both detectors in ingoing 
null paths corresponding to the Boulware, Unruh, and the Hartle-Hawking vacua. The properties of the
entanglement harvesting from the Boulware vacuum, as expected, remain the same like the outgoing scenario. Whereas the ingoing-Unruh situation follows the identical outcome of outgoing-Hartle-Hawking case and therefore we refrain to comment again. Similar situation also arises for ingoing-Hartle-Hawking scenario. Finally, the de-Sitter universe also yields the identical inferences for the ingoing trajectories as the outgoing ones.

We would like to provide a few final comments: 

\begin{itemize}
\item This article looks for entanglement harvesting from the conformal vacuums with Unruh-DeWitt 
detectors in null trajectories. We have considered the $(1+1)$ and $(1+3)$ dimensional de Sitter 
and $(1+1)$ dimensional Schwarzschild backgrounds as they are conformally flat (see 
\cite{book:Birrell}). One should notice that the $(1+3)$ dimensional Schwarzschild black hole 
spacetime is not conformally flat. It is possible to study the effects of quantum field theory in 
regions near the event horizon and asymptotic infinity in a $(1+3)$ dimensional Schwarzschild 
background, where the quantum field perceives the effective spacetime to be conformally flat. This 
effective background does not comply with our current investigation as the null trajectories in our 
study traverse the whole region outside of the event horizon to spatial infinity. However, the 
$(1+1)$ dimensional Schwarzschild spacetime is considered a solution of the two-dimensional 
Einstein-Dilation theory, dimensionally reduced from the Einstein theory (see the discussion above 
Eq. (\ref{eq:metric-sch})). Thus our results may retain some signatures of higher dimensions, but we 
do not have any conclusive evidence at this stage. Therefore, we believe it will be naive to readily 
comment on the features related to higher dimensional spacetimes just by investigating the same in 
$(1+1)$ dimensions.

\item Note that working with the linear coupling model in $(1+1)$ dimensions has its shortcomings. 
The presence of infrared (IR) divergence in the position space representation of the Wightman 
function makes considering the linear coupling model dependent on the IR cutoff in most cases. For 
instance, one should note that working with linear couplings with detectors switched on for a finite 
time; one cannot avoid the contributions of the IR cutoff. It compels one to consider the derivative 
coupling model provided in \cite{Tjoa:2022oxv}, which gives the Wightman function corresponding to 
typical Hadamard asymptotics. However, in a few cases, it has been observed that the contributions 
from the IR cutoff may vanish for infinite switching. The familiar one is the infinite time 
detector transition probabilities which are free of IR cutoff due to the appearance of Dirac delta 
distribution with a positive argument (for more examples, see \cite{Chowdhury:2021ieg}). Moreover, 
this infinite switching model has also been traditionally considered in various related previous 
works. In this spirit, we consider the same in our work. In order to circumvent the issues from the 
IR divergence, we assessed the necessary quantities for fixed field mode frequency $(\omega_{k})$, 
giving transition probability corresponding to a specific frequency. Similar attempts have also been 
taken in \cite{Scully:2017utk} to calculate the detector response.

\item Here the possibility of entanglement harvesting has been checked through the condition 
(\ref{eq:cond-entanglement}). Note that this condition depends on both the local 
($\mathcal{I}_{j_{\omega_{k}}}$) and non-local terms ($\mathcal{I}_{\varepsilon_{\omega_{k}}}$). 
This condition says that the entanglement harvesting is possible only when 
$|\mathcal{I}_{\varepsilon_{\omega_{k}}}| 
-(\mathcal{I}_{A_{\omega_{k}}}\,\mathcal{I}_{B_{\omega_{k}} })^{1/2}$ is positive. Notably, the 
concurrence, which has been considered here for the measure of entanglement harvesting, depends on 
the aforesaid difference. Moreover, $\mathcal{I}_{j_{\omega_{k}}}$ denotes the individual detector 
transition probability. Therefore, the competition between the non-local terms and the local terms 
plays a vital role in the measurement of harvesting. In this sense, the perception of the particle 
by the individual detectors has a significant role in this phenomenon.
It should also be noted that the local terms $\mathcal{I}_{j_{\omega_{k}}}$ denoting 
individual detector transition probabilities in all of the above cases do not depend on the distance 
$d$. Therefore, one concludes that the occurrence of the oscillations in the concurrence with 
respect to the distance $d$ is due to the nonlocal entangling term 
$|\mathcal{I}_{\varepsilon_{\omega_{k}}}|$.

\item So far, the examples we have considered in our study are the Schwarzschild background with an 
event horizon and the de Sitter background without an event horizon. In both cases, we have 
perceived the occurrence of entanglement harvesting shadow. Now since $(1+1)$ dimensional 
Schwarzschild background near the event horizon behaves like a Rindler frame, one can expect similar 
results in the later frame as well. Below we show two such instances of which one will mimic the 
$(1+1)$ dimensional de Sitter, and the other will mimic the Schwarzschild case.

First, we consider the line element in Rindler coordinates as given by $ds^2 = 
e^{2a\xi}(-d\eta^2+d\xi^2)$. In this case, if the Minkowski vacuum is taken as the conformal vacuum, 
then the null paths $\eta-\xi=d$ followed by detector $A$ and $\eta-\xi=0$ followed by detector $B$ 
will provide the necessary integrals for estimating the concurrence. These integrals look similar to 
those in $(1+1)$ de Sitter spacetime. Therefore one could expect similar features in concurrence. 

Second, we consider the line element in Rindler coordinates as given by $ds^2 = -2ax\, 
dt^2+dx^2/(2ax)$, which is reminiscent of the near horizon $(1+1)$ dimensional Schwarzschild 
spacetime. Here one can construct a tortoise-like coordinate $(dx_{\star}=dx/(2ax))$ transformation 
which makes the metric conformally flat. In these coordinates $(t, x_{\star})$, the defined vacuum 
for the fields will be identical to the Boulware vacuum. Therefore we consider our detectors to be 
moving in null paths in the Eddington-Finkelstein-like coordinates for the Rindler one; then, the 
situation will be quite identical to the already obtained results for the $(1+1)$ Schwarzschild 
background.

It may be interesting to note that the entanglement shadow regions occur for both the black hole and 
the de Sitter spacetimes. Also, we found that the same can occur for the Rindler spacetime as well. 
A few things can be noted related to these examples, like de Sitter universe does not contain an 
event horizon, whereas Rindler spacetime is curvature less. Therefore the appearance of the 
entanglement shadow might be dependent on the choices of our paths and the background field vacua.

\item In $(1+1)$ dimensional de Sitter background, we have observed that with increasing field mode frequency $\overline{\omega_{k}}$ the length of the shadow region decreases. Moreover, for a finite value of $\overline{ \omega_{k}}$ (in a range $\overline{ \omega_{k}}\in [10^{-4},10^{4}]$) the lower kinks in the curves remain below the zero concurrence line. Therefore, in this case, we always have shadow regions. Hence, one cannot find a parameter space for which entanglement harvesting is always or never possible. In the $(1+1)$ Schwarzschild and $(1+3)$ de Sitter background spacetimes, we found that it is not always point-like entanglement shadows. The occurrence of entanglement harvesting shadow regions is possible in relatively low frequency $\overline{\omega_{k}}$ regimes. With increasing $\overline{\omega_{k}}$, these shadow regions decrease and become shadow points. However, increasing the frequency further does not make these shadow points go away, i.e., we keep getting entanglement harvesting shadow points for larger and larger $\overline{\omega_{k}}$ values. In this regard, we have considered the frequency range up to $10^{4}$. However, the similarity between $(1+1)$ Schwarzschild and $(1+3)$ de Sitter may occur in different frequency ranges.

The above observations show that $(1+1)$ dimensional Schwarzschild and $(1+3)$ dimensional de Sitter backgrounds have a close resemblance between them in terms of the shadows. In contrast, the $(1+1)$ dimensional de Sitter case does not show such similarity. The possible reason behind this may be as follows. Note that the concurrence is dependent on both the non-local entangling term  $\mathcal{I}_{\varepsilon_{\omega_{k}}}$ and the local term $\mathcal{I}_{j_{\omega_{k}}}$. Therefore, the shadow happens only if $\sqrt{\mathcal{I}_{A_{\omega_{k}}} \mathcal{I}_{B_{\omega_{k}}}}\ge |\mathcal{I}_{\varepsilon_{\omega_{k}}}|$. Interestingly, in $(1+1)$ dimensional de Sitter case $\mathcal{I}_{j_{\omega_{k}}}$ are independent of $\omega_{k}$, see Eq. (\ref{eq:Ij-DS-1p1-5}). While in the other two spacetimes, the same is $\omega_{k}$ dependent, see Eqs. (\ref{eq:Ij-Sch-BV-6}) and (\ref{eq:Ij-DS-1p3-4}). It implies that the subtracted term $\sqrt{\mathcal{I}_{A_{\omega_{k}}} \mathcal{I}_{B_{\omega_{k}}}}$ in the concurrence changes for $(1+1)$ Schwarzschild and $(1+3)$ de Sitter cases as one changes $\omega_{k}$. On the other hand, for the $(1+1)$ de Sitter case, the same does not happen. Hence, it may be the reason that in the first scenario, we have a change in shadow from regions to points as one increases $\overline{\omega_{k}}$. Since this $\omega_{k}$ is a property of the field, the above two scenarios mostly depend on the particular field mode with which the detectors are interacting. Therefore, we feel that these observations are based on the underlying properties of the quantum fields rather than the background geometry. Of course, further investigation is needed to provide a concrete reason in favour of the above scenarios. Also, it would be interesting to provide a detailed analysis in finding a condition that will dictate the critical value of $\overline{\omega_{k}}$ above which the shadow region becomes a point.

\end{itemize}

\begin{acknowledgments}
SB would like to thank the Indian Institute of Technology Guwahati (IIT Guwahati) for financial support.
DB would like to acknowledge Ministry of Education, Government of India for providing financial support for his research via the PMRF May 2021 scheme. The research of BRM is partially supported by a START-UP RESEARCH GRANT (No. SG/PHY/P/BRM/01) from the Indian Institute of Technology Guwahati, India. We thank the anonymous referee for the crucial suggestions that have helped to improve the manuscript.
\end{acknowledgments}

\appendix

\section{Evaluation of the integrals $\mathcal{I}^{W}_{\varepsilon_{\omega_{k}}}$ and 
$\mathcal{I}^{R}_{\varepsilon_{\omega_{k}}}$ in Schwarzschild black hole 
spacetime}\label{Appn:IW-IR-Sch}

\subsection{Boulware vacuum}\label{Appn:IW-IR-BV}

We introduce regulator of the form $(y_{A}y_{B})^{\epsilon}\,e^{-\epsilon 
(y_{A}+y_{B})}$ to evaluate the integral $\mathcal{I}^{W}_{\varepsilon_{\omega_{k}}}$ from Eq. 
(\ref{eq:IeW-Sch-BV-3}) corresponding to two detectors in outgoing null trajectories in an $(1+1)$ 
dimensional Schwarzschild black hole spacetime. In particular, this integral is evaluated as
\begin{eqnarray}\label{eq:IeW-Sch-BV-4}
 \mathcal{I}^{W}_{\varepsilon_{\omega_{k}}} &=& \frac{e^{i d \omega_{k}}\left(4 \rh^2 
\omega_{k}^2+9 \epsilon ^2\right)}{\rh^2 \left(4 
\omega_{k}^2-\Delta E^2\right)-2 i \Delta E \rh \epsilon +\epsilon ^2}\, \nonumber\\
~&& (\epsilon +i 
\rh (2 \omega_{k}-\Delta E))^{-\epsilon 
+2 i \rh (\omega_{k}-\Delta E)} \nonumber\\
~&& (\epsilon -i \rh (\Delta E+2 
\omega_{k}))^{-\epsilon -2 i \rh (\Delta E+\omega_{k})} \nonumber\\
~&& \Gamma (\epsilon -2 i \rh 
(\omega_{k}-\Delta E)) \Gamma (2 i \rh (\omega_{k}+\Delta E)+\epsilon 
)~.\nonumber\\
\end{eqnarray}
On the other hand, using same regulator of the form $(z_{A}z_{B})^{\epsilon}\,e^{-\epsilon 
(z_{A}+z_{B})}$ with small positive real parameter $\epsilon$ the integral 
$\mathcal{I}^{R}_{\varepsilon_{\omega_{k}}}$ from Eq. (\ref{eq:IeR-Sch-BV-4}) is evaluated as
\begin{widetext}
\begin{eqnarray}\label{eq:IeR-Sch-BV-5}
\mathcal{I}^{R}_{\varepsilon_{\omega_{k}}} &=& -\frac{e^{-i d \omega_{k}} \left(-1+e^{2 i d 
\omega_{k}}\right)}{\left(4 
\omega_{k}^2-\Delta E^2\right) \rh^2-2 i \Delta E \epsilon  \rh+\epsilon 
^2}\, (i \rh (2 \omega_{k}-\Delta E)+\epsilon )^{2 i \rh 
(\omega_{k}-\Delta E)-\epsilon } (\epsilon -i \rh (2 \omega_{k}+\Delta E))^{-2 i \rh 
(\omega_{k}+\Delta E)-\epsilon } \nonumber\\
~&& \times~\left(4 \rh^2 \omega_{k}^2+9 \epsilon ^2\right) \Gamma 
(\epsilon -2 
i \rh (\omega_{k}-\Delta E)) \Gamma (2 i \rh (\omega_{k}+\Delta E)+\epsilon )\nonumber\\
~&& +\,\Bigg[-\frac{4 i e^{i d \omega_{k}} (i \rh (2 \omega_{k}-\Delta E)+\epsilon )^{-2 (2 i \rh 
\Delta E+\epsilon )}}{2 \rh (\omega_{k}+\Delta E)-i \epsilon }\nonumber\\
~&& \times\, _2F_1\left(2 (2 i \rh \Delta E+\epsilon ),2 i 
\rh (\omega_{k}+\Delta E)+\epsilon ;2 i \rh (\omega_{k}+\Delta E)+\epsilon 
+1;\frac{2 \rh \omega_{k}+\rh \Delta E+i \epsilon }{2 \rh 
\omega_{k}-\rh \Delta E-i \epsilon }\right)\nonumber\\
~&& +\frac{4 e^{-i d 
\omega_{k}} (2 i \rh \Delta E+\epsilon ) (\epsilon -i \rh (2 \omega_{k}+\Delta E))^{-2 (2 i 
\rh \Delta E+\epsilon )} }{(2 \rh (\omega_{k}-\Delta E)+i \epsilon ) (\rh (2 
\omega_{k}+\Delta E)+i \epsilon )}\nonumber\\
~&& \times\, _2F_1\left(\epsilon -2 i \rh 
(\omega_{k}-\Delta E),4 i \rh \Delta E+2 \epsilon +1;-2 i \rh 
(\omega_{k}-\Delta E)+\epsilon +1;\frac{2 \rh \omega_{k}-\rh \Delta E-i \epsilon }{2 \rh 
\omega_{k}+\rh \Delta E+i \epsilon }\right)
\nonumber\\
~&& -\frac{4 e^{-i d \omega_{k}} (2 i \rh \Delta 
E+\epsilon ) (\epsilon -i \rh (2 \omega_{k}+\Delta E))^{-4 i \rh \Delta E-2 \epsilon -1} }{-2 i \rh 
(\omega_{k}-\Delta E)+\epsilon 
+1}\nonumber\\
~&& \times\, 
_2F_1\left(-2 i \rh (\omega_{k}-\Delta E)+\epsilon +1,4 i \rh \Delta E+2 \epsilon +1;-2 i \rh 
(\omega_{k}-\Delta E)+\epsilon +2;\frac{2 \rh \omega_{k}-\rh \Delta E-i \epsilon }{2 \rh 
\omega_{k}+\rh \Delta E+i \epsilon }\right)\nonumber\\
~&& -\frac{2 e^{i d \omega_{k}} (2 \rh \Delta E-i \epsilon ) (i \rh (2 \omega_{k}-\Delta 
E)+\epsilon )^{-4 i \rh \Delta E-2 (\epsilon +1)} (4 \rh \Delta E-i (2 \epsilon +1)) }{2 i \rh 
(\omega_{k}+\Delta E)+\epsilon 
+1}\nonumber\\
~&& \times\, _2F_1\left(2 
(2 i \rh \Delta E+\epsilon +1),2 i \rh (\omega_{k}+\Delta E)+\epsilon +1;2 i \rh (\omega_{k}+\Delta 
E)+\epsilon +2;\frac{2 \rh \omega_{k}+\rh \Delta E+i \epsilon }{2 \rh 
\omega_{k}-\rh \Delta E-i \epsilon }\right)\nonumber\\
~&& -\frac{4 i e^{i d \omega_{k}} (i \rh (2 \omega_{k}-\Delta E)+\epsilon )^{-4 i \rh 
\Delta E-2 \epsilon -1} (2 i \rh \Delta E+\epsilon ) }{2 \rh (\omega_{k}+\Delta E)-i \epsilon 
}\nonumber\\
~&& \times\, _2F_1\left(2 i \rh (\omega_{k}+\Delta 
E)+\epsilon ,4 i \rh \Delta E+2 \epsilon +1;2 i \rh (\omega_{k}+\Delta E)+\epsilon +1;\frac{2 
\rh \omega_{k}+\rh \Delta E+i \epsilon }{2 \rh \omega_{k}-\rh 
\Delta E-i \epsilon }\right)\nonumber\\
~&& +\frac{4 e^{i d \omega_{k}} (i \rh (2 \omega_{k}-\Delta E)+\epsilon )^{-4 i 
\rh \Delta E-2 \epsilon -1} (2 i \rh \Delta E+\epsilon ) }{2 i \rh (\omega_{k}+\Delta E)+\epsilon 
+1}\nonumber\\
~&& \times\, 
_2F_1\left(2 i \rh (\omega_{k}+\Delta E)+\epsilon +1,4 i \rh \Delta E+2 \epsilon +1;2 i \rh 
(\omega_{k}+\Delta E)+\epsilon +2;\frac{2 \rh \omega_{k}+\rh \Delta E+i \epsilon }{2 \rh 
\omega_{k}-\rh \Delta E-i \epsilon }\right)\nonumber\\
~&& +\frac{8 i (\epsilon -i \rh \Delta E)^{-4 i \rh \Delta E-2 \epsilon } \, _2F_1(2 
i \rh \Delta E+\epsilon ,2 (2 i \rh \Delta E+\epsilon );2 i \rh 
\Delta E+\epsilon +1;-1) \sin (d \omega_{k})}{2 i \rh \Delta E+\epsilon 
}\nonumber\\
~&& +\frac{8 i (\epsilon -i \rh \Delta E)^{-4 i \rh \Delta E-2 \epsilon 
-1} (2 i \rh \Delta E+\epsilon )  
\sin (d \omega_{k})}{2 i \rh \Delta E+\epsilon +1}\nonumber\\
~&& \times~\, _2F_1(2 i \rh \Delta E+\epsilon 
+1,4 i \rh \Delta E+2 \epsilon +1;2 i \rh \Delta E+\epsilon +2;-1)\nonumber\\
~&& +\frac{4 (2 \rh 
\Delta E-i \epsilon ) (\epsilon -i \rh \Delta E)^{-4 i \rh 
\Delta E-2 \epsilon -1} (4 \rh \Delta E-i (2 \epsilon +1)) \sin (d \omega_{k})}{(\rh \Delta E+i 
\epsilon ) (2 i \rh \Delta E+\epsilon +1)}\nonumber\\
~&& \times\, _2F_1(2 i 
\rh \Delta E+\epsilon +1,2 (2 i \rh \Delta E+\epsilon +1);2 i 
\rh \Delta E+\epsilon +2;-1)\nonumber\\
~&& -\frac{4 e^{-i d \omega_{k}} (\epsilon -i 
\rh (2 \omega_{k}+\Delta E))^{-4 i \rh \Delta E-2 \epsilon }}{\epsilon -2 i \rh (\omega_{k}-\Delta 
E)}\nonumber\\
~&& \times\, 
_2F_1\left(\epsilon -2 i \rh (\omega_{k}-\Delta E),2 (2 i \rh \Delta E+\epsilon );-2 i \rh 
(\omega_{k}-\Delta E)+\epsilon +1;\frac{2 \rh \omega_{k}-\rh \Delta E-i \epsilon }{2 \rh 
\omega_{k}+\rh \Delta E+i \epsilon }\right)\nonumber\\
~&& -\frac{2 e^{-i d \omega_{k}} (2 \rh \Delta E-i \epsilon ) (\epsilon -i \rh (2 
\omega_{k}+\Delta E))^{-2 (2 i \rh \Delta E+\epsilon )} (4 \rh 
\Delta E-i (2 \epsilon +1))}{(\rh (2 \omega_{k}+\Delta E)+i 
\epsilon )^2 (-2 i \rh (\omega_{k}-\Delta E)+\epsilon +1)}\nonumber\\
~&& \times\, _2F_1\left(-2 i \rh (\omega_{k}-\Delta E)+\epsilon +1,2 (2 i \rh 
\Delta E+\epsilon +1);-2 i \rh (\omega_{k}-\Delta E)+\epsilon +2;\frac{2 \rh \omega_{k}-\rh \Delta 
E-i \epsilon }{2 \rh \omega_{k}+\rh \Delta E+i \epsilon }\right)\nonumber\\
~&& -\frac{8 (\epsilon -i \rh 
\Delta E)^{-2 (2 i \rh \Delta E+\epsilon )} \, _2F_1(2 i \rh \Delta E+\epsilon ,4 i \rh \Delta E+2 
\epsilon +1;2 i \rh \Delta E+\epsilon +1;-1) \sin (d \omega_{k})}{\rh \Delta E+i \epsilon }\Bigg] 
\Gamma (4 i \rh \Delta E+2 \epsilon )\nonumber\\
~&& +\,\frac{2 (\epsilon -i \rh \Delta E)^{-2 (2 i \rh \Delta E+\epsilon 
)} \left(-8 i \rh^2 \Delta E^2+4 \rh \epsilon  \Delta E+5 i \epsilon ^2\right) \Gamma (2 i \rh 
\Delta E+\epsilon )^2 \sin (d \omega_{k})}{(\rh \Delta E+i \epsilon )^2}\nonumber\\
~&& +\,\frac{8 (\epsilon -i \rh 
\Delta E)^{-2 (2 i \rh \Delta E+\epsilon )} \Gamma (2 i \rh 
\Delta E+\epsilon ) \Gamma (2 i \rh \Delta E+\epsilon +1) \sin (d 
\omega_{k})}{\rh \Delta E+i \epsilon }~.
\end{eqnarray}
\end{widetext}
For both $\mathcal{I}^{W}_{\varepsilon_{\omega_{k}}}$ and 
$\mathcal{I}^{R}_{\varepsilon_{\omega_{k}}}$ we have considered the evaluation for $\Delta E^{A} = 
\Delta E^{B} = \Delta E$. Here the functions $\Gamma(x)$ and $~_2F_1(x)$ respectively denote 
the \emph{Gamma functions} and the \emph{Hypergeometric functions}.

\subsection{Unruh vacuum}\label{Appn:IW-IR-Unruh}

One can introduce regulator of the form $(y_{A}y_{B})^{\epsilon}\,e^{-\epsilon 
(y_{A}+y_{B})}$ to evaluate the integral $\mathcal{I}^{W}_{\varepsilon_{\omega_{k}}}$ from Eq. 
(\ref{eq:IeW-Sch-UV-1}) corresponding to detectors in outgoing null trajectories in an $(1+1)$ 
dimensional Schwarzschild black hole spacetime with Unruh vacuum. In particular, this integral 
has the same expression as provided in Eq. (\ref{eq:IeW-Sch-BV-4}).

%
\begin{figure}[h]
\centering
 \includegraphics[width=0.85\linewidth]{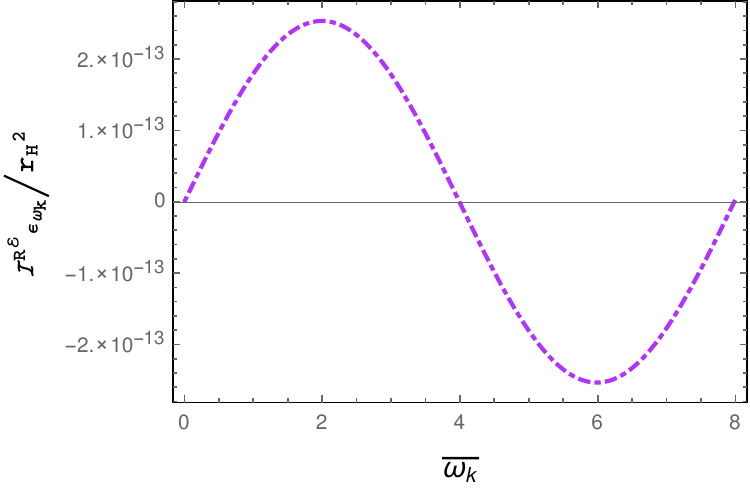}
 \caption{The quantity $\mathcal{I}^{R^{\mathcal{E}}}_{\varepsilon_{\omega_{k}}}$ is plotted with 
respect to $\overline{\omega}_{k}$ for fixed $d/\rh=1$, $\overline{\Delta E} = 
1$.}
 \label{fig:IeRE-Schwarzschild-d1-dwk}
\end{figure}

On the other hand, utilizing the same regulator of the form $(y_{A}y_{B})^{\epsilon}\,e^{-\epsilon 
(y_{A}+y_{B})}$ one can also proceed to evaluate the integral 
$\mathcal{I}^{R}_{\varepsilon_{\omega_{k}}}$ from Eq. (\ref{eq:IeR-Sch-UV-1}). In particular, one 
can observe that this integral is different from Eq. (\ref{eq:IeR-Sch-BV-3}) of the Boulware case in 
only the last term with a factor of $-2i\sin\{\omega_{k}2\rh( 1-\exp{(-d/2\rh)})\}$. Let us term 
this quantity to be $\mathcal{I}^{R^{\mathcal{E}}}_{\varepsilon_{\omega_{k}}}$ and evaluate this 
integral
\begin{eqnarray}\label{eq:IeR-Sch-UV-2}
 \mathcal{I}^{R^{\mathcal{E}}}_{\varepsilon_{\omega_{k}}} &=& -2i\sin\Big\{\omega_{k}2\rh\big( 
1-e^{-d/2\rh}\big)\Big\} \int_{\rh}^{\infty} 
dr_{A}\, \frac{r_{A}+\rh}{r_{A}-\rh}\nonumber\\
~&& \times\int_{\rh}^{r_{A}} 
dr_{B}\, \frac{r_{B}+\rh}{r_{B}-\rh}~ e^{i\{\Delta 
E^{B} r_{B}+\Delta 
E^{A} (r_{A}+d)\}}~ \nonumber\\
~&& \times~\bigg(\frac{r_{B}}{\rh}-1\bigg)^{2i\rh\Delta 
E^{B}} \bigg(\frac{r_{A}}{\rh}-1\bigg)^{2i\rh\Delta 
E^{A}}~.
\end{eqnarray}
This integral can be evaluated with regulator $(y_{A}y_{B})^{\epsilon}\,e^{-\epsilon 
(y_{A}+y_{B})}$ and is obtained as
\begin{widetext}
\begin{eqnarray}\label{eq:IeR-Sch-UV-3}
 \mathcal{I}^{R^{\mathcal{E}}}_{\varepsilon_{\omega_{k}}} &=& -2\sin\Big\{\omega_{k}2\rh\big( 
1-e^{-d/2\rh}\big)\Big\} (\epsilon -i \Delta E \rh)^{-4 i \Delta E 
\rh-2 \epsilon -1} \nonumber\\
~&& \Bigg[2 \bigg\{-\frac{2 (\Delta E \rh+i \epsilon ) \, _2F_1(2 
i \rh \Delta E+\epsilon ,2 (2 i \rh \Delta E+\epsilon );2 i \rh 
\Delta E+\epsilon +1;-1)}{\epsilon +2 i \Delta E \rh}\nonumber\\
~&& -2 i \, _2F_1(2 i 
\rh \Delta E+\epsilon ,4 i \rh \Delta E+2 \epsilon +1;2 i \rh 
\Delta E+\epsilon +1;-1)\nonumber\\
~&& +\frac{(4 i \Delta E \rh+2 \epsilon +1) (2 
\Delta E \rh-i \epsilon ) \, _2F_1(2 i \rh \Delta E+\epsilon +1,2 (2 i 
\rh \Delta E+\epsilon +1);2 i \rh \Delta E+\epsilon 
+2;-1)}{(\Delta E \rh+i \epsilon ) (2 \Delta E \rh-i (\epsilon 
+1))}\nonumber\\
~&& -\frac{2 i (2 \Delta E \rh-i \epsilon ) \, _2F_1(2 i \rh \Delta E+\epsilon +1,4 i 
\rh \Delta E+2 \epsilon +1;2 i \rh \Delta E+\epsilon +2;-1)}{2 \Delta E 
\rh-i (\epsilon +1)}\bigg\} \Gamma (4 i \rh \Delta E+2 \epsilon )\nonumber\\
~&& +\frac{\left(8 \Delta 
E^2 \rh^2+4 i \Delta E \rh \epsilon -5 \epsilon ^2\right) \Gamma (2 i \rh \Delta 
E+\epsilon )^2}{\Delta E \rh+i \epsilon }+4 i \Gamma (2 i \rh \Delta E+\epsilon 
+1) \Gamma (2 i \rh \Delta E+\epsilon )\Bigg]~.
\end{eqnarray}
\end{widetext}
One can numerically plot this quantity for similar parameter values for which the plots of 
$\mathcal{I}^{R}_{\varepsilon_{\omega_{k}}}$ are performed and observe that 
$\mathcal{I}^{R^{\mathcal{E}}}_{\varepsilon_{\omega_{k}}}$ is many order lower, see. Fig. 
\ref{fig:IeRE-Schwarzschild-d1-dwk}.

\section{Heaviside step function corresponding to the Unruh 
modes}\label{Appn:Stepfn-simplification-Unruh}

The Kruskal null coordinates $(V,\,U)$ are related to the null coordinates $(v,\,u)$ 
by the relation $V = 2\rh\,e^{v/2\rh}$ and $U = -2\rh\,e^{-u/2\rh}$. Whereas these coordinates $(v,\,u)$ are 
again related to the Schwarzschild time and the tortoise coordinates as $v= \ts+\rstar$ and $u= 
\ts-\rstar$. We have already stated that while dealing with modes represented in terms of the 
Kruskal coordinates one should consider the Kruskal time $T_{K} = (U+V)/2$. This time is represented 
in terms of the Schwarzschild time and the tortoise coordinates as $T_{K} = 
2\rh\,e^{\rstar/2\rh}\,\sinh{(\ts/2\rh)}$. We have considered that Alice and Bob denoted 
respectively by the detectors $A$ and $B$ are both moving along outgoing null trajectories. However, 
for Alice the outgoing null path is  $t_{s_{A}}-r_{\star_{A}} = d$, while for Bob the path is 
$t_{s_{B}}-r_{\star_{B}} = 0$. With these conditions let us search for the situation when the 
Heaviside step function $\theta(T_{K_A}-T_{K_B})$ will be non zero.

For the above mentioned Heaviside step function to be non zero one must have $T_{K_A}\ge T_{K_B}$. 
In terms of the Schwarzschild time and the tortoise coordinate and using the appropriate 
prescription of the null paths for the detectors $A$ and $B$ this condition becomes
\begin{eqnarray}\label{eq:cond-NZ-ThetaFn-Unruh}
 e^{(2r_{\star_{A}}+d)/2\rh}-e^{r_{\star_{B}}/\rh} \ge e^{-d/2\rh}-1\,.
\end{eqnarray}
Here we have considered $d$ to be positive and real. In that case the maximum value of the right hand side of the above inequality is zero and therefore the above will be automatically satisfied if one has 
\begin{equation}
e^{(2r_{\star_{A}}+d)/2\rh}-e^{r_{\star_{B}}/\rh} \ge 0~.
\end{equation}
This can be re-expressed as $e^{(r_{\star_{A}}- r_{\star_{B}})/\rh} \ge e^{-(d/2r_H)}$. Now again as $0\le e^{-(d/2r_H)}\leq 1$, the required condition will be satisfied if
one has 
\begin{equation}
e^{(r_{\star_{A}}- r_{\star_{B}})/\rh} \ge 1~.
\end{equation}
This basically implies that one must have $r_{\star_{A}} \ge r_{\star_{B}}$ or $r_{A} \ge r_{B}$. Note that this very condition has been considered in our main analysis.

\section{Evaluation of the integrals $\mathcal{I}^{W}_{\varepsilon_{\omega_{k}}}$ and 
$\mathcal{I}^{R}_{\varepsilon_{\omega_{k}}}$ in de Sitter spacetime}\label{Appn:IW-IR-DS}

\subsection{$(1+1)$ dimensions}\label{Appn:IW-IR-DS-1p1}

With the introduction of regulator of the form $(z_{A}z_{B})^{\epsilon}\,e^{-\epsilon 
(z_{A}+z_{B})}$ the integral $\mathcal{I}^{W}_{\varepsilon_{\omega_{k}}}$ from Eq. 
(\ref{eq:IeW-DS-1p1-2}) corresponding to two detectors in outgoing null trajectories in an $(1+1)$ 
dimensional de Sitter spacetime can be evaluated to be
\begin{eqnarray}\label{eq:IeW-DS-1p1-3}
 \mathcal{I}^{W}_{\varepsilon_{\omega_{k}}} &=& e^{-i d \omega_{k}} \Gamma \big(\epsilon -i 
\alpha_d \Delta E\big)^2 (\epsilon -2 i \alpha_d \omega_{k})^{-\epsilon +i 
\alpha_d \Delta E} \nonumber\\
~&& ~~~~~~~\times~(\epsilon +2 i \alpha_d \omega_{k})^{-\epsilon +i 
\alpha_d \Delta E}~.
\end{eqnarray}
On the other hand, using same regulator of the form $(z_{A}z_{B})^{\epsilon}\,e^{-\epsilon 
(z_{A}+z_{B})}$ with small positive real parameter $\epsilon$ the integral 
$\mathcal{I}^{R}_{\varepsilon_{\omega_{k}}}$ from Eq. (\ref{eq:IeR-DS-1p1-1}) is evaluated as
\begin{widetext}
\begin{eqnarray}\label{eq:IeR-DS-1p1-2}
 \mathcal{I}^{R}_{\varepsilon_{\omega_{k}}} &=& \Gamma (\epsilon -i \alpha_d \Delta E) \Bigg[\Gamma 
(2 \epsilon -2 i \alpha_d \Delta E) \bigg\{e^{i d \omega_{k}} (\epsilon +2 i \alpha_d 
\omega_{k})^{-2 \epsilon +2 i \alpha_d \Delta E} \,\nonumber\\
~&&  _2\tilde{F}_1\left(\epsilon -i \alpha_d \Delta 
E,2 (\epsilon -i \alpha_d \Delta E);-i \alpha_d \Delta E+\epsilon +1;\frac{4 i \omega_{k} 
\alpha_d}{2 i \omega_{k} \alpha_d+\epsilon }-1\right) -e^{-i d \omega_{k}} (\epsilon -2 i 
\alpha_d \omega_{k})^{-2 \epsilon +2 i \alpha_d \Delta E} \, \nonumber\\
~&& _2\tilde{F}_1\left(\epsilon -i 
\alpha_d \Delta E,2 (\epsilon -i \alpha_d \Delta E);-i \alpha_d \Delta E+\epsilon +1;\frac{4 
\omega_{k} \alpha_d}{2 \omega_{k} \alpha_d+i \epsilon }-1\right)\bigg\} -i \epsilon ^{-2 
\epsilon +2 i \alpha_d \Delta E} \sin (d \omega_{k}) \Gamma (\epsilon -i 
\alpha_d \Delta E)\Bigg]~.\nonumber\\
\end{eqnarray}
\end{widetext}
For both $\mathcal{I}^{W}_{\varepsilon_{\omega_{k}}}$ and 
$\mathcal{I}^{R}_{\varepsilon_{\omega_{k}}}$ we have considered the evaluation for $\Delta E^{A} = 
\Delta E^{B} = \Delta E$. Here the functions $\Gamma(x)$ and $~_2\tilde{F}_1(x)$ respectively denote 
the \emph{Gamma functions} and the \emph{regularized Hypergeometric functions}.

\subsection{$(1+3)$ dimensions}\label{Appn:IW-IR-DS-1p3}

With the introduction of regulator of the form $(z_{A}z_{B})^{\epsilon}\,e^{-\epsilon 
(z_{A}+z_{B})}$ the integral $\mathcal{I}^{W}_{\varepsilon_{\omega_{k}}}$ from Eq. 
(\ref{eq:IeW-DS-1p3-2}) corresponding to two detectors in outgoing null trajectories in an $(1+3)$ 
dimensional de Sitter spacetime can be evaluated to be
\begin{eqnarray}\label{eq:IeW-DS-1p3-3}
&& \mathcal{I}^{W}_{\varepsilon_{\omega_{k}}} = e^{-i d k_{x}} \Gamma (-i \alpha_d 
\Delta E+\epsilon +1)^2 \nonumber\\
~&& \times~(\epsilon -i \alpha_d (k_{x}+\omega_{k}))^{i 
\alpha_d \Delta E-\epsilon -1} \nonumber\\
~&& (\epsilon +i \alpha_d 
(k_{x}+\omega_{k}))^{i \alpha_d \Delta E-\epsilon -1} + e^{i d k_{x}} \Gamma 
(-i \alpha_d \Delta E+\epsilon +1)^2 \nonumber\\
~&& (\epsilon -i \alpha_d 
(k_{x}-\omega_{k}))^{i \alpha_d \Delta E-\epsilon -1} (i \alpha_d 
k_{x}-i \alpha_d \omega_{k}+\epsilon )^{i \alpha_d \Delta E-\epsilon 
-1}~.\nonumber\\
\end{eqnarray}
On the other hand, using same regulator of the form $(z_{A}z_{B})^{\epsilon}\,e^{-\epsilon 
(z_{A}+z_{B})}$ with small positive real parameter $\epsilon$ the integral 
$\mathcal{I}^{R}_{\varepsilon_{\omega_{k}}}$ from Eq. (\ref{eq:IeR-DS-1p3-1}) is evaluated as
\begin{widetext}
\begin{eqnarray}\label{eq:IeR-DS-1p3-2}
 \mathcal{I}^{R}_{\varepsilon_{\omega_{k}}} &=& \frac{e^{-i d k_{x}}}{-i 
\alpha_d \Delta E+\epsilon +1}\, \Gamma (-2 i \alpha_d \Delta 
E+2 \epsilon +2) \Bigg[(\epsilon -i \alpha_d (k_{x}-\omega_{k}))^{-2 (\epsilon +1)+2 i \alpha_d 
\Delta E} \nonumber\\
~&& \times\, _2F_1\left(-i \alpha_d \Delta E+\epsilon +1,-2 i \alpha_d \Delta E+2 
\epsilon +2;-i 
\alpha_d \Delta E+\epsilon +2;\frac{k_{x} \alpha_d-\omega_{k} 
\alpha_d-i \epsilon }{k_{x} \alpha_d-\omega_{k} \alpha_d+i \epsilon 
}\right)\nonumber\\
~&& -e^{2 i d k_{x}} (\epsilon +i \alpha_d (k_{x}-\omega_{k}))^{-2 (\epsilon 
+1)+2 i \alpha_d \Delta E} \nonumber\\
~&& \times\, _2F_1\left(-i \alpha_d \Delta E+\epsilon +1,-2 i \alpha_d \Delta E+2 
\epsilon +2;-i \alpha_d \Delta E+\epsilon +2;\frac{k_{x} \alpha_d-\omega_{k} \alpha_d+i 
\epsilon }{k_{x} \alpha_d-\omega_{k} \alpha_d-i \epsilon }\right)\Bigg]\nonumber\\
~&& + e^{-i d k_{x}} \Gamma (-i \alpha_d 
\Delta E+\epsilon +1) \Gamma (-2 i \alpha_d \Delta E+2 \epsilon +2) 
\Bigg[e^{2 i d k_{x}} (\epsilon +i \alpha_d (k_{x}+\omega_{k}))^{-2 (\epsilon +1)+2 
i \alpha_d \Delta E} \nonumber\\
~&& \times\, _2\tilde{F}_1\left(-i \alpha_d \Delta E+\epsilon +1,2 (-i \alpha_d 
\Delta E+\epsilon +1);-i \alpha_d \Delta E+\epsilon +2;\frac{(k_{x}+\omega_{k}) \alpha_d+i \epsilon 
}{(k_{x}+\omega_{k}) \alpha_d-i \epsilon }\right)\nonumber\\
~&& -(\epsilon -i \alpha_d 
(k_{x}+\omega_{k}))^{-2 (\epsilon +1)+2 i \alpha_d \Delta E} \nonumber\\
~&& \times\, 
_2\tilde{F}_1\left(-i \alpha_d \Delta E+\epsilon +1,2 (-i \alpha_d 
\Delta E+\epsilon +1);-i \alpha_d \Delta E+\epsilon 
+2;\frac{(k_{x}+\omega_{k}) \alpha_d-i \epsilon }{(k_{x}+\omega_{k}) \alpha_d+i \epsilon 
}\right)\Bigg]~.
\end{eqnarray}
\end{widetext}
Here also we have considered the specific scenario $\Delta E^{A} = \Delta E^{B} = \Delta E$ for 
evaluating both $\mathcal{I}^{W}_{\varepsilon_{\omega_{k}}}$ and 
$\mathcal{I}^{R}_{\varepsilon_{\omega_{k}}}$, and the functions $\Gamma(x)$, $~_2F_1(x)$ and 
$~_2\tilde{F}_1(x)$ respectively denote the \emph{Gamma functions}, the \emph{Hypergeometric 
functions}, and the \emph{regularized Hypergeometric functions}.

\bibliographystyle{apsrev}

\bibliography{bibtexfile}

\end{document}